\documentclass[twocolumn, twocolappendix]{aastex63}
\usepackage{amsmath, amssymb}
\usepackage{chngcntr}
\usepackage{scrextend}

\newcommand{\ha}{\hbox{H$\alpha$}}
\newcommand{\hb}{\hbox{H$\beta$}}
\newcommand{\hii}{\hbox{H\,{\sc ii}}}
\newcommand{\oiii}{\hbox{[O\,{\sc iii}]}}
\newcommand{\oii}{\hbox{[O\,{\sc ii}]}}
\newcommand{\nii}{\hbox{[N\,{\sc ii}]}}
\newcommand{\lya}{\hbox{Ly$\alpha$}}

\graphicspath{{./}{figures/}}

\shorttitle{SITELLE Cluster ELGs}
\shortauthors{Liu et al.}

\begin{document}

\title{SITELLE H$\alpha$ Imaging Spectroscopy of z$\sim$0.25 Clusters:\\ Emission Line Galaxy Detection and Ionized Gas Offset in Abell 2390 \& Abell 2465}

\email{qliu@astro.utoronto.ca, hyee@astro.utoronto.ca}

\author[0000-0002-7490-5991]{Qing Liu}
\affil{David A. Dunlap Department of Astronomy \& Astrophysics, University of Toronto, 50 St. George St., Toronto, ON M5S 3H4, Canada}
\affil{Dunlap Institute of Astronomy and Astrophysics, University of Toronto, 50 St. George St, Toronto, ON, Canada}

\author{H.K.C. Yee}
\affil{David A. Dunlap Department of Astronomy \& Astrophysics, University of Toronto, 50 St. George St., Toronto, ON M5S 3H4, Canada}

\author[0000-0003-1278-2591]{Laurent Drissen}
\affil{D\'{e}partement de physique, de g\'{e}nie physique et d'optique, Universit\'{e} Laval, Qu\'{e}bec, QC, G1V 0A6, Canada}
\affil{Centre de Recherche en Astrophysique du Qu\'{e}bec, Canada}

\author[0000-0002-0767-8135]{Suresh Sivanandam}
\affil{David A. Dunlap Department of Astronomy \& Astrophysics, University of Toronto, 50 St. George St., Toronto, ON M5S 3H4, Canada}
\affil{Dunlap Institute of Astronomy and Astrophysics, University of Toronto, 50 St. George St, Toronto, ON, Canada}

\author[0000-0002-9133-4457]{Irene Pintos-Castro}
\affil{Department of Astronomy \& Astrophysics, University of Toronto, 50 St. George St., Toronto, ON M5S 3H4, Canada}

\author[0000-0002-2250-8687]{Leo Y. Alcorn}
\affil{Department of Physics and Astronomy, York University, 4700
Keele Street, Toronto, Ontario, ON MJ3 1P3, Canada}

\author[0000-0001-5615-4904]{Bau-Ching Hsieh}
\affil{Institute of Astronomy and Astrophysics, Academia Sinica, No. 1, Section 4, Roosevelt Road, Taipei 10617, Taiwan}

\author[0000-0001-7218-7407]{Lihwai Lin}
\affil{Institute of Astronomy and Astrophysics, Academia Sinica, No. 1, Section 4, Roosevelt Road, Taipei 10617, Taiwan}

\author[0000-0001-7146-4687]{Yen-Ting Lin}
\affil{Institute of Astronomy and Astrophysics, Academia Sinica, No. 1, Section 4, Roosevelt Road, Taipei 10617, Taiwan}

\author[0000-0002-9330-9108]{Adam Muzzin}
\affil{Department of Physics and Astronomy, York University, 4700
Keele Street, Toronto, Ontario, ON MJ3 1P3, Canada}

\author[0000-0003-1832-4137]{Allison Noble}
\affil{School of Earth and Space Exploration, Arizona State University,
Tempe, AZ, 85287, USA}

\author[0000-0003-1205-1318]{Lyndsay Old}
\affil{European Space Agency, European Space Astronomy Center, Villanueva de la Cañada, E-2691, Madrid, Spain}



\begin{abstract}

Environmental effects are crucial to the understanding of the evolution of galaxies in dense environments, such as galaxy clusters. Using the large field-of-view of SITELLE, the unique imaging fourier transform spectrograph at CFHT, we are able to obtain 2D spectral information for a large and complete sample of cluster galaxies out to the infall region. We describe a pipeline developed to identify emission line galaxies (ELGs) from the datacube using cross-correlation techniques. We present results based on the spatial offsets between the emission-line regions and stellar continua in ELGs from two z$\sim$0.25 galaxy clusters, Abell 2390 and Abell 2465. We find a preference in the offsets being pointed away from the cluster center. Combining the two clusters, there is a 3$\sigma$ excess for high-velocity galaxies within the virial radius having the offsets to be pointed away from the cluster center. Assuming the offset being a proxy for the velocity vector of a galaxy, as expected from ram pressure stripping, this excess indicates that ram pressure stripping occurs most effectively during the first passage of an infalling galaxy, leading to the quenching of its star formation. We also find that, outside the virial region, the continuum-normalized {\ha} line flux for infalling galaxies with large offsets are on average lower than those with small or no measurable offset, further supporting ram pressure as a dominant quenching mechanism during the initial infall stages.

\end{abstract}


\section{Introduction} \label{sec:intro}

It is an established consensus that the evolution of galaxies is affected by their environment, especially in high density regions such as rich galaxy clusters (e.g., \citealt{2001ApJ...547..609E}, \citealt{2002MNRAS.334..673L}, \citealt{2004MNRAS.353..713K}, \citealt{2006PASP..118..517B}, \citealt{2006ApJ...642..188P}, \citealt{2009MNRAS.394.1213W},
\citealt{2011MNRAS.411..675S}, \citealt{2012ApJ...746..188M}, \citealt{2018ApJ...866..136F}, \citealt{2019ApJ...876...40P}). In galaxy clusters the interplay between galaxies and the host cluster is likely responsible for the quenching of star formation in galaxies, i.e., the suppression or cessation of star formation, aside from their internal secular evolutions. Such external effects shape the observed correlation between the star formation rate (SFR) and the environmental density (e.g., \citealt{1998ApJ...504L..75B}, \citealt{2003ApJ...584..210G}, \citealt{2010ApJ...721..193P}, \citealt{2017ApJ...847..134K}), which is closely related to the well-known morphology-density relation (e.g., \citealt{1980ApJ...236..351D}, \citealt{2003MNRAS.346..601G}, \citealt{2005ApJ...623..721P}). 

One of the breakthroughs of present-day observational techniques is 2D imaging spectroscopy, such as the Integral Field Unit (IFU) spectroscopy. Several recent, or ongoing, IFU surveys such as CALIFA (\citealt{2012A&A...538A...8S}), MaNGA (\citealt{2015ApJ...798....7B}) and SAMI (\citealt{2018MNRAS.475..716G}), have been transforming the field of extragalactic studies. In particular, observations with 2D imaging spectroscopy have unveiled 
novel and important insights about impacts of environments on galaxy evolution in high density regions
(e.g., \citealt{2013MNRAS.435.2903B}, \citealt{2017ApJ...844...48P}, \citealt{2017MNRAS.464..121S},
\citealt{2019A&A...621A..98C},
\citealt{2019MNRAS.485.2656C},
\citealt{2019arXiv191005139G},
\citealt{2019ApJ...884..156S}).

A new prospect is revealed by the state-of-the-art imaging Fourier transform spectrograph (IFTS) SITELLE
(\citealt{2019MNRAS.485.3930D}) on the Canada-France-Hawaii Telescope (CFHT). Thanks to its incomparably wide field-of-view (FOV) of $11'\times11'$, SITELLE provides a unique opportunity to study environmental effects on star formation in galaxy clusters in more details. Specifically, SITELLE allows us to simultaneously acquire spatially-resolved spectral information for a large, complete and luminosity-limited sample of emission-line galaxies (ELGs). \textnormal{Its spatial coverage extends out to the cluster infall region \textnormal{at $z\sim0.25$}.} This is unprecedented; existing spatially-resolved data on galaxy clusters are limited in either field coverage or spectral resolution. With abundant information, both spectral and spatial, encoded in emission lines, SITELLE offers an excellent opportunity to gain a better understanding of star formation in galaxy clusters and investigate the mechanisms responsible for the quenching of star formation in high density environments. 

Many possible external quenching mechanisms have been investigated in the past two decades: gas stripping, starvation, harassment, thermal evaporation, pre-processing, etc. (see \citealt{2006PASP..118..517B} for a review). Among all the competing quenching mechanisms, the dynamical interaction of the gas in the cluster galaxy with the intracluster medium (ICM), typically in the form of ram pressure stripping (\citealt{1972ApJ...176....1G}), has been considered to be one of the most efficient physical mechanisms to quench star formation activities in star-forming galaxies (SFGs) through the removal of cold gas.
For example, \cite{2014ApJ...796...65M} concluded that ram pressure stripping is the most plausible mechanism for satellite quenching by investigating the dynamics and quenching timescales of z$\sim$1 cluster galaxies in the GCLASS survey; \cite{2016A&A...596A..11B} found rapid quenching timescales accompanied with HI deficiencies in the Virgo cluster by fitting photometric bands, which indicates ram pressure as the main source of quenching in Virgo. Evidence of quenching by ram pressure has been widely found in many other observational and simulation works (e.g., \citealt{1999MNRAS.308..947A}, \citealt{1999ApJ...516..619F}, \citealt{2006MNRAS.369.1021M}, \citealt{2008ApJ...674..742B}, \citealt{2014MNRAS.438..444B}, \citealt{2015MNRAS.447..969B}, \citealt{2015MNRAS.448.1715J}, \citealt{2016A&A...596A..11B}, \citealt{2016MNRAS.463.1916F}, \citealt{2016MNRAS.461.1202J},
\citealt{2017ApJ...844...48P}, \citealt{2017ApJ...838...81Y},
\citealt{2019A&A...631A.114B}, \citealt{2019MNRAS.488.5370L}, \citealt{2019ApJ...873...52O}, \citealt{2020ApJ...903..103W}).

One prominent observational feature of the ram pressure effect is a disturbed gas morphology. The peculiar disturbance of the HI gas distribution in cluster galaxies has been observed (\citealt{1984ARA&A..22..445H}; \citealt{2007ApJ...659L.115C}, \citealt{2015MNRAS.448.1715J}, \citealt{2017ApJ...838...81Y}) as an indication of the action of ram pressure. Star formation could subsequently occur in the disturbed gas and induce ionization, when it condenses or is compressed by shock waves, leading to disturbed or asymmetric ionized gas distributions that are observable in UV/{\ha}/optical bands (\citealt{1999ASNYN...5...22K}; \citealt{2008ApJ...688..918Y}; \citealt{2010MNRAS.408.1417S}; \citealt{2018MNRAS.476.4753J}). The most extreme cases are known as ``jellyfish galaxies'' (e.g., \citealt{2014ApJ...781L..40E}, \citealt{2017ApJ...844...48P}, \citealt{2018MNRAS.476.4753J}), which are nearby SFGs exhibiting conspicuous extended ionized gas tails. Therefore, analyzing the properties of gas in cluster galaxies is promising for shedding light on the mechanism of ram pressure stripping and placing constraints on numeric simulations of their evolutions in dense environments. For example, many interesting results have been revealed by the MUSE GASP survey (\citealt{2017ApJ...844...48P}) about the properties of jellyfish galaxies and the implications on their evolution histories.

This kind of analysis is well suited to 2D spectroscopic observations using SITELLE. An ongoing project using SITELLE targeting the {\ha}+{\nii} lines in clusters at z$\sim$0.25 is being implemented to characterize the properties of ionized gas in cluster galaxies, including luminosity, morphology, kinematics, ionization states, etc., and to study how these properties correlate with the local environment or the cluster. SITELLE data have the capacity to perform a rich class of interesting science, though in this paper we limit our investigation to the ionized gas morphologies. The goals of this paper are : 1) to describe the methodology of detection and identification of ELGs from the SITELLE datacubes, and 2) to present initial results from a small sample of two clusters based on the spatial offset of ionized gas in search for the evidence of ram pressure stripping. Our conclusions are tentative yet intriguing, but this work serves as a pilot study demonstrating the prospect of SITELLE in studying ELGs in galaxy clusters. Although the \textnormal{spatial} resolution is not as high as MUSE to resolve the gas into filaments, we are able to study the ionized gas using a large and comprehensive sample out to the infall region.

The paper is organized as follows: Section \ref{sec:obs} describes the SITELLE observation and the cluster targets observed, Abell 2390 and Abell 2465. Section \ref{sec:ELG} presents our pipeline used to automatically detect and identify ELGs from the datacube. We describe the method of measuring the centroids of emission-line light and stellar continuum light and show the results based on these centroids in Section \ref{sec:offset}. Section \ref{sec:discussion} discusses the implications of the results. We summarize our findings and conclusions in Section \ref{sec:summary}. Throughout the paper, we assume a cosmology with $\rm H_0$ = 70 km s$^{-1}$ Mpc$^{-1}$, $\Omega_m$ = 0.3 and $\Omega_\Lambda$ = 0.7.

\section{Observations} \label{sec:obs}

    \begin{figure*}
      \centering
      \includegraphics[width=0.85\hsize]{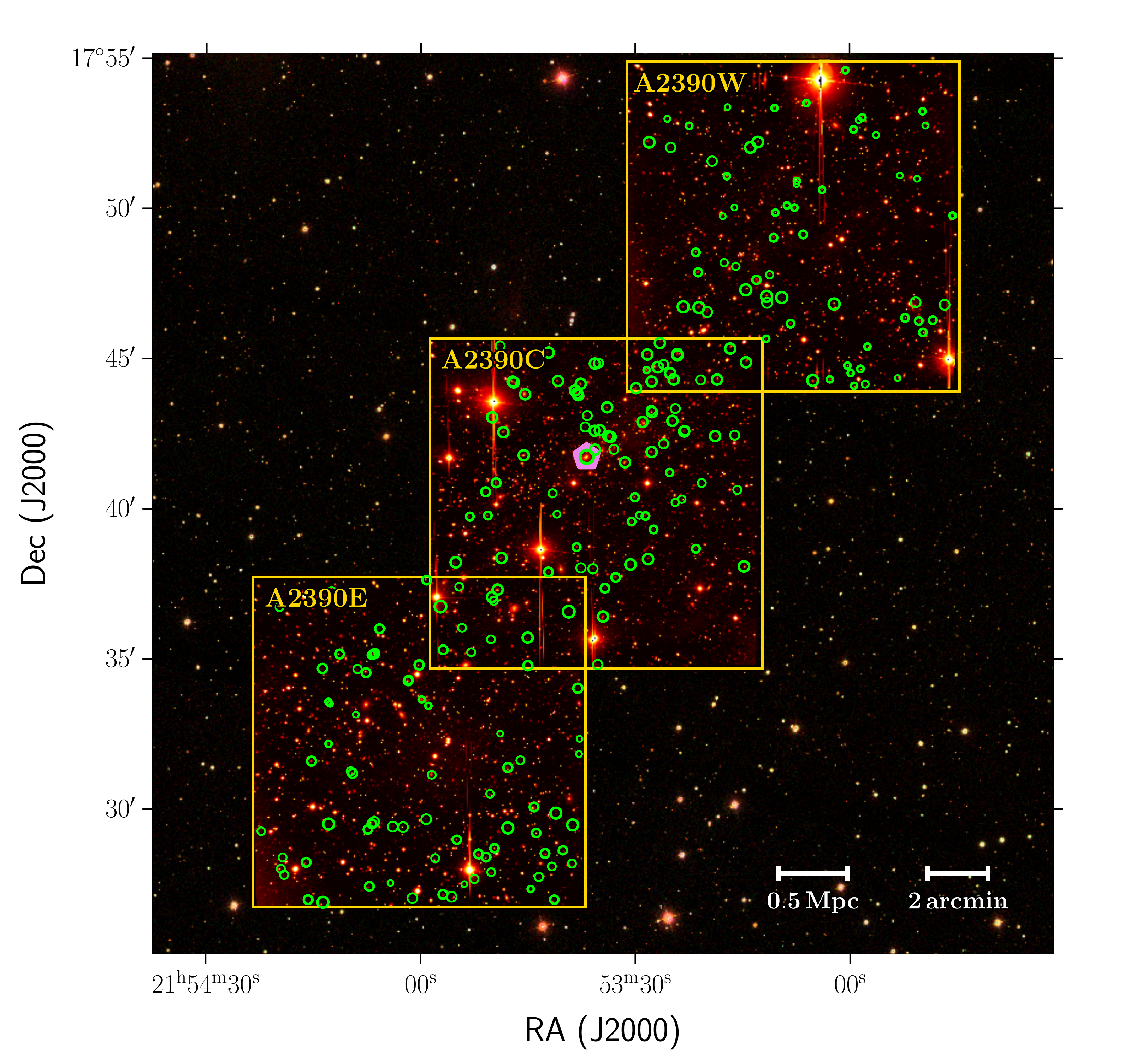}
      \caption{Deep frames of the Abell 2390 central (C) / south-east (E) / north-west (W) fields observed by SITELLE, overlaid on the SDSS DR12 mosaic. The FOV of a single field is $11'\times11'$. The deep frames are constructed using the ORCS piepline (\citealt{2015ASPC..495..327M}). Emission-Line Galaxies (ELGs) identified in Section \ref{sec:ELG} are marked by green circles. The brightest central galaxy marked by a magenta polygon is also an ELG.}
    \label{fig:cluster_A2390}
    \end{figure*}
    
    \begin{figure}
      \centering
      \resizebox{\hsize}{!}
      {\includegraphics[width=\hsize]{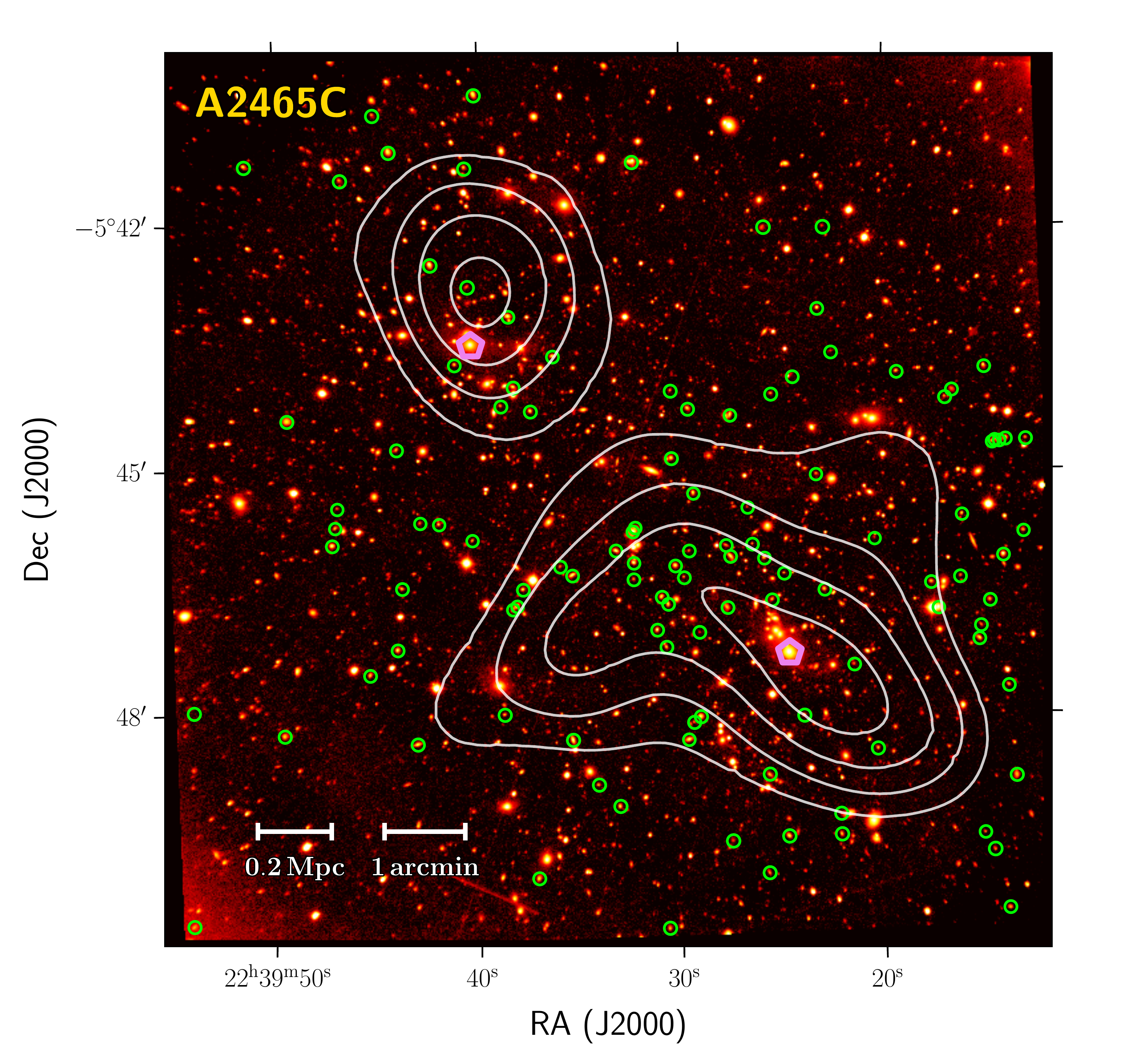}}
      \caption{Deep frame of the central field of the double cluster Abell 2465 observed by SITELLE. Identified ELGs are marked by green circles. The two brightest central galaxies are marked as purple polygons which do not have detected emission. \textnormal{White contours show a rendering of the weak lensing mass contours in Figure 7 of \cite{Wegner2017}.}}
    \label{fig:cluster_A2465}
    \end{figure}

\subsection{SITELLE}
    Our project uses SITELLE (\citealt{2019MNRAS.485.3930D}), the new imaging Fourier instrument at CFHT. With a broad working wavelength range from 3500 {\AA} to 9000 {\AA}, SITELLE is designed to focus on emission-line objects. Band-limiting filters are used to increase spectral resolution with finite scanning steps and to reduce the contamination from strong sky-lines. The equivalent spectral resolution $R$ of IFTS is determined by the total number of mirror steps of the Michelson-type interferometer, which is chosen to resolve the {\ha} and {\nii} lines and to provide spatially-resolved kinematic information. 
    
    The primary science drivers for SITELLE include nebulae and supernova remnants in the Milky Way, {\hii} and star-formation regions in nearby galaxies, and ELGs in galaxy clusters. The uniqueness of SITELLE in studying galaxy clusters is its unparalleled $11'\times11'$ FOV, allowing for IFU-like 2D spectroscopy of a large and complete, luminosity-limited ELG sample of a cluster. The SITELLE data product is a spectral datacube, with an image at each wavelength step (``channel''), after being corrected, transformed and calibrated by the ORBS\footnote{https://github.com/thomasorb/orbs} (\citealt{2015ASPC..495..327M}) pipeline. The pixel scale of SITELLE is 0.322$''$/pix, with 2048$\times$2064 pixels in total. 
    
    The optical-to-near-infrared night sky is densely occupied by OH molecular lines, leading to the difficulty of observing redshifted {\ha} lines. To reduce sky contamination, a narrow window (796 nm -- 826 nm) that is less affected by sky-lines is selected for the design of the SITELLE C4 filter, which corresponds to the redshift range of {\ha-\nii} lines at 0.21-0.25, allowing us to detect ELGs in that redshift range. The window corresponds to the region where the observing efficiency of SITELLE (CCD quantum efficiency, modulation efficiency and optical transmission) is the highest (at around 800 nm). Nevertheless, there exist some sky-lines that affect the detectability and measurement of the lines. We present the subtraction of the sky in Sections \ref{sec:SkySub} and \ref{sec:fringe}.

    Detailed information about SITELLE, including the instrumentation design, the advantages and drawbacks of IFTS, science  capabilities and commission performance, can be found in \cite{2019MNRAS.485.3930D}.

\subsection{\textnormal{Target Clusters}}
    The observations for the first two targets for the survey were carried out on SITELLE in 2017 and 2018 on two well-studied galaxy clusters: Abell 2390 and Abell 2465 (hereafter, A2390 and A2465). 
    
    \begin{table}
    \centering
    \caption{Properties of the clusters observed by SITELLE.}\label{cluster_table}
    \begin{tabular}{ccc}
    \hline \hline
    Cluster & Abell 2390 \footnote{$M_{200}$, $R_{200}$ and $\sigma_{v}$ from \cite{1997ApJ...478..462C}.\label{note:A2390}} & Abell 2465 \footnote{$M_{200}$, $R_{200}$ and $\sigma_{v}$ from \cite{Wegner2011}. The values listed are for each subcluster. \label{note:A2465}} \\
    \hline
    Redshift (z) & 0.228 & 0.245\\
    R.A. ($\alpha_{2000}$) & $21^h53^m35^s$ & $22^h39^m39^s$ (NE)\\
    & & $22^h39^m25^s$ (SW)\\
    Decl. ($\delta_{2000}$) & $+17^\circ
    40'11''$ & $-5^\circ
    43'28''$ (NE)\\
    & & $-5^\circ 47'15''$ (SW)\\
    $M_{200}\,(M_{\odot})$ & $\sim 2\times10^{15}$  & $\sim 4\times10^{14}$ each\\
    $R_{200}\,(Mpc)$ & 2.1  & 1.2\\
    $\sigma_{v}\,(km/s)$ & 1100 & 763 (NE)\\
    & & 722 (SW)\\
    \hline
    \end{tabular}
    \end{table}

    \begin{table*}
    \centering
    \caption{Observed information about the individual fields.} \label{field_table}
    \begin{tabular}{ccccc}
    \hline\hline
    Field & A2390C & A2390W & A2390E & A2465C \\
    \hline
     N$_{\rm steps}$\footnote{\label{note:step}Number of mirror steps to reach  the specified R.} & 124 & 150 & 150 & 207 \\
     Exposure (s) & 11,408 & 13,800 & 13,800 & 13,662 \\
     $\Delta \lambda$ (${\AA}$) & 5.2 & 4.3 & 4.3 & 3 \\
     R & 1080 & 1300 & 1300 & 1800 \\
     Seeing($\arcsec$) & 1.1 & 1.2 & 1.3 & 1.1 \\
     RA ($\alpha_{2000}$)\footnote{\label{note:coords_field}Central coordinates of the SITELLE fields.} & $21^h53^m34.56^s$ & $21^h53^m07^s$ & $21^h53^m59.3^s$ & $22^h39^m32.5^s$ \\
     Dec ($\delta_{2000}$)\footref{note:coords_field} & $17^\circ 40'11''$ & $17^\circ 49''24''$ & $17^\circ 32'15''$ & $-5^\circ 45'22''$ \\
    \hline
    \end{tabular}
    \end{table*}

    
    A2390 is a rich massive cool-core galaxy cluster ($M_{200}\approx2\times10^{15}$ $M_\odot$) at redshift of 0.228. 
    It has been widely studied in many previous studies for the understanding of galaxy evolution in dense environment (e.g., \citealt{1996ApJS..102..289Y}, \citealt{1996ApJ...471..694A}, \citealt{2000MNRAS.318..703B}, \citealt{2005MNRAS.358..233F}). For example, \cite{1996ApJ...471..694A} found that the cluster is gradually formed by the accretion of infalling galaxies with truncation in their star formation by investigating the properties (kinematics, colors, morphologies, etc.) of a large sample of cluster galaxies using the CNOC data (\citealt{1996ApJS..102..289Y}). 
    
    Different from A2390, A2465 consists of two merging subclusters, a north-east (NE) clump A2465NE and a south-west (SW) clump A2465SW, at a redshift of 0.245 (\citealt{Wegner2011}). Members of the subclusters have been identified based on kinematic energies for the measurements of properties for each subcluster in \cite{Wegner2011}. The two subclusters have roughly equal virial masses. A series of detailed multi-wavelength analyses on A2465 can be found in \cite{ Wegner2011} and \cite{Wegner2015, Wegner2017}, where the authors found interesting results revealing its collision history and the evolution history of galaxies within it. The redshifts, positions, and physical properties of the two target clusters observed by SITELLE are summarized in Table \ref{cluster_table}.
    
    There are three pointings for A2390 and one for A2465. \textnormal{The initial observational design was to use an observation time of 4 hr per pointing, including overhead between steps, with $\sim$150 steps, producing a resolution $R\sim$1300, sufficient to resolve the {\ha-\nii} lines. However, for A2390C, which was observed during a commissioning run for the science verification for the C4 filter, only a shorter exposure of 3 hours was obtained, resulting in fewer steps and a lower $R$. The number of steps was increased for the subsequent run for A2365C, and future targets, to achieve better spectral resolution}. Observational information on the four fields including the total number of mirror steps, the exposure time, the average wavelength interval, \textnormal{the average seeing,} the approximate spectral resolution, and the coordinates of field centers are listed in Table \ref{field_table}.

\section{Emission-Line Galaxy Identification} \label{sec:ELG}
The IFTS spectral datacube is transformed from the original interferometric cube acquired by SITELLE at CFHT and calibrated with the data reduction software ORBS (\citealt{2015ASPC..495..327M}). 
This section introduces the procedures for detecting and identifying ELGs from the SITELLE spectral datacube.

\subsection{Sky-Line Subtraction} \label{sec:SkySub}
    While the C4 filter is designed to minimize the number of OH sky-lines in the spectral region, the sky continuum and a few sky-lines still dominate the spectral range of the filter. Therefore, we first subtract the sky background in each channel. The background is evaluated and subtracted in 2D using 
    the \texttt{photutils} package.
    The background is locally estimated within a $128\times128$ pix$^2$ box using a mode estimator and then subtracted from the data of each channel. The spectral axis of the datacube is clipped into 12100 -- 12550 cm$^{-1}$ and converted into wavelength corresponding to 7970 {\AA} -- 8265 {\AA}. The wavelength axis is then interpolated to be uniformly spaced in logarithmic scale. A stacked image is also created by coadding of all the channels within this spectral range for the initial source detection described in Section 3.3.

\subsection{Fringe Reduction using Low-pass Filtering} \label{sec:fringe}

    The interferometric nature of SITELLE and the presence of strong sky-lines in the C4 filter induce fringes across the field in specific channels. The fringes are superimposed on the uneven large-scale background that has been subtracted in Section \ref{sec:SkySub} . The level of contamination from these fringes depends on the channel wavelength, the number of mirror steps, and the position in the field. In general, channels around stronger sky-lines with fewer mirror steps (i.e. lower spectral resolution) present brighter fringes away from the field center. The brightness and spatial pattern of these fringes differ from channel to channel, which makes it challenging to model them from first principles. The brightness of fringes vary across the field and could be comparable to/brighter than a large number of sources, affecting the source detection, ELG identification, and potentially their measurements. We adopt an empirical method to reduce the influence of fringes by applying low-pass filtering (LPF) on each channel. Details of the LPF procedure and a figure illustrating the fringes and cleaning are presented in Appendix \ref{appendix:fringe}.

\subsection{Source Detection and Spectral Extraction} \label{sec:source}

    Although the fringes have been alleviated by the LPF procedure, small-scale fringe residuals from the sky occasionally show up in some channels because of the finite kernel size used. We apply a moving average procedure to further mitigate fringe contamination (Appendix \ref{appendix:fringe}) and construct a new datacube for source detection. We note that this \textnormal{smoothed} datacube is used only for source detection but not for analysis. \textnormal{The new} datacube is collapsed in the spectral axis into an image for the detection of ELGs (referred to below as the ``detection image'') generated by \textnormal{using} the maximum value in the spectral axis at each spatial position, taking advantage of the 2D spectroscopy. Using the maxima in wavelength enhances the possibility of detecting weak narrow peaks. We compare the output candidate list \textnormal{obtained from} the detection image with the one \textnormal{from} the \textnormal{mean} image \textnormal{of all channels} and it proves that the detection image constructed \textnormal{in such way} offers a more complete sample that includes line emitters with weak or zero continuum.
    
    The python software \texttt{photutils} is used to detect and deblend sources. We adopt a detection signal-to-noise ratio (S/N) threshold of 2.5, a 5 pixel connection, a multi-thresholding level number of 64, and a local peak contrast of 0.01. The threshold is locally estimate based on a 2D background estimate using the \texttt{SExtractorBackground} class as in Section \ref{sec:SkySub}. A segmentation map is created in this process. An integrated spectrum of each detected source is extracted from the low-pass-filtered datacube using the corresponding segmentation in the map. \textnormal{These spectra are used for source identification and redshift measurement.} In total, the detection algorithm locates $\sim$2000 sources in A2390C field, $\sim$1500 sources in A2390E/W fields, and $\sim$1200 sources in A2465C field.

\subsection{Removal of Continuum}
    \label{sec:cont_removal}
    For the purpose of automatically identifying ELGs using templates that only contain emission lines, the continua of sources must be subtracted to obtain the residual emission lines. This is accomplished using Gaussian Process (GP) regression, typically used for fitting curves in a non-parametric manner.
    The advantage of using GP regression is its flexibility, which allows for slight variations in the continuum such as a mild gradient and is able to smoothly tackle possible irregularity around the filter edges. We use the \texttt{GaussianProcessRegressor} realization in the python package \texttt{scikit-learn}. The kernel is chosen to be a combination of a radial basis function (RBF) kernel with a bandwidth of 100 {\AA} (such that it is wider than the width of {\ha-\nii} lines) which accounts for the continuum component, and a white kernel accounting for the noise. Possible emission that is one $\sigma$ above the median value of the spectrum is iteratively replaced with the median. In the case of a few bright sources, the sharp drops around the filter edges make them poorly fitted by smooth GP kernels and are masked in the fitting. \textnormal{This procedure is applied on each of the integrated spectra extracted in Section \ref{sec:source}}. 

\subsection{Construction of Line Template}
    We use the cross-correlation technique to identify ELGs from detected sources. It is done by cross-correlating the normalized residual spectra with a library of emission-line templates. While the observation is designed to capture the {\ha-\nii} lines in the galaxy cluster, emission lines other than {\ha-\nii} (e.g., \oiii$\lambda$5007 at z$\sim$0.64, \oii$\lambda$3727 at z$\sim$1.2) from background sources may fall into the filter as well. To construct a robust ELG sample for cluster members, it is critical to distinguish these lines from the target {\ha+\nii} lines. Below we describe three sets of emission-line templates used to perform cross-correlation.
    
    The primary set includes the {\ha} line and the {\nii$\lambda\lambda$6548,6584} lines, with a fixed line ratio between {\nii$\lambda$6584} and {\nii$\lambda$6548} of 3:1 (\citealt{1989Msngr..58...44A}). The line ratio between {\ha} and {\nii$\lambda$6584} varies from 1:1 to 8:1 uniformly in logarithmic scale. The second set consists of the {\oiii$\lambda\lambda$4959,5007} doublet, with the line ratio of {\oiii$\lambda$5007} to {\oiii$\lambda$4959} varying from 2:1 to 4:1. The {\hb} line is not included in the template set, considering that in most cases it is located out of the filter or at the filter's blue edge in the presence of the {\oiii} doublet. As a result, the inclusion of {\hb} might reduce the optimal S/N of a spectrum with only the {\oiii} doublet in the cross-correlation process, and therefore it is removed from the template. The third set contains a single line, which represents several possible origins including an {\oii$\lambda$3727} line at z$\sim$1.2, an isolated {\oiii$\lambda$5007} line with an undetected weak {\oiii$\lambda$4959} or with {\oiii$\lambda$4959} outside of the blue band limit, a {\ha} line with undetected {\nii} (e.g., from metal-poor galaxies, or with low S/N), an isolated {\hb} at z$\sim$0.66 with the {\oiii$\lambda$4959} outside of the red band limit, and possibly a bright {\lya} emitter at z$\sim$5.5.
    
    The template lines are broadened into Gaussian profiles by convolution:
    \begin{align}
    T(\lambda) &= \left[\sum_i \delta(\lambda-\lambda_{0,i}) \right] * \text{Gaussian}(\lambda_{0,i}, \sigma_{line}) \nonumber \\
    &= \sum_i
      A_i
      \exp \left[ - \frac{(\lambda-\lambda_{0,i})^2}{2\sigma_{line}^2}
      \right] \;,
    \end{align}
    with the normalization $A_i$ being the line ratios, $\lambda_{0,i}$ being the line center for the $i$-th line component, and $\sigma_{line}$ being the common line width. The line width ranges from $\sigma_{min}=\Delta\lambda$, which is the spectral resolution, to $\sigma_{max}=\sqrt{\sigma_{min}^2+\sigma_{gal}^2}$, which corresponds to a galactic velocity dispersion $\sigma_{gal}=$300 km/s. It is noteworthy that the intrinsic line shape of SITELLE is a sinc-gaussian resulting from an instrumental sinc line shape with Gaussian broadening (\citealt{2019MNRAS.485.3930D}). However, in practice, introducing sinc-gaussian in a template adds another degree of freedom and sometimes unintentionally magnifies the noise. This is possibly due to the lower S/N in the data as the observation was obtained in the verification run. In the C4 filter potential sky-line residuals also make it noisier than observations using bluer SITELLE filters. Consider the fact that using Gaussian templates in cross-correlation would not practically reduce the efficiency of ELG identification and narrow-band centroids in this study, we opt to use Gaussian line templates. Nevertheless, we caution the readers that when carrying out line fitting procedures to obtain delicate information such as line kinematics and their uncertainties, the use of sinc-gauss functions is more ideal (see documents of ORCS\footnote{http://celeste.phy.ulaval.ca/orcs-doc} and \citealt{2015ASPC..495..327M}) for SITELLE.
    
    \begin{figure*}
      \centering
      \resizebox{0.9\hsize}{!}{\includegraphics{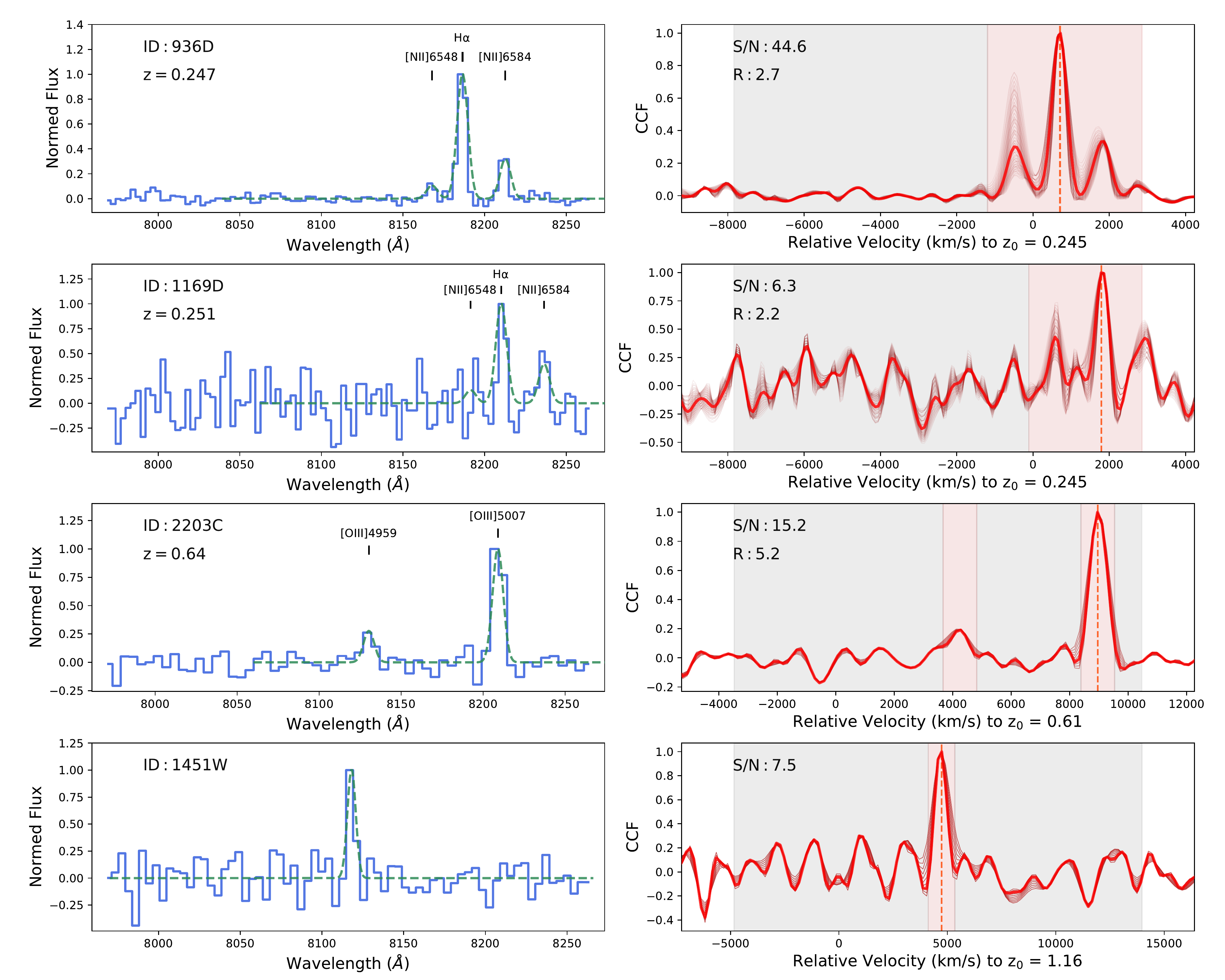}}
      \caption{Examples of continua-removed integrated spectra (left column) and the corresponding cross-correlation results (right column). From top to bottom: high S/N {\ha}+{\nii}, low S/N {\ha}+{\nii}, {\oiii$\lambda\lambda$4959,5007} doublets, and a likely isolated line (cross-correlated by {\oii}). In the left column, the continuum-subtracted spectra are shown in blue, whereas the green dashed lines indicate the matched templates; the redshift determined from cross-correlations are shown for the top three rows. In the right column, cross-correlation functions (CCF) are plotted versus velocity shifts with reference to the systematic redshift of the cluster. Shaded areas indicate the range used for computing the signal (light red) and noise (gray). The thick red line shows CCF of the best-matched template. The thin red lines show CCF from individual templates.}
    \label{fig:Line_CC}
    \end{figure*}

\subsection{Cross-correlation with Template} \label{sec:ELG:CC}

    Cross-correlation, also known as matched filtering, is a technique widely used in signal processing that is applied for the optimization of the S/N by cross-correlating the observed signal with templates:
    \begin{equation}
    \begin{split}
    CCF(v\leftrightarrow\delta\lambda) 
    &= \left[F\star T\right](\delta\lambda) \\
    &= \int _{\lambda_1}^{\lambda_2}{{F(\lambda)}}T(\lambda+\delta\lambda)\,d\lambda \,,	
    \end{split}
    \end{equation}
    where $F(\lambda)$ is the input continuum-subtracted spectrum within the filter range $(\lambda_1, \lambda_2)$, and CCF is the cross-correlation function. By convention, CCF is in units of velocity $v$, which is converted from wavelength difference $\delta\lambda$. The residual spectra is normalized and linearly interpolated with twice as many points as the original spectra in log scale for a higher precision in peak matching. In essence, the cross-correlation procedure accomplishes two jobs: the detection of a potential signal, and the localization of its position (redshift). This is run on all the extracted integrated spectra obtained above.
    
    The S/N of the CCF, $(S/N)_{cc}$, is calculated as follows:
    \begin{equation}
    (S/N)_{cc} = \frac{CCF_{max}}{\sigma_{cc}} \,,
    \end{equation}
    where $CCF_{max}$ is the peak of the CCF and $\sigma_{cc}$ is the rms noise calculated in the region $5\,\sigma_{line}$ away from any of the line centers. A wider signal range ($8\,\sigma_{line}$) is used if the matched line ratio {\ha}/{\nii$\lambda$6584} is lower than 3:1 in high S/N cases ($(S/N)_{cc}>50$) in order to better match bright broad lines. Regions near the filter edges ($\sim$20\AA) are clipped when calculating $(S/N)_{cc}$.
    $(S/N)_{cc}$ gives an estimate of the credibility of emission, but it does not place constraints on the line width and ratio. 
    
    We further use a significance parameter $R$ defined as:
    \begin{equation}
    R = \xi \cdot \gamma = \frac{CCF_{max}}{CCF_{max}^{(2)}}\cdot \frac{(S/N)_{cc}}{(S/N)_{cc,max}} \,,   
    \end{equation}
    which multiplies the peak contrast $\xi$, defined as the ratio of the highest peak to the second highest peak of the CCF, by the credibility $\gamma$, defined as the ratio of the computed $(S/N)_{cc}$ and the highest $(S/N)_{cc}$ among all templates. The best matched template is chosen as the one with maximum R. The parameter R is motivated by experiments suggesting that maxmizing $R$ typically returns better estimates for the line ratio and the line width $\sigma_{line}$ than simply using $(S/N)_{cc,max}$, while retaining the accuracy of the matched redshift. Examples of CCF are shown in the right panels of Figure \ref{fig:Line_CC} with the corresponding residual spectra shown in the left panels.
    
    With the aid of visually inspecting a number of spectra and their CCF, we apply the following S/N criteria to select potential ELG candidates for detection:
    \begin{itemize}
    \item $(S/N)_{cc, \rm{H\alpha}}>5$ or $(S/N)_{cc, \rm{OIII}}>5$\,,
    \item $(S/N)_{cc, single}>3$\,,
    \end{itemize}
    where $(S/N)_{cc, \rm{H\alpha}}$, $(S/N)_{cc, \rm{OIII}}$ and $(S/N)_{cc, single}$ are the resultant $(S/N)_{cc}$ using the three template sets. 
    We then calculate quick estimates of $W_{em}$, the equivalent width of the total emission using a 10 $\sigma_{line}$ wide window, for the purpose of reducing false detections. An uncertainty of $W_{em}$, $\sigma(W_{em})$, is roughly obtained by perturbing spectra 250 times with the continuum rms. We require the candidate to have $W_{em}>5\AA$ and $W_{em}>\sigma(W_{em})$ to be considered as a detection.
    We further require the candidate to be located at least 20 pixels away from the field edge and its matched peak at least 20 {\AA} away from the filter edge.
    
    Through all the steps above, we achieve a list of ELGs in each of the fields, with 88/80/72 from the A2390C/A2390E/A2390W field and 111 from the A2465C field. Repeated detections in overlapping field coverage of A2390 are cross-matched and cleaned when combining detections among fields. Finally, a visual check is done on all the candidates to screen out dubious detections and to confirm strong {\oiii} emitters. In total, 194 ELGs are found in A2390 and 110 ELGs are found in A2465. ELGs identified in the A2390 and A2465C are marked in Figure \ref{fig:cluster_A2390} and Figure \ref{fig:cluster_A2465}, respectively, which includes line emitters in the background. In A2390, the central field has the most ELGs, while A2465C shows an excess close to the center of the south-west subcluster. The prevalence of ELGs found in A2465C suggest enhanced star formation occurring in the merging clusters, as observed and discussed in \cite{Wegner2011} and \cite{Wegner2015}. 
    \textnormal{The average 5$\sigma$ detection limit is $F_
    {\ha}\sim 3/2/4/2 \times$10$^{-16}$ ergs/s/cm$^2$ in A2390C/A2390E/A2390W/A2465C}, estimated by taking the median of the flux of S/N$>$5 objects rescaled to S/N=5 with the flux noise measured using Monte Carlo simulations (see Section \ref{sec:Haflux}). The average 5$\sigma$ detection limit is $F_
    {\ha}\sim 2 \times$10$^{-16}$ ergs/s/cm$^2$. Adopting the conversion of \cite{2012ARA&A..50..531K}, this corresponds to SFR $\sim 0.2 M_{\odot}$/yr, or $\sim 0.8 M_{\odot}$/yr if 1 mag dust extinction is applied.

\section{Analysis of Spatial Offset of Ionized Gas} \label{sec:offset}

With the acquired ELG list, a simple yet interesting analysis is to investigate the connection between the spatial offset of ionized gas from its parent galaxy and the cluster center in search for ram pressure stripping effects. We present in this section some initial results based on the centroid offsets between emission-line regions and stellar continua in the two clusters observed by SITELLE. 

In Section \ref{sec:cen_measure} we introduce the methodology of measuring the centroids of ionized gas and stellar continuum. We then measure the centroid offset and the difference angle between the \textit{emission-to-continuum vector} and the \textit{cluster-centric vector} following the analysis of \cite{2010MNRAS.408.1417S}. We show the centroid analysis results of A2390 and A2465 in Section \ref{sec:angle_dist}.

\subsection{Centroid and Angle Measurements} \label{sec:cen_measure}

    \begin{figure*}
      \centering
      {\includegraphics[width=0.9\hsize]{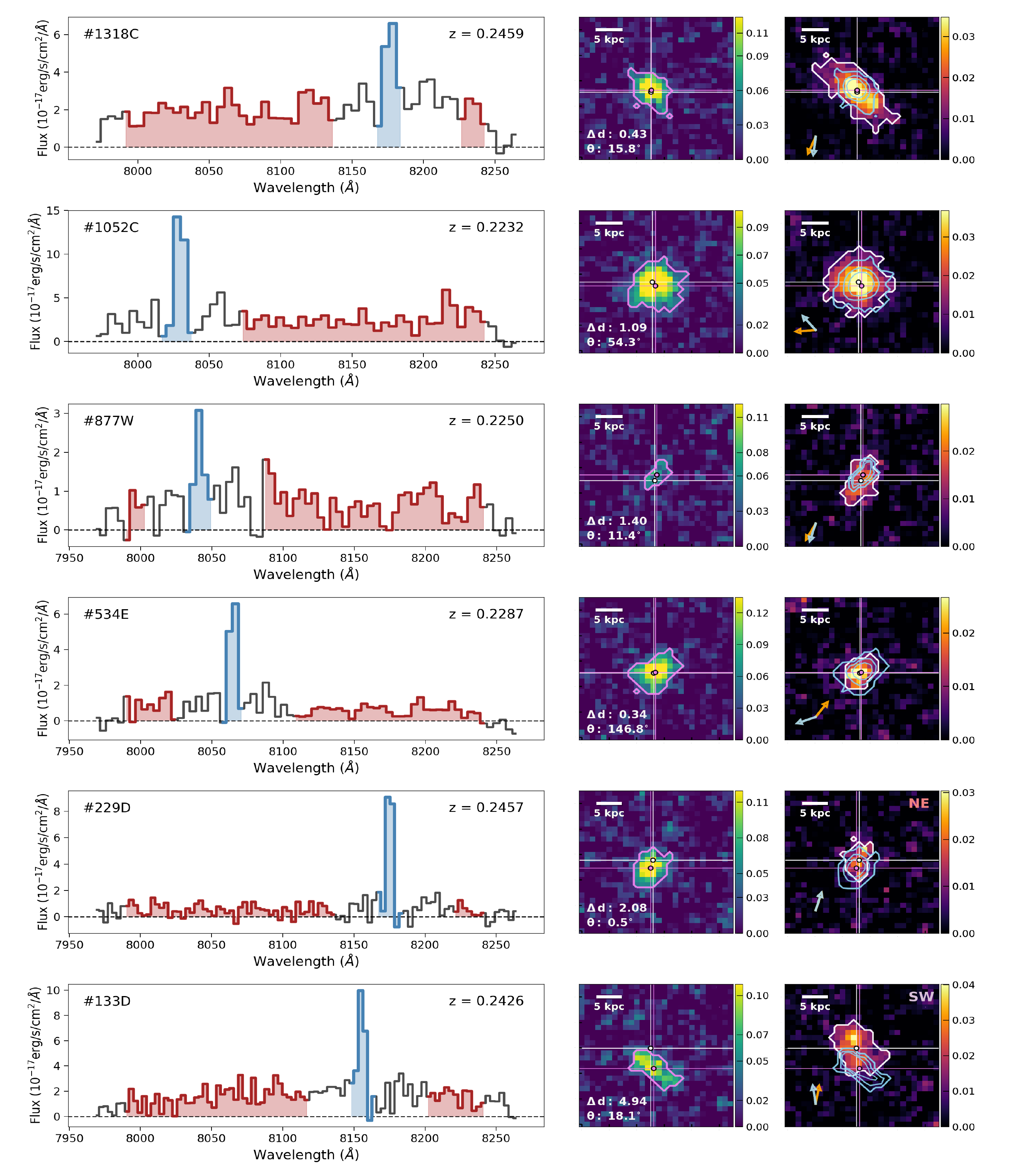}}
      \caption{Example centroid measurements from the A2390C/W/E (top four rows) and A2465C fields (bottom two rows). For each row, the spectrum of the detected ELG is shown in the left panel with the channels used for the creations of the emission/continuum image shown in red/blue. The middle panel shows the {\ha} emission image and the right panel shows the continuum image, on which the centroid measurements are performed. The boundary of the distribution of ionized gas emission ($I_E$) / the stellar continuum ($I_C$) and the light-weighted centroid are plotted on each postage stamp in magenta (for emission) / white (for continuum). The emission-line and continuum centroids are indicated as magenta and white dots \textnormal{with crosses}, respectively. The \textnormal{30\%-60\%-90\%} level of the \textnormal{{\ha}} emission distribution is overlaid as blue contours on the continuum image after a mild smoothing by a 3x3 pix$^2$ Gaussian kernel. The blue arrow indicates the emission-to-continuum offset vector $\mathbf{d}$ and the orange arrow indicates the direction to the cluster center. Measured $\Delta\,d$ and $\theta_d$ are in units of \textnormal{kpc} and degrees, respectively. \textnormal{Galaxies with $\theta_d>1$ kpc are used for analysis in Sec. \ref{sec:angle_dist}.} The image scale is in units of 10$^{-17}$erg/s/cm$^2/{\AA}$/pix$^2$.}
    \label{fig:Centroid_Measure}
    \end{figure*}
    
    \begin{figure}
      \centering
      \resizebox{\hsize}{!}{\includegraphics{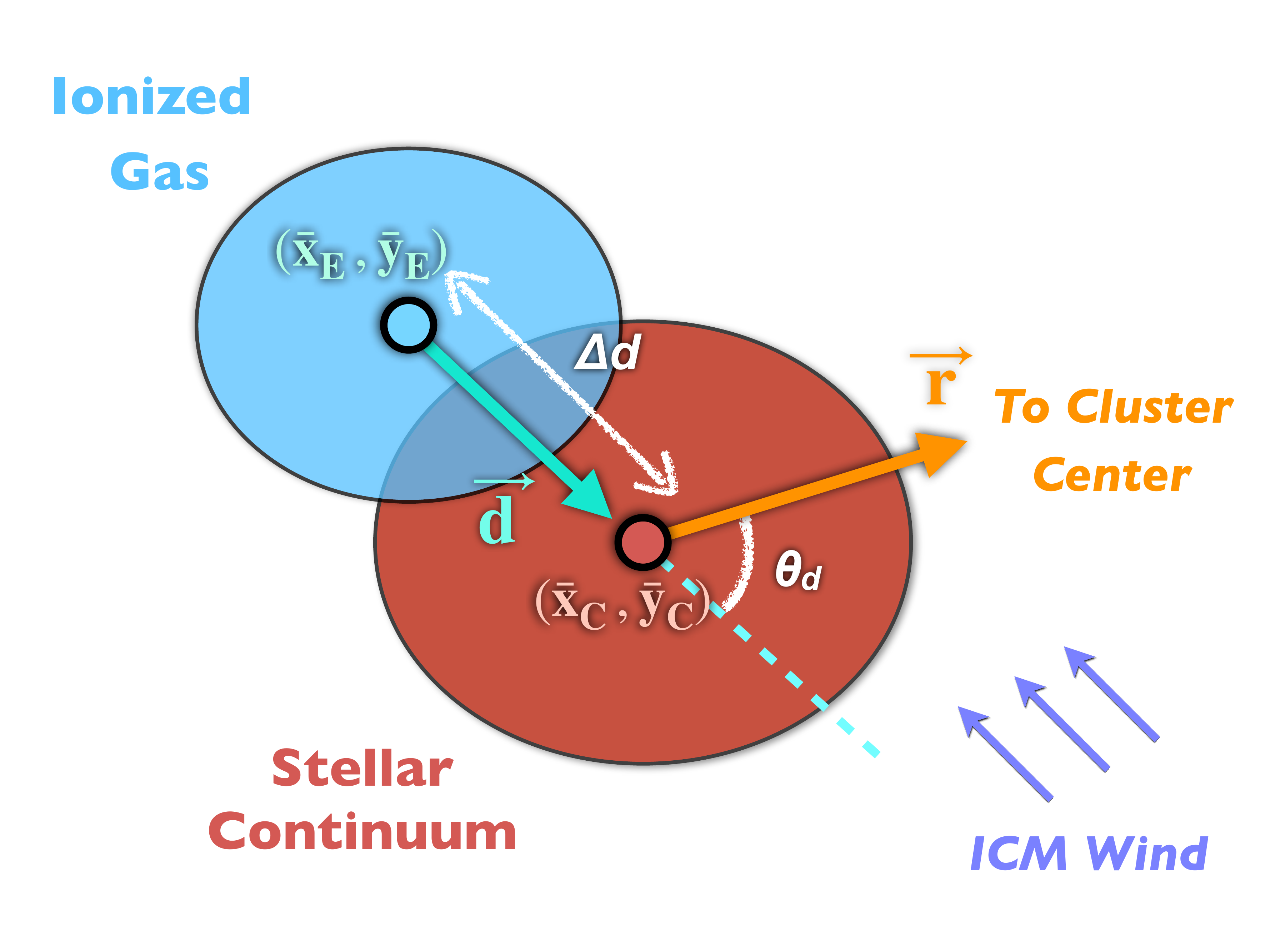}}
      \caption{Schematic illustrating the centroid analysis. The emission-line region emission ($\bar{x}_E, \bar{y}_E$) and the stellar-continuum centroids ($\bar{x}_C, \bar{y}_C$) are measured from the emission image $I_E$ and continuum image $I_C$. The centroid offset $\Delta d$ and difference angle $\theta_d$ between the offset vector $\mathbf{d}$ and cluster-centric vector $\mathbf{r}$ are measured based on the centroid positions. The purple arrows illustrate the expected direction of the ICM wind arising from the movement of the galaxy in the ICM halo.}
    \label{fig:measure_schem}
    \end{figure}
    
    To construct an initial sample of cluster galaxies for each cluster, we first remove objects with a velocity difference of more than 4$\sigma_v$ ($\Delta z \gtrsim$ 0.015) from the systematic velocity of the cluster. We then measure the centroid of the ionized gas and the centroid of the stellar continuum for each cluster galaxy as follow:

    For each galaxy, a thumbnail datacube centered on the object is created. The continuum image, $I_C(x,y)$, is constructed by taking the mean value of channels in the continuum range at each position $(x,y)$ after an iterative 3$\sigma$-clipping. The continuum range is defined as the region further than $15\times(1+z_{cc})$ {\AA} away from the {\nii} lines, but at least 20 {\AA} away from the filter edges.
    The emission image, $I_E(x,y)$, is constructed in a narrow-band likewise by taking the mean of the few channels within $\pm\,5\times(1+z_{cc})$ {\AA} around the matched {\ha} peak. The continuum image is subtracted from it to obtain the final emission image.
    
    
    Another source detection is performed on $I_C(x,y)$ and $I_E(x,y)$ using a detection S/N threshold of 2.5, with nearby sources masked using the segmentation map in Section \ref{sec:source}. The detection of emission or continuum could fail for a portion of objects in the case of object being : (1) a faint target with low S/N; (2) a background source with weak or no detectable continuum; (3) a false positive peak composed of noise; and (4) close to contaminants such as stars or diffraction spikes. For objects with successful detections in both emission and continuum, we then measure the flux-weighted centroids $(\bar{x}, \bar{y})$ for the ionized gas or the stellar continuum through
    \begin{equation}
        (\bar{x}, \bar{y}) = \left(\frac{\sum\,x_i \cdot I(x_i,y_i)}{\sum\,I(x_i,y_i)}, \frac{\sum\,y_i \cdot I(x_i,y_i)}{\sum\,I(x_i,y_i)}\right)\,,
    \end{equation}
    where the sum is performed on all pixels within the segmentation of the distribution \{$x_i$, $y_i$\} from the source detection for the emission image I$_E$(x,y) or the continuum image I$_C$(x,y). Example spectra, continuum images, and emission images are presented in Figure \ref{fig:Centroid_Measure}.
    
    \begin{figure}
      \centering
      \resizebox{0.95\hsize}{!}{\includegraphics{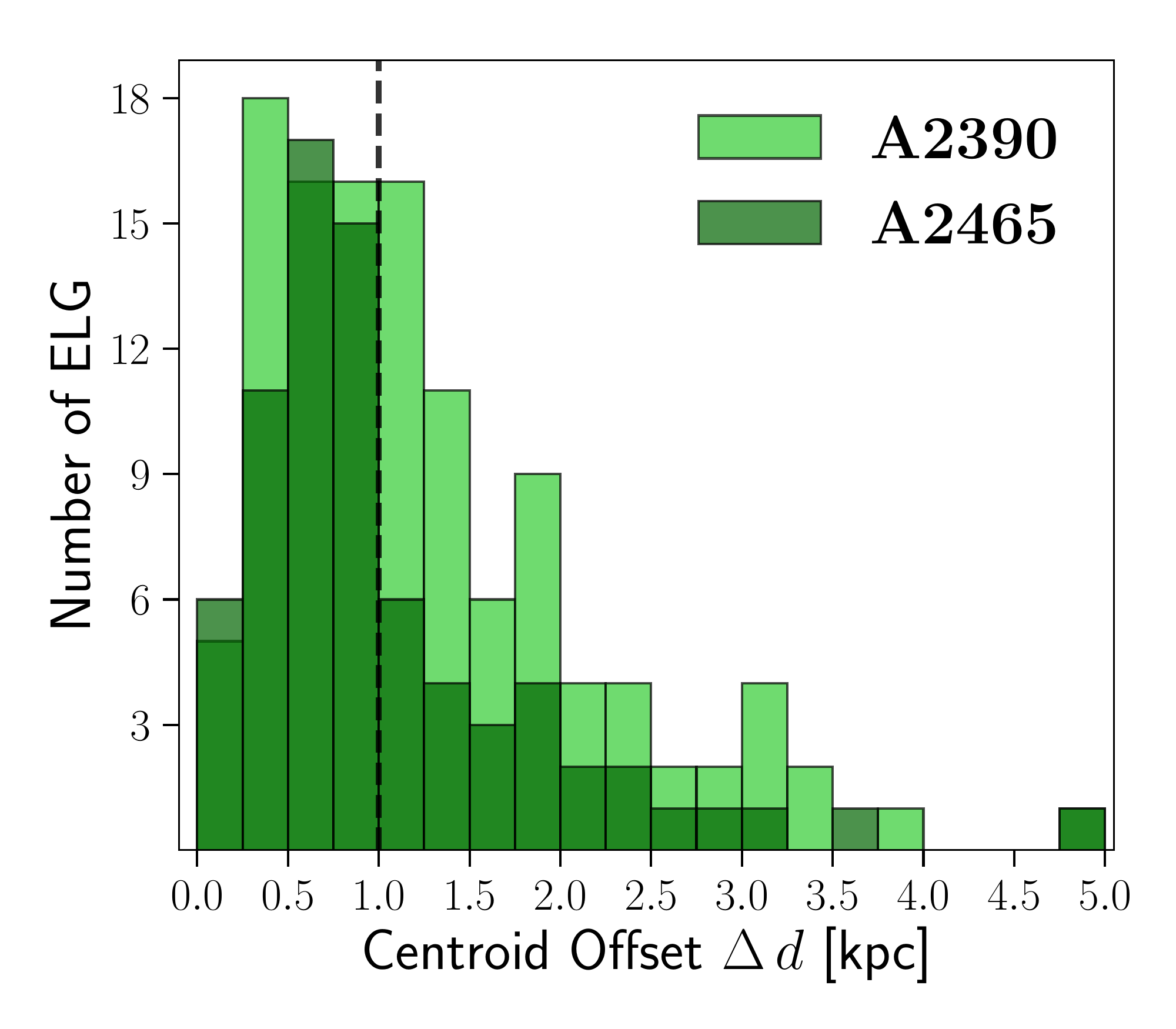}}
      \caption{The distributions of light-weighted centroid offset $\Delta\,d$ between the emission and the stellar continuum for ELGs identified in A2390 (left, combining three fields) and the central field of A2465 (right). The 1kpc threshold used is indicated by the black dashed line.}
    \label{fig:centroid_offset}
    \end{figure}
    
    

    With the emission and continuum centroids, we then measure the offset vector $\mathbf{d}$, defined as the vector from the emission-line centroid to the continuum centroid:
    \begin{equation}
        \mathbf{d} = ({\bar{x}}_\mathrm{E}-{\bar{x}}_\mathrm{C},\, {\bar{y}}_\mathrm{E}-{\bar{y}}_\mathrm{C})\,.
    \end{equation}
    We measure the centroid offset $\Delta d$, using the flux-weighted centroids:
    \begin{equation}
        \Delta d = \sqrt{({\bar{x}}_\mathrm{E} - {\bar{x}}_\mathrm{C})^2+({\bar{y}}_\mathrm{E}-{\bar{y}}_\mathrm{C})^2}\,,
    \end{equation}
    and the difference angle between $\mathbf{d}$ and the cluster-centric vector $\mathbf{r}$, a vector pointing from the galaxy to the brightest central galaxy (BCG):
    \begin{equation}
    \theta_d = \measuredangle(\mathbf{d}, \mathbf{r})\,,
    \end{equation}
    
    Given that the offset is much smaller than the distance to the BCG, the object center measured in the primary source detection is adopted as the galaxy center. An illustration of the definitions of quantities in the measurements is shown in Figure \ref{fig:measure_schem}. The \textnormal{uncertainty} propagation is presented in Appendix \ref{appendix:uncertainty_prop}. To remove measurements with larger uncertainties, we require the measured offset to have $\Delta\,d > 3\,\sigma_{\Delta d}$ and exclude objects near the field edges (distance $<$ 100 pix). These criteria reject $\sim$38\% of the sample on average in the four fields. We obtain 117 ELGs in A2390 and 75 ELGs in A2465 after the rejection. 
    
    Figure \ref{fig:centroid_offset} shows the distributions of $\Delta d$ for A2390 and A2465. \textnormal{The centroid offsets are in general relatively small. An empirical assessment for the centroid measurements is presented in Appendix \ref{appendix:assessment} showing that these relatively small {$\Delta d$} can be measured robustly.} The distributions of $\Delta d$ extend from 0 to $\sim$5 kpc. Their peak positions deviate from zero, suggesting the existence of some impact on the spatial distribution of ionized gas relative to the stellar continuum, which one would expect from ram pressure stripping. However, as discussed in Section \ref{sec:gas_offset_rps}, the offset alone cannot be fully attributed to the impact of ram pressure as it may also be caused by strong outflow or tidal force. It may also be affected by large star-forming clumps and dust geometry in the disks. Therefore, in the next section we explore whether such an offset has a connection with the host cluster, i.e., it is caused by environmental factors. Internal mechanisms such as outflows are expected to be uncorrelated with the cluster center.
    We note that, alternatively, one can measure the morphological centroids of the regions without weighting by light. The measurements and results are presented in Appendix \ref{appendix:res_morph_centroid}, where the two methods produce similar results.

\subsection{Distribution of Difference Angles in A2390 and A2465} \label{sec:angle_dist}

    In the search for evidence of ram pressure effects, it is of great interest to further investigate the correlation between the emission spatial offset and the cluster center, i.e. the distribution of the difference angles, because the strength and direction of the ram pressure from the ICM on the galaxy are expected to be closely linked to the infalling galaxy's orbit and position in the cluster.
    Because small spatial offsets are prone to be affected by measurement uncertainties, we further limit our sample to ELGs with a cutoff in the measured offset. We choose a physical length of 1 kpc \textnormal{that is higher than the peak values of distributions of $\Delta d$} to be the sample threshold, which corresponds to $\sim$0.85 pix in A2390 and $\sim$0.8 pix in A2465. \textnormal{The results are qualitatively the same with a slightly higher or lower threshold adopted. Examples of $\Delta d$ measurements are shown in Figure 4 where the contours and centroids for {\ha} emission and continuum are displayed. It is interesting to note that the ELGs with $\Delta d$ $<$ 1kpc show {\ha} emission being largely symmetric with the continuum, while those with larger $\Delta d$ display a clear non-symmetry.}
    
    Finally, we obtain 55 ELGs from A2390 and 24 ELGs from A2465 out of the parent ELG samples. Hereafter, these ELGs with relatively large emission-to-continuum offsets ($\Delta d$) are referred to as the selected ELGs.
    
    \subsubsection{Abell 2390} \label{sec:A2390}
    
\begin{figure}
      \centering
      \includegraphics[width=\hsize]{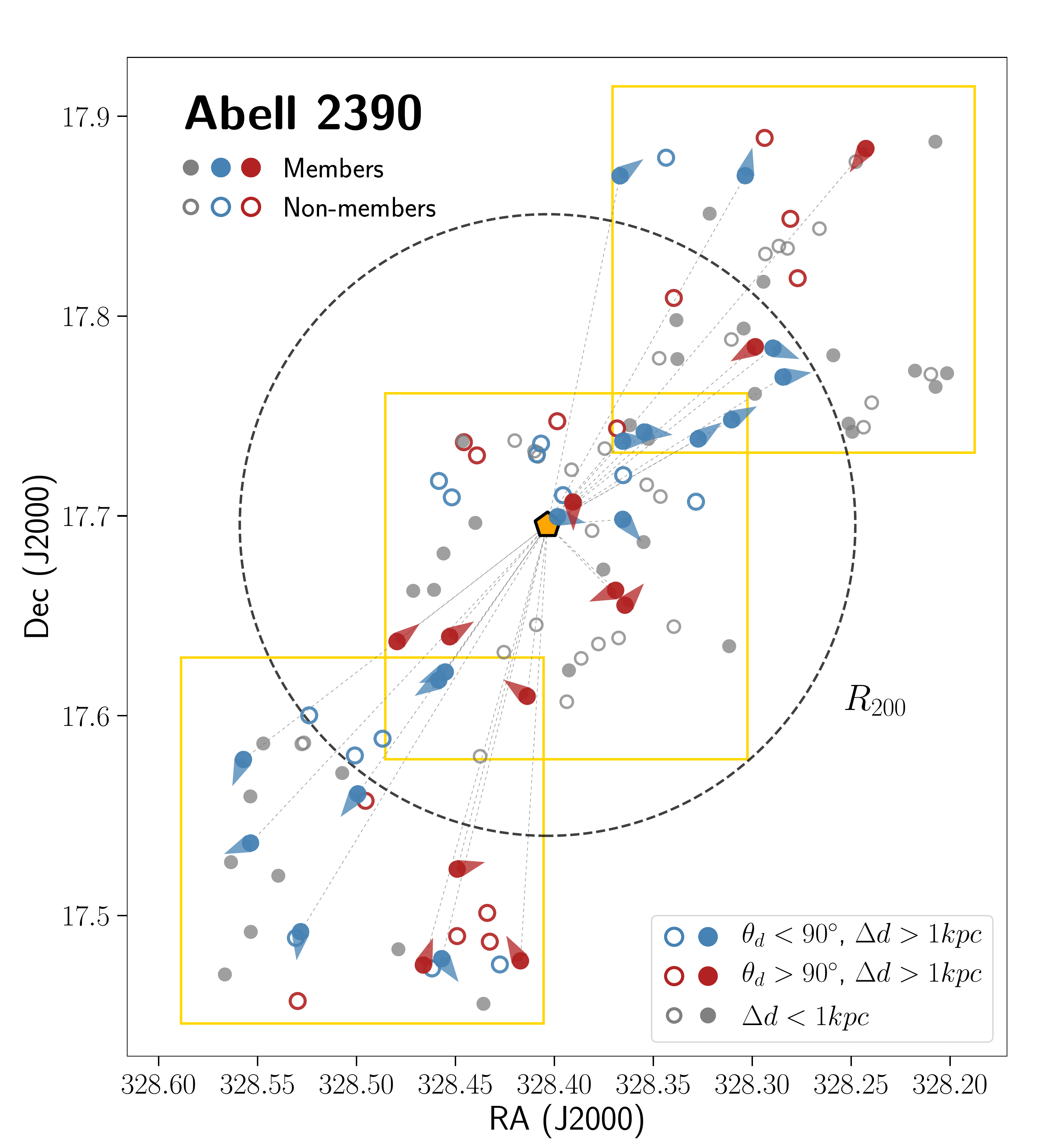}
      \caption{Location of the selected ELGs ($\Delta d>$ 1 kpc) in A2390. Blue symbols represent selected ELGs ($\Delta d>$ 1 kpc) in the cluster with $\theta_d<90^\circ$, while red symbols represent selected ELGs in the cluster with $\theta_d>90^\circ$. For display purpose, colored arrows indicate the opposite directions of the difference vector $\mathbf{d}$, i.e. from continuum to emission. ELGs with velocities above the escape curve (non-cluster members) are shown as open circles. ELGs with small $\Delta d$ ($<$ 1kpc) are marked in gray (solid: cluster members; open: non-cluster members). The BCG is shown as the central orange polygon. ELGs with large centroid uncertainties or near field edges are not shown. The virial radius $R_{200}$ is indicated by the black dashed circle. The yellow squares indicate the footprints of the SITELLE fields.}
      \label{fig:ELG_PA_A2390}
    \end{figure}

\begin{figure}
      \centering
      \includegraphics[width=\hsize]{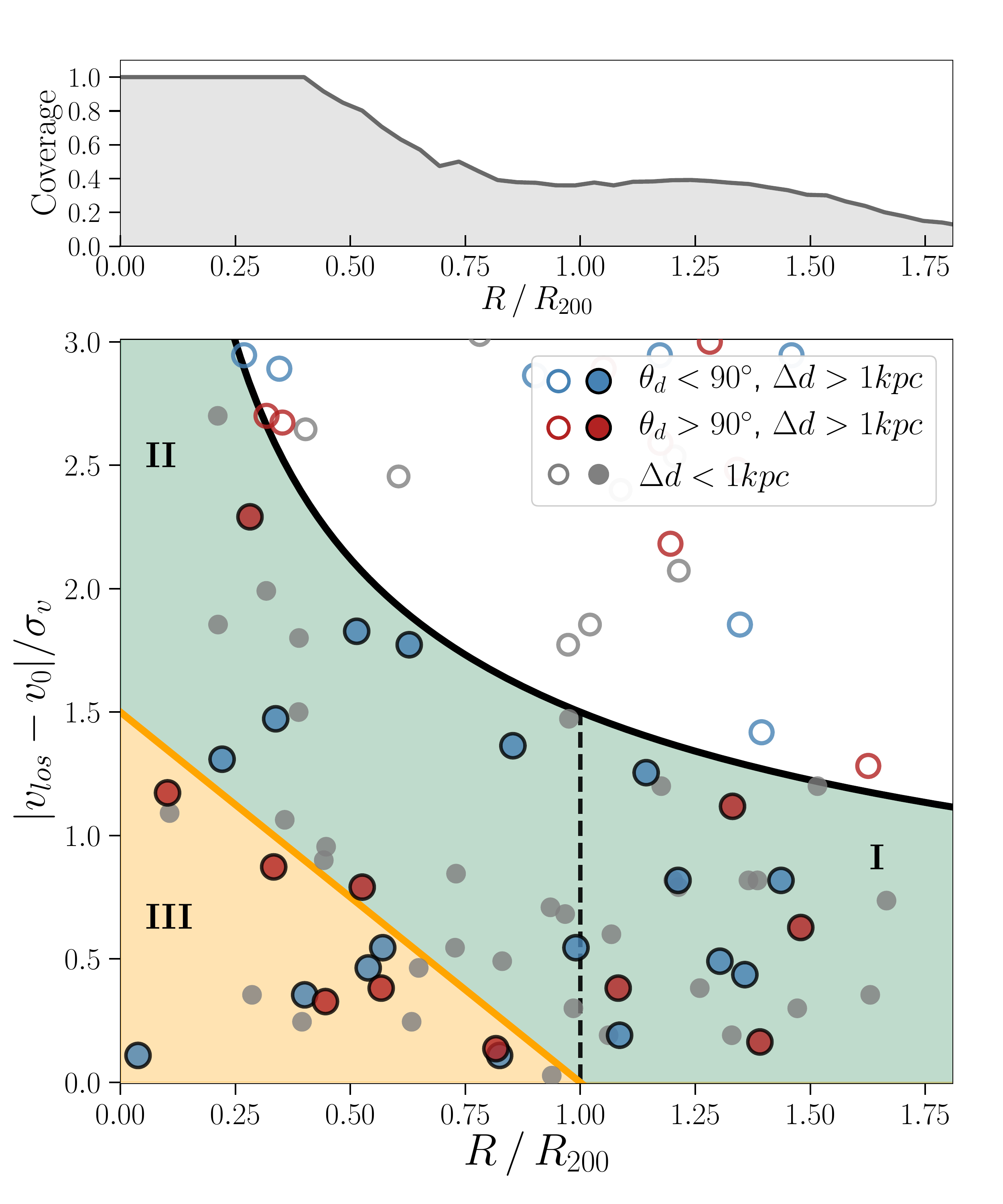}
      \caption{The main panel shows the ELGs in Figure \ref{fig:ELG_PA_A2390} on the position-velocity phase space diagram of A2390, with the same markers and color coding. The distances to the cluster center and the velocities relative to the cluster velocity of galaxies are normalized by $R_{200}$ and $\sigma_{v}$. The black solid curve corresponds to $|v_{los}/v_{200}|$ $\sim$ 1.5$|(R/R_{200})^{-1/2}|$, adopted as the escape velocity boundary. Cluster members that fall below the escape velocity curve are shown as solid markers. The cluster region is divided into three sub-regions: the outer non-virial region I, the inner non-virial region II containing galaxies near orbit pericenters, and the ``virialized'' region III (\citealt{2015MNRAS.448.1715J}). The upper subpanel shows the fractional field coverage for A2390.} 
      \label{fig:phase_A2390}
    \end{figure}

    Figure \ref{fig:ELG_PA_A2390} shows positions of the cluster ELGs in the A2390 fields and Figure \ref{fig:phase_A2390} presents their locations on the phase space diagram. The phase space diagram plots the distribution of the projected positions and velocities of galaxies, normalized by $R_{200}$ and the cluster velocity dispersion, respectively. This is often used as an indication of the cluster galaxy's dynamical state in the cluster (e.g., \citealt{2013ApJ...768..118N},  \citealt{2014ApJ...796...65M}, \citealt{2015MNRAS.448.1715J}, \citealt{2017ApJ...843..128R}, \citealt{2017ApJ...838...81Y}). We use the phase space diagram to identify cluster members, i.e., galaxies bound to the gravitational field of the host cluster, by excluding ELGs falling above the escape velocity boundary (black solid line in Figure \ref{fig:phase_A2390}). 
    
    In Figure \ref{fig:ELG_PA_A2390}, colored symbols mark the selected ELG sample ($\Delta d>$ 1 kpc), where the blue solid circles indicate those with $\theta_d\leq 90^{\circ}$ and red solid circles indicate those with $\theta_d>90^{\circ}$. The BCG itself is a strong ELG (\citealt{2000AJ....119.1123H}), which has been excluded in our analysis. The direction of the ionized gas offset relative to the stellar continuum (opposite to $\mathbf{d}$) measured with flux-weighted centroids is indicated by arrows. The rest of the cluster ELGs with small offset detected ($\Delta d>$ 1 kpc) are shown as gray solid circles. Non-cluster members are marked as open symbols, with the same color coding as the others. There are more cluster ELGs with ionized gas offsets pointing away from the cluster center than those pointing toward the cluster center in the NW and SE fields, while they are comparable in the central field. It is notable that there appears to be a small group of cluster ELGs in the north-west of the cluster center having gas offset away from the cluster center. Indeed, two galaxy groups have been identified in that region using the photometric redshift in \cite{2009ApJ...698...83L}.
    
    \begin{figure}
      \centering
      \resizebox{0.9\hsize}{!}{\includegraphics{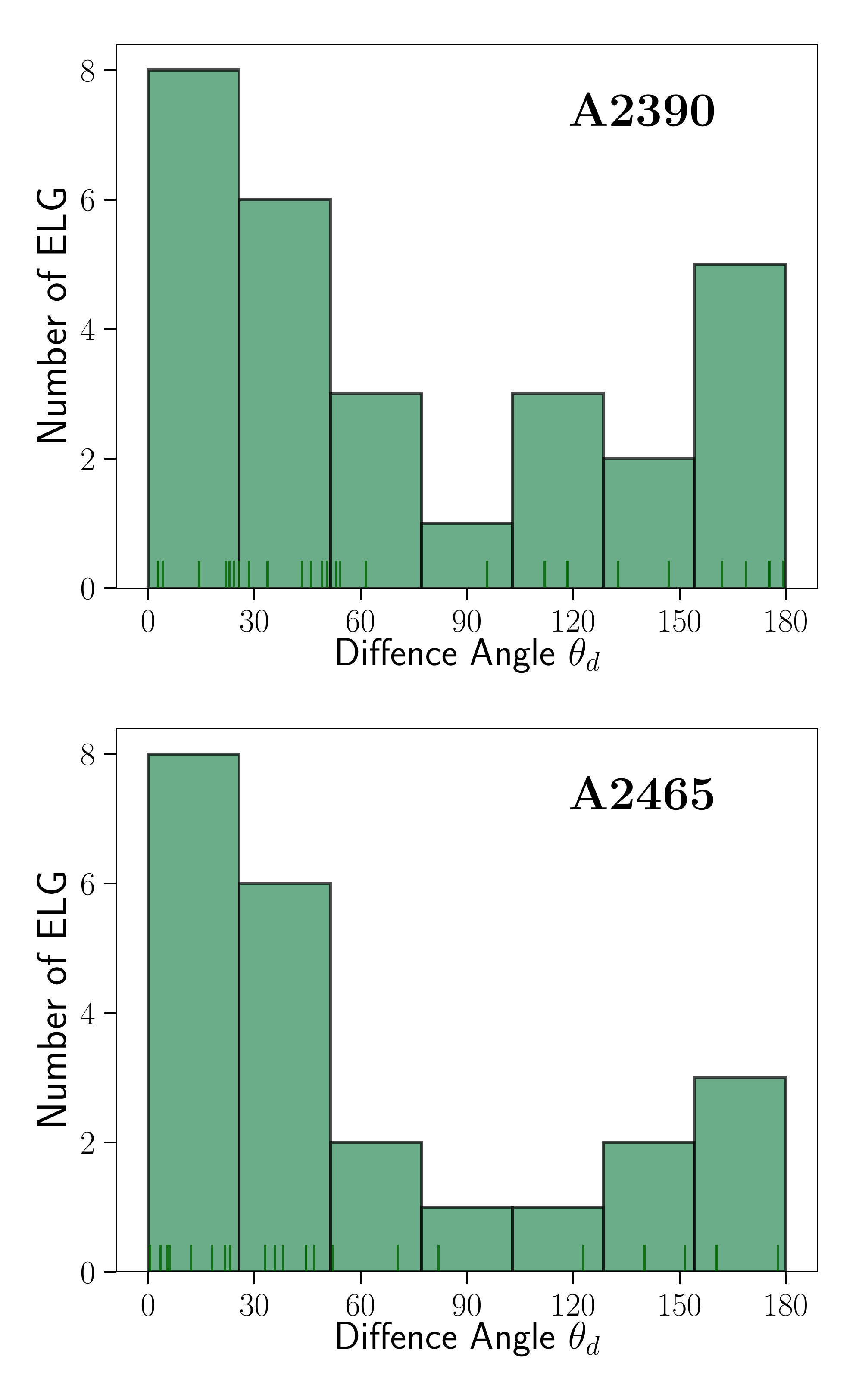}}
      \caption{Histogram of difference angles measured from light-weighted centroids ($\theta_d$) for cluster member ELGs in A2390 (top) and A2465 (bottom) with $\Delta\,d$ $>$ 1 kpc. The cluster membership is determined from the phase space diagram (Figure \ref{fig:phase_A2390}). The bin width is chosen to be larger than $\sigma_{\theta_d}$. Individual data points are shown as small sticks at the bottoms of the histograms.}
    \label{fig:diff_angle_hist}
    \end{figure}
    
    In Figure \ref{fig:phase_A2390}, galaxies are color coded in the same manner as Figure \ref{fig:ELG_PA_A2390}. To avoid visual bias on the actual densities of galaxies, we plot the fractional field coverage in the upper small panel: the cluster is fully covered out to 0.4 R$_{200}$ and it maintains a $\sim$25\% spatial coverage until it drops to zero beyond $\sim$1.7 R$_{200}$. We further divide the in-cluster region (below the escape velocity boundary) on the phase space diagram into three sub-regions, according to the typical path followed by an infalling galaxy during its virialization (see Section \ref{sec:dist_offset_phase}): 
    (I) galaxies outside the cluster virial radius; 
    (II) high-velocities galaxies within the virial radius; 
    (III) low-velocities galaxies within the virial radius. 
    The distribution of the selected ELGs on the phase space diagram reveals that the ratio of those with offsets away from the cluster center (blue symbols) to those toward the center (red symbols) tends to increase from low to high velocity. This is most significant in region II -- a simple number count yields 6:2 in region II, compared with 6:4 in region I and 5:5 in region III. We will further discuss the physical interpretation in Section \ref{sec:dist_offset_phase} below.
    
    The left histogram in Figure \ref{fig:diff_angle_hist} presents the distributions of $\theta_d$ for the selected cluster member ELGs of A2390. Several patterns can be revealed: first, the distributions of the difference angles show distinct deviations from uniformity. 
    We perform the Kolmogorov-Smirnov (K-S) test to test the uniformity of the distributions of $\theta_d$. Based on the p-values calculated, we can reject the uniformity of $\theta_d$ at a 95\% confidence level ($p=0.03$). 
    As discussed below, alternative triggers of ionized gas offset other than ram pressure are expected to result in a more or less flat distribution.
    Second, the distribution of $\theta_d$ shows a clear peak at the $0^\circ$ end. The histogram suggests that the emission offset is found to be preferentially pointed away from the cluster center. This is expected from the effect of ram pressure stripping, considering the infalling and quenching process of cluster galaxies (see discussion in Section \ref{sec:infall_history}). Finally, There is a hint of another peak at the $180^\circ$ end, suggesting the emission offsets to be toward the cluster center. The possible peak mainly consists of objects in region III that suffer significantly from projection effect or might in the backsplash stage of their infall (see discussion in Section \ref{sec:excess_offset}). Further observations with a larger cluster sample are needed to confirm its presence.
    
    In general, the distribution of difference angles in A2390 shows that the spatial offsets of ionized gas in cluster ELGs have a preference that is correlated to the cluster center, indicating the impact of ram pressure stripping from the ICM on galactic gas reservoirs.

    \subsubsection{Abell 2465}
    \label{sec:A2465}
    
\begin{figure}
      \centering
      \includegraphics[width=\hsize]{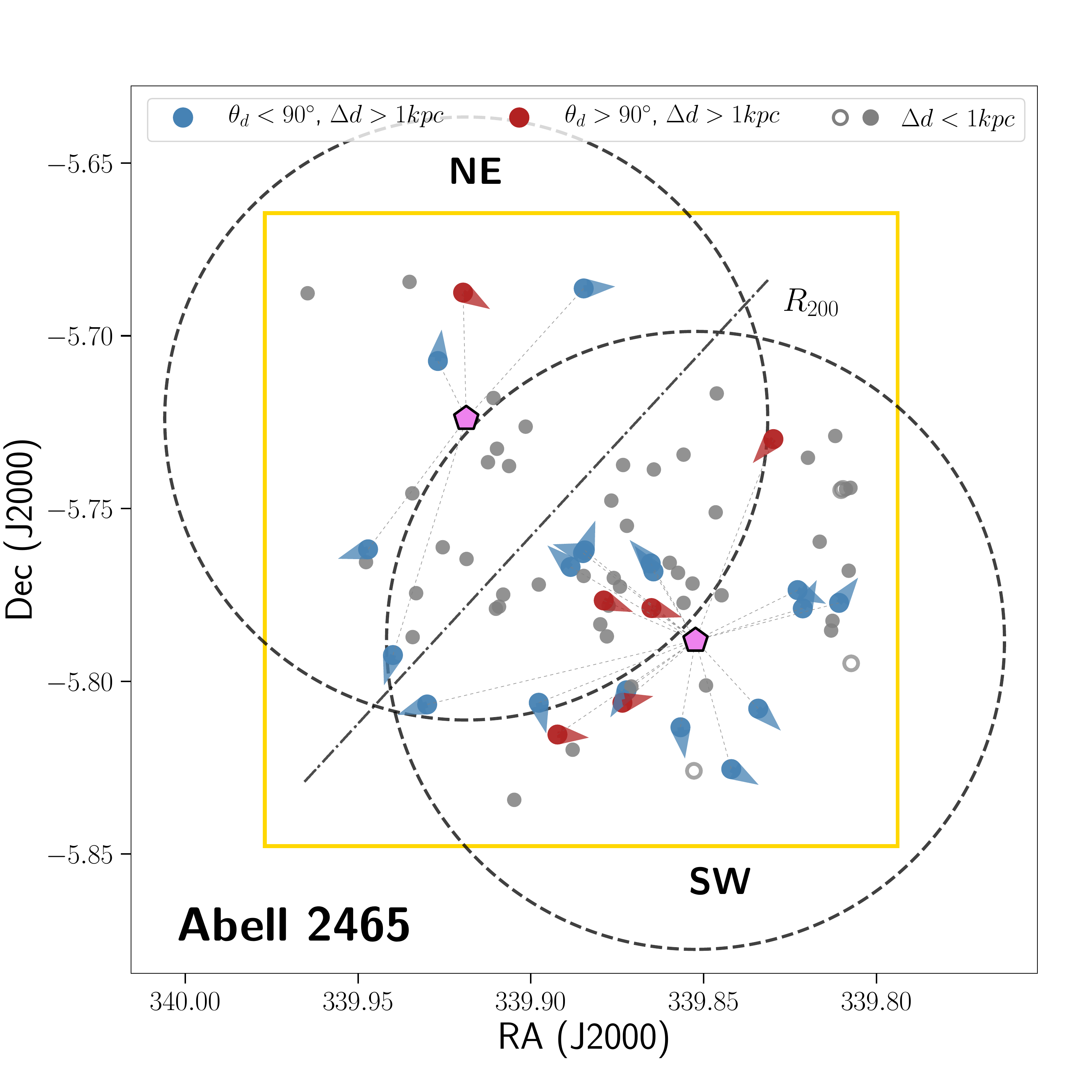}
      \caption{Locations of the selected ELGs ($\Delta d>$ 1 kpc) in A2465. Markers are colored coded in the same manner as Figure \ref{fig:ELG_PA_A2390}. ELGs with large centroid uncertainties or near field edges are not shown. The black dashed circles indicate their virial radii. The dotted-dashed line marks the empirical boundary adopted for cluster member classification. The yellow frame indicates the field footprint.}
      \label{fig:ELG_PA_A2465}
    \end{figure}

\begin{figure}
      \centering
      \includegraphics[width=\hsize]{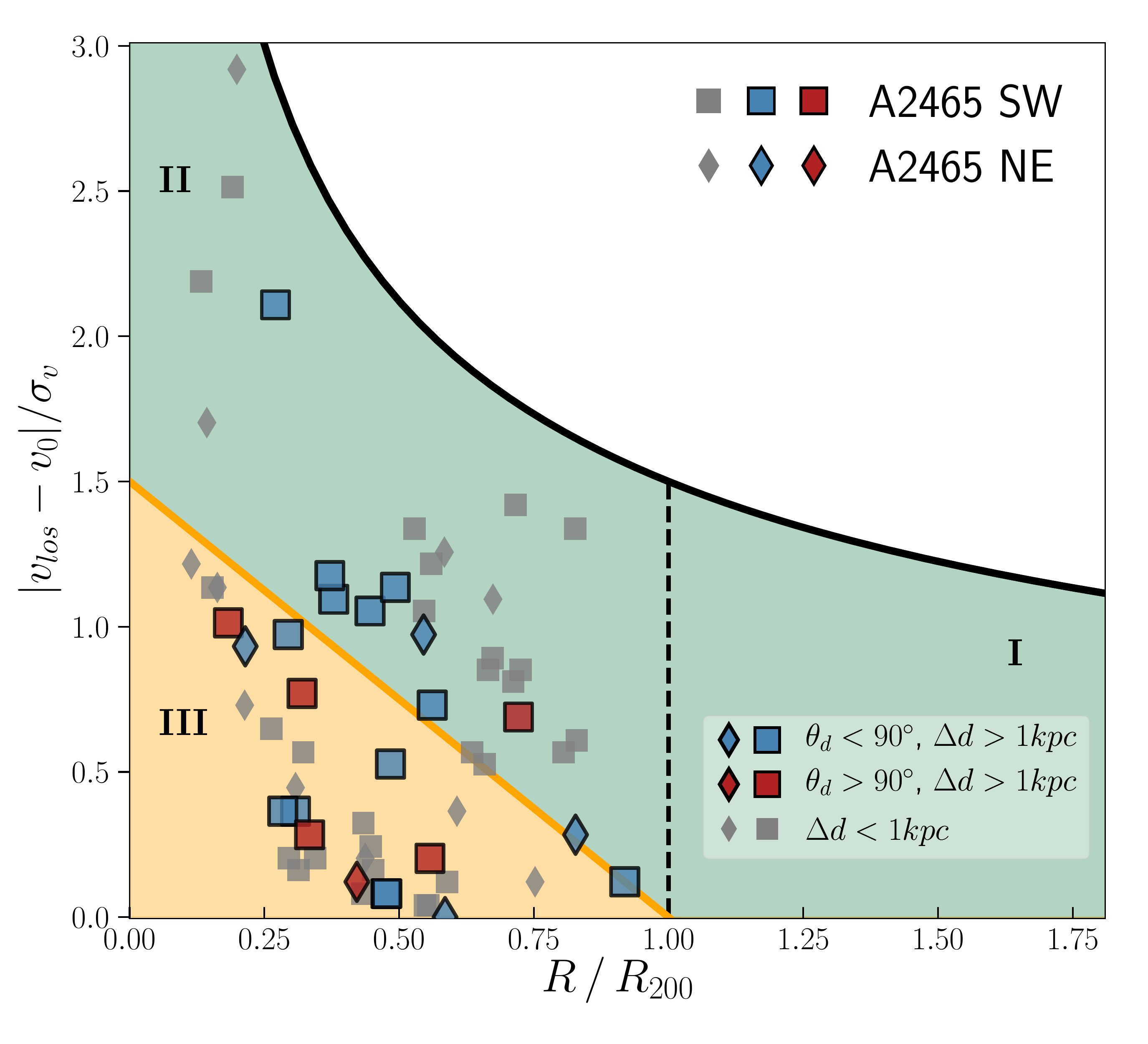}
      \caption{ELGs in Figure \ref{fig:ELG_PA_A2390} on the position-velocity phase space diagram of A2465. Markers are color coded in the same way as in Figure \ref{fig:ELG_PA_A2390}. The black solid curve corresponds to $|v_{los}/v_{200}|$ $\sim$ 1.5$|(R/R_{200})^{-1/2}|$, adopted as the escape boundary. Cluster members below the escape velocity curve are shown as solid markers. The region division is the same as in Figure \ref{fig:phase_A2390}. To match with Figure \ref{fig:phase_A2390}, region I is shown although with no data coverage.}
      \label{fig:phase_A2465}
    \end{figure}

    Figure \ref{fig:ELG_PA_A2465} shows the position of ELGs identified in the double cluster A2465; while Figure \ref{fig:phase_A2465} presents their locations in the phase space diagram. Symbols and their color coding follow the same as Figures \ref{fig:ELG_PA_A2390} and \ref{fig:phase_A2390}. The membership discrimination between the two sub-clusters is simplified by using a straight line that divides the double cluster \footnote{In specific, the linear division is trained based on sub-cluster membership presented in \cite{Wegner2011} obtained from kinematics using support vector machine.} because the two components have comparable masses. It can be observed from Figure \ref{fig:ELG_PA_A2465} that there are more ELGs in the south-west sub-cluster, with the majority located at the border of the two sub-clusters. However, a large fraction of ELGs at the collision border do not present significant systematic emission offset, possibly caused by the complex condition of ICM imprinted by shocks and/or galaxy kinematics there, while projection effects may also play a role.
    
    The right panel of Figure \ref{fig:diff_angle_hist} presents the distribution of $\theta_d$ for cluster member ELGs in A2465 with $\Delta d>$ 1 kpc. Similar to A2390, the distribution of $\theta_d$ shows deviation from a uniform angle distribution, with a major peak at $0^\circ$ and a hint of a weak peak at $180^\circ$.
    The K-S test indicates the distribution is significantly different from the uniform distribution at the confidence level of 95\%. 
    
    The result in A2465 is consistent with the scenario where ionized gas spatial offset is induced by ram pressure during the infall of gas-rich galaxies into the cluster (see further discussion in Section \ref{sec:dist_offset_phase}). However, it is not clear whether such effect is enhanced or suppressed by the collision of the two clusters.

\section{Discussion} \label{sec:discussion}
    
In this section, we discuss the implications of the observed spatial offsets of ionized gas, specifically on the non-uniform distribution of difference angles and the excess in region II on the cluster phase space diagram. We then propose a scenario which qualitatively explains the observation.

\subsection{Ionized Gas Centroid Offset: Ram Pressure Stripping in the Act}
    \label{sec:gas_offset_rps}
    It is natural to infer that under ram pressure, the gas offset in a galaxy moving through the ICM generally follows the opposite direction to its velocity relative to the ICM, given the pressure follows (\citealt{1972ApJ...176....1G}):
    \begin{equation}
        P_{ram} = \rho_{\rm ICM} v_{gal}^2\,,
    \end{equation}
    where $\rho_{\rm ICM}$ is the local ICM density and $v_{gal}$ is the relative velocity. Due to the projection effect, it is hard to determine velocities in 3D and connect them to the 2D gas offsets. However, observations have revealed that recent infall gas-rich galaxies in clusters affected by ram pressure are mostly on highly radial orbits (\citealt{2007ApJ...659L.115C}, \citealt{2010MNRAS.408.1417S}, \citealt{2017ApJ...837..126V}, \citealt{2018MNRAS.476.4753J}; see also \citealt{2014ApJ...781L..40E} where such effect is observed on galaxies with tangential infall orbits.). This suggests a connection between the projected offset of ionized gas emission in these galaxies and their directions to the cluster center. In fact, \cite{2010MNRAS.408.1417S} observed that ionized gas tails of gas-stripping galaxies in the nearby Coma cluster predominantly point opposite to the cluster center using UV+{\ha} data. \cite{2017ApJ...837..126V} found that ram pressure stripped SFGs out to 0.5 virial radius in intermediate redshift clusters preferentially show radial {\ha} offsets away from the cluster center. Using the IllustrisTNG simulations, \cite{2019MNRAS.483.1042Y} investigated the directions of gas tails of jellyfish galaxies in galaxy clusters in 3D. They found a nearly flat distribution with slow drops in the $0^{\circ}$ and $180^{\circ}$ ends for the angle between gas tail and direction to the host center. This is not unexpected, as, unlike in 2D, in 3D only galaxies with purely radial orbits would show gas tails aligned with the direction to the cluster center. However, they did observe a tight correlation between the directions of gas tails and the 3D velocity vector, which should also hold in 2D projection, allowing us to use the offset vector as a proxy for the galaxy's 2D projected velocity vector.

    Besides ram pressure stripping, other physical processes might serve as potential explanations for the ionized gas offset such as:
    
    (1) Outflows: strong outflows driven by active galactic nuclei (AGNs) or starbursts have been observed in many active galaxies, which could also lead to significant gas offset (e.g., \citealt{2017ApJ...847...41C}, \citealt{2019MNRAS.490.3025R}). They are considered as an important source of ICM heating to explain the cooling problem in galaxy clusters (e.g., \citealt{2005ApJ...630..740B}). Indeed, the high-velocity {\ha} bump in the the north-west of the BCG of A2390 is likely to be driven by an AGN. Because we have \textnormal{included galaxies with nuclear activities} in our ELG sample, it cannot be ruled out that a portion of the spatial offsets of ionized gas in our ELG sample are caused by outflows. \textnormal{In fact, recent studies based on simulations and observations found that galactic nuclear activities can be triggered by ram pressure stripping due to enhanced accretion onto the central black hole (e.g. \citealt{2017Natur.548..304P}, \citealt{2018MNRAS.476.3781R}, \citealt{2020ApJ...895L...8R}). However, there has been little evidence of preference in outflow directions in galaxy clusters. Because outflow is an internal phenomenon controlled by the central black hole in the nuclei region, its subsequent interaction with the gas in the host galaxy is likely independent of the global environment. Therefore, gas offsets induced by outflows are not expected to have a correlation with the cluster center as observed.} Further analysis on the kinematics and ionization conditions can discriminate gas offsets caused by outflows and that from ram pressure.
    
    (2) Tidal disruption: tidal forces from nearby encounters may contribute to the observed disturbance of the gas. For some ELGs in our sample, we indeed observe neighboring objects that are possibly interacting with them. However, pure tidal stripping have similar gravitational effects on gas and stars, leaving no offset between them.
    On the other hand, results from simulations (e.g., \citealt{2006MNRAS.369.1021M}; \citealt{2008MNRAS.389.1405K}; \citealt{2013MNRAS.435.2713V}) have demonstrated that ram pressure stripping is in fact more efficient in the presence of tidal interactions. This is because the gas pulled out of the galactic potential is more likely to be stripped away by the ICM wind and local density $\rho_{\rm ICM}$ is increased by compression due to merger shocks. A larger sample is needed to construct a sample of mergers with reference to non-mergers to investigate the difference in their ionized gas distributions.
    
    \textnormal{Other systematics such as the presence of AGNs and dust cause dilution effects in our centroid measurements that decreases the large $\Delta d$ sample size and increases the uncertainties, but are unlikely to change the observed patterns.} Therefore, we conclude that the non-uniform distribution of the spatial offset of the ionized gas with respect to the stellar disk in our observation is primarily contributed by ram pressure stripping. In comparison, the long-term interplay between a galaxy and the host cluster, such as strangulation or thermal evaporation, would typically leave behind an undisturbed, symmetric gas distribution.

\subsection{The Distributions of Ionized Gas Offset in Phase Space}
    \label{sec:dist_offset_phase}
    As discussed above, we are able to use the offset vector $\mathbf{d}$ as a proxy for the projected velocity of an infalling galaxy, in light of ram pressure stripping acting as the main driver producing the non-uniformity that we observe in distributions of $\theta_{d}$. Combining with the infall history of a typical gas-rich cluster galaxy in the phase space diagram, the distributions of ELGs with large ionized gas offset in phase space offer evidence of ram pressure stripping as a quenching mechanism in the act.
    
    \subsubsection{Infall History of a Cluster Galaxy on the Phase Space Diagram}
    \label{sec:infall_history}
    
    \begin{figure}
      \centering
      \includegraphics[width=\hsize]{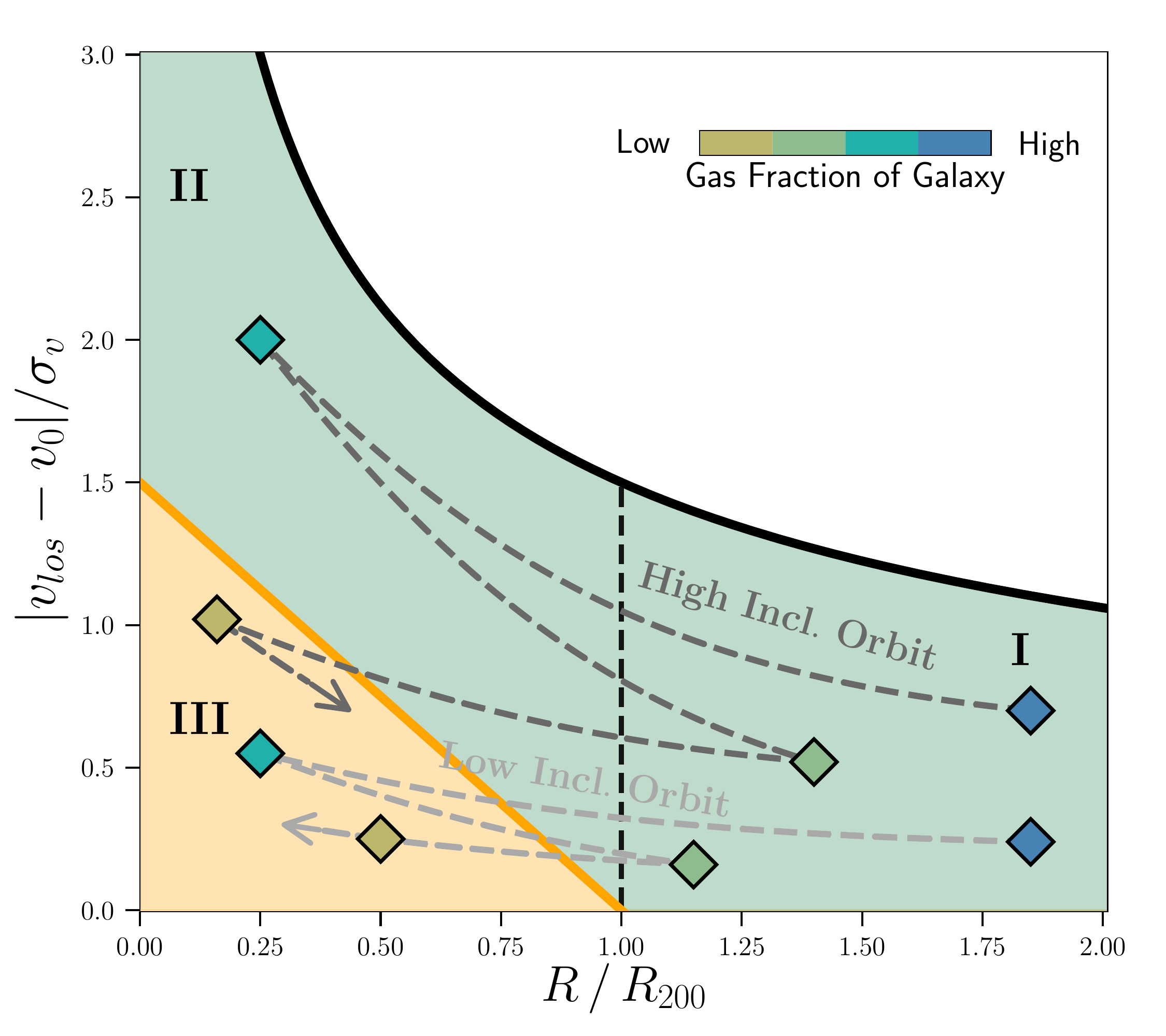}
      \caption{Schematic of the trajectories of an infalling gas-rich cluster galaxy on the velocity-position phase space diagram. The dark dashed line shows the typical path in which it moves from region I to region II, undergoes a few turnarounds, and eventually being virialized in region III. The light dashed line shows the path of a similar galaxy but with a low inclination (nearly face-on) orbit. It spends most of the time in region III from its first passage due to projection effect. The color coding illustrates the gas fraction of the galaxy, with the dark/light-green points indicating orbit pericenters/apocenters. The black dashed line indicates $R_{200}$. The figure is adapted from Figure 1 in \cite{2017ApJ...843..128R} (note the original plot is in 3D).}
      \label{fig:Phase_Schematic}
    \end{figure}
    
    The trajectory of an infalling galaxy can be traced on the cluster phase space diagram. The dark dashed curve in Figure \ref{fig:Phase_Schematic} illustrates the typical infall history of a cluster galaxy.
    
    Generally, an incoming SFG is first found in region I with a high gas fraction. During its first infall into the potential well of the cluster, the orbit can be highly radial (e.g. \citealt{2017ApJ...837..126V}, \citealt{2018MNRAS.476.4753J}). It will increase its velocity as it approaches the orbit pericenter, moving toward the upper left of the phase space diagram (region II). The local ICM density also increases as the galaxy penetrates into the cluster core. Consequently, the ram pressure from head-on ICM winds increases dramatically. It compresses and strips the cold gas, while having little effect on the stellar components. The SFR may temporarily rise due to the compression of gas (e.g., \citealt{1999ApJ...516..619F}, \citealt{2017ApJ...844...48P}, \citealt{2018ApJ...866L..25V}). As star formation proceeds, the neutral gas offset propagates to ionized gas, leading to the difference in the spatial distribution of emission and continuum.
    Post pericenter, it will then turn around with decreasing velocity and move out of the central region, referred to as the ``backsplash'' stage (e.g., \citealt{2005MNRAS.356.1327G}, \citealt{2014A&A...564A..85M}). As a result, the backsplash region overlaps with the lower-velocity part of the first infall region. The galaxy then enters its subsequent infall but with a lower gas fraction. Such process may happen multiple times until the radial velocity is dissipated as the galaxy spirals in. Eventually, the galaxy falls into the virialized region, with little or no gas reservoir left to maintain its star formation.
    
    The fact that cluster galaxies follow typical paths in the course of their sinking into the cluster potential leads to their occupations of certain regions on the phase space diagram. However, the condition becomes more complex when orbit orientation and projection effects are taken into account. The light dashed line in Figure \ref{fig:Phase_Schematic} illustrates the path of an infalling galaxy with a low inclination angle orbit, hence, with low line-of-sight velocity. While it still takes a few turnarounds to achieve the virialization of the orbit, the galaxy is not able to move up to region II in the phase space diagram due to the viewing angle of the orbit. As a consequence, region III also contains galaxies in their first passage. Another concern is that an infalling galaxy physically outside the virial radius may be projected to be within $R_{200}$ on the phase space diagram. For an extreme case, a galaxy following an orbit along the line of sight could have strong emission-lines detected if there is ongoing star formation, but it would barely move in the radial axis. However, in this case the gas offset is also projected, which makes it more likely to have a small $\Delta d$ and thus be excluded from the selected sample. 
    
    \subsubsection{Excess of Gas Offset Vectors toward the Cluster Center near Orbit Pericenters at First Infall}
    \label{sec:excess_offset}
    
    The number ratios of blue to red symbols in region II (6:2 in A2390, and 8:1 in A2465 combining two subclusters) suggest that first-infalling galaxies have a preference for their ionized gas offsets to point away from the cluster center when they approach the orbit pericenters. The probability to have such ratios by coincidence is 11\% for A2390, 4\% for A2465, and 1\% combining the two, assuming an equal opportunity (i.e. a binomial distribution with p=0.5) for the offset vector to point toward/away from the center. 
    
    Such preference can be attributed to the impact of the radial component of the ram pressure on the cluster galaxy. The result is consistent with the other observational evidence and simulations that suggest ram pressure stripping has substantial effects when gas-rich galaxies cross the cluster virial radius and is most effective in their first infall (e.g., \citealt{2016MNRAS.461.1202J, 2018MNRAS.476.4753J}; \citealt{2019MNRAS.488.5370L}). Because of the efficient removal of cold gas, star formation has been mostly suppressed before the incoming galaxies turn around and start their second infalls. The typical quenching timescale for these galaxies is thus suggested to be shorter than the ram pressure stripping timescale ($\lesssim 1.5$ Gyr within the virial radius, \citealt{2011MNRAS.413.2057R}) and the dynamic timescale of first infall (1$\sim$2 Gyr, \citealt{2018MNRAS.476.4753J}). Indeed, using cosmological hydrodynamical simulations, \cite{2019MNRAS.488.5370L} found most satellite galaxies are quenched within 1 Gyr during the first infall. 
    
    Starvation/strangulation may also play an important role in quenching cluster galaxies (e.g., \citealt{2013ApJ...775..126H}, \citealt{2015Natur.521..192P}) in which a cutoff of gas supply into the galaxy leads to the long-term exhaustion of gas and thus a drop in the SFR. Whether the effect is dominant for the evolution of cluster galaxies has not reached a consensus (e.g., \citealt{2014A&A...564A..67B}, \citealt{2015Natur.521..192P}). However, such process operates on a longer timescale ($>$3 Gyr, \citealt{2014A&A...564A..67B}) and does not leave systematic imprint on the gas spatial distribution. Therefore we conclude at least for cluster galaxies entering region II (which have fallen into the cluster $<$ 1$\sim$2 Gyr ago), ram pressure stripping is the predominant quenching mechanism, although it is likely that halting the gas supply also catalyzes the process. Combining with other information such as the SFR, and gas fraction, and gas kinematics, etc., future analyses using large samples of clusters would be promising in distinguishing the contributions of different quenching mechanisms.
    
    Roughly equal numbers of blue (offset away from the center) and red (offset toward the center) symbols are observed in the lower-velocity regions (regions I and III). 
    If a galaxy showing an offset toward the center has already entered and left region II and returned to region I (Figure \ref{fig:Phase_Schematic}), ram pressure stripping is expected to remove most of the gas and lower the level of star formation. However, they may still have detectable emission lines, which indicates the existence of ionized gas. Several origins may explain their detected line emission and gas offsets: (1) they have low orbit inclinations, i.e., being more face-on (the light dashed trajectory in Figure \ref{fig:Phase_Schematic}), which leaves a larger projected gas offset, and thus have higher chances to be included in the selected sample during the ``backsplash'' stage. (2) They could be low-velocity infalling galaxies with large impact parameters. The lower incoming velocities cause the ram pressure to be low near orbit pericenters. (3) They might contain face-on high-velocity fly-bys that spend less time near the pericenters. For a fly-by, the gas is only partially stripped by ram pressure after the passage close to the cluster core, but the disturbance could be prominent due to its high velocity. (4) They contain contamination from outflows, which presumably do not have a directional correlation with the cluster. Projection effects make it challenging to disentangle the infalling histories of low-velocity cluster members as galaxies with different dynamics and initial conditions mix together. But we note that for the first three cases, the galaxy never enters, or spends very little time, in the high-velocity region on the 2D phase space diagram, and therefore would not affect our conclusions in region II.
    

    Finally, it is noteworthy that although the result of A2465 is consistent with A2390, the infalling history discussed in Section \ref{sec:infall_history} may not fully apply to A2465. This is because most interpretations on the phase space diagram here assume a single, undisturbed, cluster, which may not be fully applicable to merging clusters due to the possibly more complicated ICM condition and galaxy dynamics. In addition, a selection effect exists due to the filter bandwidth, which disables the sampling of receding objects in A2465 with velocity higher than 1$\sigma_{v}$. Both may contribute the scarcity of ELGs with high velocity ($>$1.5 $\sigma_{v}$) in Figure \ref{fig:phase_A2465}. Our current data of A2465 are also restricted within $\sim$1 Mpc. As pointed by \cite{2017ApJ...837..126V}, detailed knowledge about the dynamics of cluster mergers is needed for in-depth analysis, which is beyond the range of this work. Nevertheless, it is promising to gain more insightful and definitive results with the aid of further data and comparisons with simulations.
    
    \subsection{Additional Support for Quenching by Ram Pressure Stripping} \label{sec:Haflux}
    
    To find additional evidence of ram pressure stripping affecting the star formation in cluster galaxies, we investigate the line flux. Here we only look into A2390 for two reasons:(1) our current A2465 data do not have coverage in the outer region, and (2) A2390 is more of a typical \textnormal{rich} cluster compared with the merging double cluster A2465. The {\ha} emission-line flux $F_{\ha}$ is obtained from the residual spectra yielded in Section \ref{sec:cont_removal} by a trapezoidal integral of $\pm 5\,\sigma_{line}$ around the peak of the {\ha} line, where $\sigma_{line}$ is the best-matched template line width. To normalize the emission-line flux by galaxy mass, we divide it by the mean continuum times a fixed wavelength range of 250$\AA$. The normalized line flux $F_{{\ha}, n}$ \textnormal{can be used as} a proxy for the specific star formation rate (sSFR), defined as the SFR per stellar mass, \textnormal{under the circumstance where the ionization is caused by star formation. For a composite ionization of star formation and AGN, $F_{{\ha}, n}$ should be treated as upper limits of star formation. $F_{{\ha}, n}$ also has a dependency on the mass-to-light ratio (and thus the star formation history) of the galaxy, but the dependency is minor, and we use it here as a first-order estimate.} 
    
    \begin{figure}
      \centering
      \includegraphics[width=\hsize]{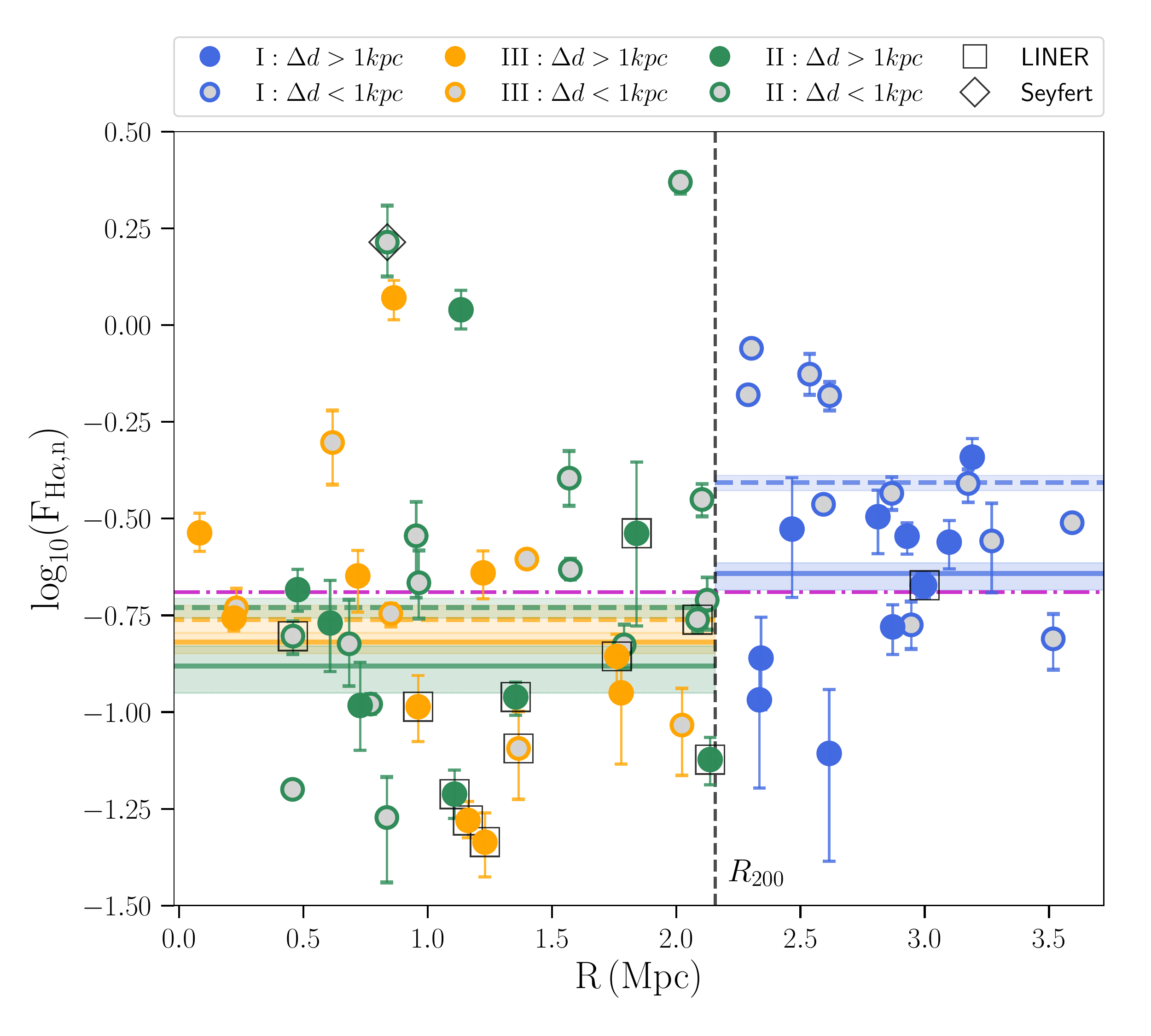}
      \caption{Emission-line flux normalized by continuum vs distance to the cluster center for cluster member ELGs in A2390. ELGs with large/small (projected) ionized gas offset $\Delta\,d$ are shown as colored/gray-filled symbols. The face/edge-colors mark ELGs in different regions on the phase space diagram (Figure \ref{fig:phase_A2390}). The colored solid/dashed lines indicate the central locations of distributions of the colored/gray symbols with the corresponding color. The colored bands indicate the 1$\sigma$ (16\%-84\%) uncertainty of the mean. \textnormal{Galaxy with Seyfert/LINER features are marked in diamonds/squares}. The magenta dashed-dotted line shows the central location of the distribution of all markers. The vertical black dashed line marks $R_{200}$.}
      \label{fig:line_flux_A2390}
    \end{figure}
    
    The continuum-normalized line flux is plotted in Figure \ref{fig:line_flux_A2390} versus the distance to the cluster center. ELGs with larger $\Delta\,d$ are shown by solid colored symbols, whereas those with \textnormal{smaller or undetected} $\Delta\,d$ are represented by gray-filled symbols. Blue, green, orange colors represent ELGs in the low-velocity outer region (region I), the high-velocity inner region (region II), and the virialized region (region III), respectively, as indicated in Figure \ref{fig:phase_A2390}. The 1$\sigma$ error bars are calculated numerically by Monte Carlo simulations, where the spectra are perturbed 250 times using the rms noise of the continua. We use the biweighted location, which is a robust statistic to represent the central location of a distribution. The biweighted locations of solid/open symbols are shown as colored solid/dashed lines with the corresponding colors. The central location of all markers is shown as the magenta dashed-dotted line. The shaded bands indicate the 1$\sigma$ uncertainties of the central locations, calculated by Monte Carlo simulations where each data point is randomly perturbed based on its uncertainty. \textnormal{For the initial identification of galaxies with AGN features, we adopt a classification based on the WHAN diagram (\citealt{2011MNRAS.413.1687C}) using line ratios obtained from cross-correlation. We label them in Figure \ref{fig:line_flux_A2390} but do not exclude them from the sample, given that those with LINER features present simultaneous ionization from star formation in the disk and Seyferts are rare. AGNs in our fields will be investigated in detail in future work.} 
    
    Several interesting patterns are revealed in Figure \ref{fig:line_flux_A2390}. In the outer region (region I), ELGs with smaller $\Delta\,d$ have higher $F_{{\ha}, n}$ on average ($\sim$0.2 dex) than those with larger $\Delta\,d$. The significance is above 3$\sigma$. This is also true for samples in region II - larger $\Delta\,d$ objects have lower $F_{{\ha}, n}$, although the significance (at 1.9$\sigma$) is not high considering the uncertainties. Region III appears to have statistically consistent average $F_{{\ha}, n}$ for large and small $\Delta\,d$ objects. Furthermore, comparing objects within $R_{200}$ (region II \& III, green \& orange symbols) with objects outside $R_{200}$ (region I, blue symbols), they have on average lower $F_{{\ha}, n}$ ($\sim$0.2/0.3 dex for large/small $\Delta\,d$ objects), although a small fraction of objects within $R_{200}$ appear to show strong lines.
    
    The lower average line flux for large $\Delta\,d$ objects in region I could be the result of them undergoing ram pressure stripping, with the star formation being suppressed due to gas removal. Some of these could have already passed through the pericenter of their infall orbit, moving out to the backsplash stage. Others may be in their first infall, where ram pressure stripping has just been turned on not long ago. On the other hand, objects with smaller $\Delta\,d$ are likely incoming galaxies that have not experienced significant, if any, ram pressure stripping. They might be farther away from the cluster but projected to be closer in 2D, or with smaller impact parameters and thus less affected by the ICM winds. Again, we note that galaxies with small $\Delta\,d$ could also have suffered from ram pressure stripping, but with their small $\Delta\,d$ being the result of projection effect. However, the lower emission line flux observed in larger $\Delta\,d$ ELGs clearly serves as additional evidence of ram pressure stripping being responsible for quenching.
    
    As discussed, the large $\Delta\,d$ objects in region II are likely experiencing strong ram pressure stripping. This effect can consistently explain the difference in line flux compared with small $\Delta\,d$ objects. However, the composition of the small $\Delta\,d$ objects is actually complex: it may contain galaxies (1) in their first passages where gas removal through ram pressure stripping is in progress; (2) in the backsplash stage, having survived from the dramatic gas removal; and (3) with prominent gas offset in 3D, but projected to be small. This may explain the larger scatter in $F_{{\ha}, n}$ for small $\Delta\,d$ objects. In the first case, the galaxies have higher line fluxes as they are in the beginning stage of having their star formation undergoing suppression, while in the latter two cases, the level of star formation is expected to be comparable or lower. In the last case, some galaxies that appear within $R_{200}$ might actually fall outside it in 3D due to projection and thus would have a smaller amount of star formation suppression. It would be interesting to discriminate and study them using other tracers such as gas fractions.
    
    In region III, large and small $\Delta\,d$ objects have consistent line fluxes. This suggests that either the offset is unassociated with ram pressure stripping (e.g. from outflows), or the offsets have different origins but the galaxies are projected to be there. Galaxies in region III suffer from all kinds of projection effects discussed above, making them hard to interpret in the phase space alone. 
    
    The few objects inside $R_{200}$ observed with exceptionally high $F_{{\ha}, n}$ suggest that a small fraction of cluster galaxies suffered from ram pressure stripping could have enhanced star formation, or nuclei activities, due to gas compression by shocks. The enhancement is suggested to have short timescales because the fraction of such objects is low. Such enhancement has been reported by many previous studies (e.g. \citealt{1999ApJ...516..619F}, \citealt{2008A&A...481..337K}, \citealt{2014MNRAS.438..444B}). \textnormal{More recently, \cite{2018ApJ...866L..25V} found enhanced SFR in a statistically significant sample of jellyfish galaxies from GASP. However, there is a difference in sample construction given the low occurrence of jellyfish galaxies per cluster, whereas our sample mostly consists of mild cases of galaxies undergoing ram pressure stripping.}
    
    Finally, the lower average line flux in regions interior to $R_{200}$ serves as an indication of environmental quenching, suggesting that the sSFR is dependent on the cluster-centric radius. One of the most competitive mechanisms is ram pressure stripping, as discussed above, while other quenching mechanisms (starvation, thermal evaporation, etc.) could take charge when ram pressure stripping is not efficient enough.

\section{Summary} \label{sec:summary}

    An ongoing survey using the IFTS SITELLE at CFHT targeting the {\ha}-{\nii} lines in clusters at z$\sim$0.25 is being carried out to study properties of ELGs (morphology, kinematics, abundance, SFR, etc.) and quenching mechanisms in dense environments. We have developed a pipeline using the cross-correlation technique to detect and identify ELGs from the datacube acquired by SITELLE. So far, we have obtained a list of ELG samples in two galaxy clusters, Abell 2390 and Abell 2465, with 194 ELGs from three fields of A2390 and 110 ELGs from the central field of A2465. The spatially-resolved spectroscopy allows us to separate the emission-line distribution from the ionized gas and the continuum distribution from the stellar populations. We conduct a centroid analysis by comparing the centroids of the emission-line and continuum distributions for the ELGs. We measure the centroid offset $\Delta\,d$ and the difference angle $\theta_d$ between the emission-to-continuum offset vector $d$ and the cluster-centric vector to investigate the correlation between spatial offsets of ionized gas in cluster galaxies and their positions in the galaxy cluster. 
    
    Based on the distributions of $\Delta\,d$ and $\theta_d$, we find (1) ELGs in A2390 and A2465 have $\Delta\,d$ extending from 0 to around 4 kpc with the peak of the offset distribution deviating from zero, implying some systematic impact causing the spatial offset of ionized gas. (2) Looking into ELGs with larger emission-line spatial offsets ($\Delta\,d>$ 1 kpc), the distributions of $\theta_d$ in A2390 and A2465 clearly deviate from a uniform distribution. The peak of $\theta_d$ toward $0^{\circ}$ in our result is consistent with previous work that ionized gas preferentially points away from the cluster center (e.g. \citealt{2010MNRAS.408.1417S}, \citealt{2017ApJ...837..126V}), although we also observe a hint of a minor peak in the $180^{\circ}$, possibly caused by backsplash or projection effects. The p-values from KS tests reject the uniform hypothesis at 95\% confidence. This serves as an evidence of ram pressure stripping playing an important role in shaping and removing the gas reservoirs of cluster galaxies, because other mechanisms such as outflows and tidal stripping are expected to have no directional relation to the cluster center while ram pressure stripping impacts radially in the course of the infall of cluster galaxies. 
    
    We further investigate their distributions on the position-velocity phase space diagram, using the offset vector $\mathbf{d}$ as a proxy for the 2D projected velocity vector. We divide the phase space diagram into three regions according to the infall history of a cluster galaxy: (I) the outskirts region outside the virial radius, (II) the high-velocity region inside the virial radius, and (III) the low-velocity region inside the virial radius. We look into the ratio of ELGs with large $\Delta\,d$ having $\theta_d<90^{\circ}$ and $\theta_d>90^{\circ}$ (i.e. with emission offset vectors pointing away and toward the cluster center, respectively). Combining the statistics from both clusters, we find a 3$\sigma$ excess of galaxies in region II with $\theta_d<90^{\circ}$, i.e., galaxies approaching or close to the orbit pericenters in their first infalls. The preference suggests star formation in these galaxies is mostly suppressed before they turn around and start their second infalls. This is consistent with conclusions from some numerical studies and observations of individual gas-stripped cluster galaxies that galaxies are quenched at first infall under ram pressure stripping after they penetrate into the ICM halo. In addition, we find that the continuum-normalized line flux of ELGs with large $\Delta\,d$ in region I (and region II, but with a lower significance), where galaxies are at the beginning of their first infalls or undergoing backsplash, is lower on average than those with small $\Delta\,d$. This supports the scenario that ram pressure stripping is a dominant mechanism in the suppression of star formation. No significant difference, however, is found in region III, where projection effects may mix galaxies with different dynamic states together.
    
    Our study is the first analysis of such types of observations and demonstrates the uniqueness and promise of panoramic, wide-field 2D spectroscopy on galaxy clusters using SITELLE. Data from a larger sample of clusters and further detailed analysis will shed more light on the environmental effects and star formation quenching mechanisms operating on cluster galaxies.

\smallskip

\section*{Acknowledgement}
We thank the referee for the useful review that helped improve this manuscript. Based on observations obtained at the Canada-France-Hawaii Telescope (CFHT) which is operated from the summit of Maunakea by the National Research Council of Canada, the Institut National des Sciences de l'Univers of the Centre National de la Recherche Scientifique of France, and the University of Hawaii. The observations at the Canada-France-Hawaii Telescope were performed with care and respect from the summit of Maunakea, which is a significant cultural and historic site.
Based on observations obtained with SITELLE, a joint project between Université Laval, ABB-Bomem, Université de Montréal, and the CFHT with funding support from the Canada Foundation for Innovation (CFI), the National Sciences and Engineering Research Council of Canada (NSERC), Fond de Recherche du Québec - Nature et Technologies (FRQNT) and CFHT.
Q.L. is supported by an Ontario Trillium Awards. HY’s research is supported by an NSERC Discovery Grant and grants from the Arts and Science Faculty at the University of Toronto. L.D.’s research is supported by an NSERC discovery grant.

\software{SExtractor \citep{1996A&AS..117..393B}, photutils \citep{2016ascl.soft09011B}, ORBS \citep{2015ASPC..495..327M}, ORCS \citep{2015ASPC..495..327M}, astropy \citep{astropy:2013, astropy:2018}, numpy \citep{harris2020array}, matplotlib \citep{Hunter:2007}, scikit-learn \citep{scikit-learn}, SAOImage DS9 \citep{2003ASPC..295..489J}}



\bibliography{export-bibtex}

\begin{thebibliography}{}
\expandafter\ifx\csname natexlab\endcsname\relax\def\natexlab#1{#1}\fi
\providecommand{\url}[1]{\href{#1}{#1}}

\bibitem[{{Abadi} {et~al.}(1999){Abadi}, {Moore}, \&
  {Bower}}]{1999MNRAS.308..947A}
{Abadi}, M.~G., {Moore}, B., \& {Bower}, R.~G. 1999, \mnras, 308, 947

\bibitem[{{Abraham} {et~al.}(1996){Abraham}, {Smecker-Hane}, {Hutchings},
  {Carlberg}, {Yee}, {Ellingson}, {Morris}, {Oke}, \&
  {Rigler}}]{1996ApJ...471..694A}
{Abraham}, R.~G., {Smecker-Hane}, T.~A., {Hutchings}, J.~B., {et~al.} 1996,
  \apj, 471, 694

\bibitem[{{Acker} {et~al.}(1989){Acker}, {K{\"o}ppen}, {Samland}, \&
  {Stenholm}}]{1989Msngr..58...44A}
{Acker}, A., {K{\"o}ppen}, J., {Samland}, M., \& {Stenholm}, B. 1989, The
  Messenger, 58, 44

\bibitem[{{Astropy Collaboration} {et~al.}(2013){Astropy Collaboration},
  {Robitaille}, {Tollerud}, {Greenfield}, {Droettboom}, {Bray}, {Aldcroft},
  {Davis}, {Ginsburg}, {Price-Whelan}, {Kerzendorf}, {Conley}, {Crighton},
  {Barbary}, {Muna}, {Ferguson}, {Grollier}, {Parikh}, {Nair}, {Unther},
  {Deil}, {Woillez}, {Conseil}, {Kramer}, {Turner}, {Singer}, {Fox}, {Weaver},
  {Zabalza}, {Edwards}, {Azalee Bostroem}, {Burke}, {Casey}, {Crawford},
  {Dencheva}, {Ely}, {Jenness}, {Labrie}, {Lim}, {Pierfederici}, {Pontzen},
  {Ptak}, {Refsdal}, {Servillat}, \& {Streicher}}]{astropy:2013}
{Astropy Collaboration}, {Robitaille}, T.~P., {Tollerud}, E.~J., {et~al.} 2013,
  \aap, 558, A33

\bibitem[{{Astropy Collaboration} {et~al.}(2018){Astropy Collaboration},
  {Price-Whelan}, {Sip{H{o}}cz}, {G{"u}nther}, {Lim}, {Crawford}, {Conseil},
  {Shupe}, {Craig}, {Dencheva}, {Ginsburg}, {Vand erPlas}, {Bradley},
  {P{'e}rez-Su{'a}rez}, {de Val-Borro}, {Aldcroft}, {Cruz}, {Robitaille},
  {Tollerud}, {Ardelean}, {Babej}, {Bach}, {Bachetti}, {Bakanov}, {Bamford},
  {Barentsen}, {Barmby}, {Baumbach}, {Berry}, {Biscani}, {Boquien}, {Bostroem},
  {Bouma}, {Brammer}, {Bray}, {Breytenbach}, {Buddelmeijer}, {Burke},
  {Calderone}, {Cano Rodr{'i}guez}, {Cara}, {Cardoso}, {Cheedella}, {Copin},
  {Corrales}, {Crichton}, {D'Avella}, {Deil}, {Depagne}, {Dietrich}, {Donath},
  {Droettboom}, {Earl}, {Erben}, {Fabbro}, {Ferreira}, {Finethy}, {Fox},
  {Garrison}, {Gibbons}, {Goldstein}, {Gommers}, {Greco}, {Greenfield},
  {Groener}, {Grollier}, {Hagen}, {Hirst}, {Homeier}, {Horton}, {Hosseinzadeh},
  {Hu}, {Hunkeler}, {Ivezi{'c}}, {Jain}, {Jenness}, {Kanarek}, {Kendrew},
  {Kern}, {Kerzendorf}, {Khvalko}, {King}, {Kirkby}, {Kulkarni}, {Kumar},
  {Lee}, {Lenz}, {Littlefair}, {Ma}, {Macleod}, {Mastropietro}, {McCully},
  {Montagnac}, {Morris}, {Mueller}, {Mumford}, {Muna}, {Murphy}, {Nelson},
  {Nguyen}, {Ninan}, {N{"o}the}, {Ogaz}, {Oh}, {Parejko}, {Parley}, {Pascual},
  {Patil}, {Patil}, {Plunkett}, {Prochaska}, {Rastogi}, {Reddy Janga},
  {Sabater}, {Sakurikar}, {Seifert}, {Sherbert}, {Sherwood-Taylor}, {Shih},
  {Sick}, {Silbiger}, {Singanamalla}, {Singer}, {Sladen}, {Sooley},
  {Sornarajah}, {Streicher}, {Teuben}, {Thomas}, {Tremblay}, {Turner},
  {Terr{'o}n}, {van Kerkwijk}, {de la Vega}, {Watkins}, {Weaver}, {Whitmore},
  {Woillez}, {Zabalza}, \& {Astropy Contributors}}]{astropy:2018}
{Astropy Collaboration}, {Price-Whelan}, A.~M., {Sip{H{o}}cz}, B.~M., {et~al.}
  2018, aj, 156, 123

\bibitem[{{Bah{\'e}} \& {McCarthy}(2015)}]{2015MNRAS.447..969B}
{Bah{\'e}}, Y.~M., \& {McCarthy}, I.~G. 2015, \mnras, 447, 969

\bibitem[{{Balogh} \& {Morris}(2000)}]{2000MNRAS.318..703B}
{Balogh}, M.~L., \& {Morris}, S.~L. 2000, \mnras, 318, 703

\bibitem[{{Balogh} {et~al.}(1998){Balogh}, {Schade}, {Morris}, {Yee},
  {Carlberg}, \& {Ellingson}}]{1998ApJ...504L..75B}
{Balogh}, M.~L., {Schade}, D., {Morris}, S.~L., {et~al.} 1998, \apjl, 504, L75

\bibitem[{{Bekki}(2014)}]{2014MNRAS.438..444B}
{Bekki}, K. 2014, \mnras, 438, 444

\bibitem[{{Bertin} \& {Arnouts}(1996)}]{1996A&AS..117..393B}
{Bertin}, E., \& {Arnouts}, S. 1996, \aaps, 117, 393

\bibitem[{{Boselli} {et~al.}(2008){Boselli}, {Boissier}, {Cortese}, \&
  {Gavazzi}}]{2008ApJ...674..742B}
{Boselli}, A., {Boissier}, S., {Cortese}, L., \& {Gavazzi}, G. 2008, \apj, 674,
  742

\bibitem[{{Boselli} {et~al.}(2014){Boselli}, {Cortese}, {Boquien}, {Boissier},
  {Catinella}, {Gavazzi}, {Lagos}, \& {Saintonge}}]{2014A&A...564A..67B}
{Boselli}, A., {Cortese}, L., {Boquien}, M., {et~al.} 2014, \aap, 564, A67

\bibitem[{{Boselli} \& {Gavazzi}(2006)}]{2006PASP..118..517B}
{Boselli}, A., \& {Gavazzi}, G. 2006, \pasp, 118, 517

\bibitem[{{Boselli} {et~al.}(2016){Boselli}, {Roehlly}, {Fossati}, {Buat},
  {Boissier}, {Boquien}, {Burgarella}, {Ciesla}, {Gavazzi}, \&
  {Serra}}]{2016A&A...596A..11B}
{Boselli}, A., {Roehlly}, Y., {Fossati}, M., {et~al.} 2016, \aap, 596, A11

\bibitem[{{Boselli} {et~al.}(2019){Boselli}, {Epinat}, {Contini},
  {Abril-Melgarejo}, {Boogaard}, {Pointecouteau}, {Ventou}, {Brinchmann},
  {Carton}, {Finley}, {Michel-Dansac}, {Soucail}, \&
  {Weilbacher}}]{2019A&A...631A.114B}
{Boselli}, A., {Epinat}, B., {Contini}, T., {et~al.} 2019, \aap, 631, A114

\bibitem[{{Bradley} {et~al.}(2016){Bradley}, {Sipocz}, {Robitaille},
  {Tollerud}, {Deil}, {Vin{\'\i}cius}, {Barbary}, {G{\"u}nther}, {Bostroem},
  {Droettboom}, {Bray}, {Bratholm}, {Pickering}, {Craig}, {Pascual}, {Greco},
  {Donath}, {Kerzendorf}, {Littlefair}, {Barentsen}, {D'Eugenio}, \&
  {Weaver}}]{2016ascl.soft09011B}
{Bradley}, L., {Sipocz}, B., {Robitaille}, T., {et~al.} 2016, {Photutils:
  Photometry tools}, , , ascl:1609.011

\bibitem[{{Brough} {et~al.}(2013){Brough}, {Croom}, {Sharp}, {Hopkins},
  {Taylor}, {Baldry}, {Gunawardhana}, {Liske}, {Norberg}, {Robotham}, {Bauer},
  {Bland-Hawthorn}, {Colless}, {Foster}, {Kelvin}, {Lara-Lopez},
  {L{\'o}pez-S{\'a}nchez}, {Loveday}, {Owers}, {Pimbblet}, \&
  {Prescott}}]{2013MNRAS.435.2903B}
{Brough}, S., {Croom}, S., {Sharp}, R., {et~al.} 2013, \mnras, 435, 2903

\bibitem[{{Br{\"u}ggen} {et~al.}(2005){Br{\"u}ggen}, {Ruszkowski}, \&
  {Hallman}}]{2005ApJ...630..740B}
{Br{\"u}ggen}, M., {Ruszkowski}, M., \& {Hallman}, E. 2005, \apj, 630, 740

\bibitem[{{Bundy} {et~al.}(2015){Bundy}, {Bershady}, {Law}, {Yan}, {Drory},
  {MacDonald}, {Wake}, {Cherinka}, {S{\'a}nchez-Gallego}, {Weijmans}, {Thomas},
  {Tremonti}, {Masters}, {Coccato}, {Diamond-Stanic}, {Arag{\'o}n-Salamanca},
  {Avila-Reese}, {Badenes}, {Falc{\'o}n-Barroso}, {Belfiore}, {Bizyaev},
  {Blanc}, {Bland-Hawthorn}, {Blanton}, {Brownstein}, {Byler}, {Cappellari},
  {Conroy}, {Dutton}, {Emsellem}, {Etherington}, {Frinchaboy}, {Fu}, {Gunn},
  {Harding}, {Johnston}, {Kauffmann}, {Kinemuchi}, {Klaene}, {Knapen},
  {Leauthaud}, {Li}, {Lin}, {Maiolino}, {Malanushenko}, {Malanushenko}, {Mao},
  {Maraston}, {McDermid}, {Merrifield}, {Nichol}, {Oravetz}, {Pan}, {Parejko},
  {Sanchez}, {Schlegel}, {Simmons}, {Steele}, {Steinmetz}, {Thanjavur},
  {Thompson}, {Tinker}, {van den Bosch}, {Westfall}, {Wilkinson}, {Wright},
  {Xiao}, \& {Zhang}}]{2015ApJ...798....7B}
{Bundy}, K., {Bershady}, M.~A., {Law}, D.~R., {et~al.} 2015, \apj, 798, 7

\bibitem[{{Carlberg} {et~al.}(1997){Carlberg}, {Yee}, \&
  {Ellingson}}]{1997ApJ...478..462C}
{Carlberg}, R.~G., {Yee}, H.~K.~C., \& {Ellingson}, E. 1997, \apj, 478, 462

\bibitem[{{Chambers} {et~al.}(2016){Chambers}, {Magnier}, {Metcalfe},
  {Flewelling}, {Huber}, {Waters}, {Denneau}, {Draper}, {Farrow}, {Finkbeiner},
  {Holmberg}, {Koppenhoefer}, {Price}, {Rest}, {Saglia}, {Schlafly}, {Smartt},
  {Sweeney}, {Wainscoat}, {Burgett}, {Chastel}, {Grav}, {Heasley}, {Hodapp},
  {Jedicke}, {Kaiser}, {Kudritzki}, {Luppino}, {Lupton}, {Monet}, {Morgan},
  {Onaka}, {Shiao}, {Stubbs}, {Tonry}, {White}, {Ba{\~n}ados}, {Bell},
  {Bender}, {Bernard}, {Boegner}, {Boffi}, {Botticella}, {Calamida},
  {Casertano}, {Chen}, {Chen}, {Cole}, {Deacon}, {Frenk}, {Fitzsimmons},
  {Gezari}, {Gibbs}, {Goessl}, {Goggia}, {Gourgue}, {Goldman}, {Grant},
  {Grebel}, {Hambly}, {Hasinger}, {Heavens}, {Heckman}, {Henderson}, {Henning},
  {Holman}, {Hopp}, {Ip}, {Isani}, {Jackson}, {Keyes}, {Koekemoer}, {Kotak},
  {Le}, {Liska}, {Long}, {Lucey}, {Liu}, {Martin}, {Masci}, {McLean}, {Mindel},
  {Misra}, {Morganson}, {Murphy}, {Obaika}, {Narayan}, {Nieto-Santisteban},
  {Norberg}, {Peacock}, {Pier}, {Postman}, {Primak}, {Rae}, {Rai}, {Riess},
  {Riffeser}, {Rix}, {R{\"o}ser}, {Russel}, {Rutz}, {Schilbach}, {Schultz},
  {Scolnic}, {Strolger}, {Szalay}, {Seitz}, {Small}, {Smith}, {Soderblom},
  {Taylor}, {Thomson}, {Taylor}, {Thakar}, {Thiel}, {Thilker}, {Unger},
  {Urata}, {Valenti}, {Wagner}, {Walder}, {Walter}, {Watters}, {Werner},
  {Wood-Vasey}, \& {Wyse}}]{2016arXiv161205560C}
{Chambers}, K.~C., {Magnier}, E.~A., {Metcalfe}, N., {et~al.} 2016, arXiv
  e-prints, arXiv:1612.05560

\bibitem[{{Chung} {et~al.}(2007){Chung}, {van Gorkom}, {Kenney}, \&
  {Vollmer}}]{2007ApJ...659L.115C}
{Chung}, A., {van Gorkom}, J.~H., {Kenney}, J. D.~P., \& {Vollmer}, B. 2007,
  \apjl, 659, L115

\bibitem[{{Cid Fernandes} {et~al.}(2011){Cid Fernandes}, {Stasi{\'n}ska},
  {Mateus}, \& {Vale Asari}}]{2011MNRAS.413.1687C}
{Cid Fernandes}, R., {Stasi{\'n}ska}, G., {Mateus}, A., \& {Vale Asari}, N.
  2011, \mnras, 413, 1687

\bibitem[{{Coenda} {et~al.}(2019){Coenda}, {Mast}, {Mart{\'\i}nez}, {Muriel},
  \& {Merch{\'a}n}}]{2019A&A...621A..98C}
{Coenda}, V., {Mast}, D., {Mart{\'\i}nez}, H.~J., {Muriel}, H., \&
  {Merch{\'a}n}, M.~E. 2019, \aap, 621, A98

\bibitem[{{Comerford} {et~al.}(2017){Comerford}, {Barrows}, {Greene}, \&
  {Pooley}}]{2017ApJ...847...41C}
{Comerford}, J.~M., {Barrows}, R.~S., {Greene}, J.~E., \& {Pooley}, D. 2017,
  \apj, 847, 41

\bibitem[{{Cortese} {et~al.}(2019){Cortese}, {van de Sande}, {Lagos},
  {Catinella}, {Davies}, {Croom}, {Brough}, {Bryant}, {Lawrence}, {Owers},
  {Richards}, {Sweet}, \& {Bland -Hawthorn}}]{2019MNRAS.485.2656C}
{Cortese}, L., {van de Sande}, J., {Lagos}, C.~P., {et~al.} 2019, \mnras, 485,
  2656

\bibitem[{{Dressler}(1980)}]{1980ApJ...236..351D}
{Dressler}, A. 1980, \apj, 236, 351

\bibitem[{{Drissen} {et~al.}(2019){Drissen}, {Martin}, {Rousseau-Nepton},
  {Robert}, {Martin}, {Baril}, {Prunet}, {Joncas}, {Thibault}, {Brousseau},
  {Mandar}, {Grand mont}, {Yee}, \& {Simard}}]{2019MNRAS.485.3930D}
{Drissen}, L., {Martin}, T., {Rousseau-Nepton}, L., {et~al.} 2019, \mnras, 485,
  3930

\bibitem[{{Ebeling} {et~al.}(2014){Ebeling}, {Stephenson}, \&
  {Edge}}]{2014ApJ...781L..40E}
{Ebeling}, H., {Stephenson}, L.~N., \& {Edge}, A.~C. 2014, \apjl, 781, L40

\bibitem[{{Ellingson} {et~al.}(2001){Ellingson}, {Lin}, {Yee}, \&
  {Carlberg}}]{2001ApJ...547..609E}
{Ellingson}, E., {Lin}, H., {Yee}, H.~K.~C., \& {Carlberg}, R.~G. 2001, \apj,
  547, 609

\bibitem[{{Fillingham} {et~al.}(2016){Fillingham}, {Cooper}, {Pace},
  {Boylan-Kolchin}, {Bullock}, {Garrison-Kimmel}, \&
  {Wheeler}}]{2016MNRAS.463.1916F}
{Fillingham}, S.~P., {Cooper}, M.~C., {Pace}, A.~B., {et~al.} 2016, \mnras,
  463, 1916

\bibitem[{{Foltz} {et~al.}(2018){Foltz}, {Wilson}, {Muzzin}, {Cooper},
  {Nantais}, {van der Burg}, {Cerulo}, {Chan}, {Fillingham}, {Surace}, {Webb},
  {Noble}, {Lacy}, {McDonald}, {Rudnick}, {Lidman}, {Demarco},
  {Hlavacek-Larrondo}, {Yee}, {Perlmutter}, \& {Hayden}}]{2018ApJ...866..136F}
{Foltz}, R., {Wilson}, G., {Muzzin}, A., {et~al.} 2018, \apj, 866, 136

\bibitem[{{Fritz} {et~al.}(2005){Fritz}, {Ziegler}, {Bower}, {Smail}, \&
  {Davies}}]{2005MNRAS.358..233F}
{Fritz}, A., {Ziegler}, B.~L., {Bower}, R.~G., {Smail}, I., \& {Davies}, R.~L.
  2005, \mnras, 358, 233

\bibitem[{{Fujita} \& {Nagashima}(1999)}]{1999ApJ...516..619F}
{Fujita}, Y., \& {Nagashima}, M. 1999, \apj, 516, 619

\bibitem[{{Gill} {et~al.}(2005){Gill}, {Knebe}, \&
  {Gibson}}]{2005MNRAS.356.1327G}
{Gill}, S. P.~D., {Knebe}, A., \& {Gibson}, B.~K. 2005, \mnras, 356, 1327

\bibitem[{{G{\'o}mez} {et~al.}(2003){G{\'o}mez}, {Nichol}, {Miller}, {Balogh},
  {Goto}, {Zabludoff}, {Romer}, {Bernardi}, {Sheth}, {Hopkins}, {Castander},
  {Connolly}, {Schneider}, {Brinkmann}, {Lamb}, {SubbaRao}, \&
  {York}}]{2003ApJ...584..210G}
{G{\'o}mez}, P.~L., {Nichol}, R.~C., {Miller}, C.~J., {et~al.} 2003, \apj, 584,
  210

\bibitem[{{Goto} {et~al.}(2003){Goto}, {Yamauchi}, {Fujita}, {Okamura},
  {Sekiguchi}, {Smail}, {Bernardi}, \& {Gomez}}]{2003MNRAS.346..601G}
{Goto}, T., {Yamauchi}, C., {Fujita}, Y., {et~al.} 2003, \mnras, 346, 601

\bibitem[{{Graham} {et~al.}(2019){Graham}, {Cappellari}, {Bershady}, \&
  {Drory}}]{2019arXiv191005139G}
{Graham}, M.~T., {Cappellari}, M., {Bershady}, M.~A., \& {Drory}, N. 2019,
  arXiv e-prints, arXiv:1910.05139

\bibitem[{{Green} {et~al.}(2018){Green}, {Croom}, {Scott}, {Cortese},
  {Medling}, {D'Eugenio}, {Bryant}, {Bland-Hawthorn}, {Allen}, {Sharp}, {Ho},
  {Groves}, {Drinkwater}, {Mannering}, {Harischand ra}, {van de Sande},
  {Thomas}, {O'Toole}, {McDermid}, {Vuong}, {Sealey}, {Bauer}, {Brough},
  {Catinella}, {Cecil}, {Colless}, {Couch}, {Driver}, {Federrath}, {Foster},
  {Goodwin}, {Hampton}, {Hopkins}, {Jones}, {Konstantopoulos}, {Lawrence},
  {Leon-Saval}, {Liske}, {L{\'o}pez-S{\'a}nchez}, {Lorente}, {Mould},
  {Obreschkow}, {Owers}, {Richards}, {Robotham}, {Schaefer}, {Sweet}, {Taranu},
  {Tescari}, {Tonini}, \& {Zafar}}]{2018MNRAS.475..716G}
{Green}, A.~W., {Croom}, S.~M., {Scott}, N., {et~al.} 2018, \mnras, 475, 716

\bibitem[{{Gunn} \& {Gott}(1972)}]{1972ApJ...176....1G}
{Gunn}, J.~E., \& {Gott}, J.~Richard, I. 1972, \apj, 176, 1

\bibitem[{{Haines} {et~al.}(2013){Haines}, {Pereira}, {Smith}, {Egami},
  {Sanderson}, {Babul}, {Finoguenov}, {Merluzzi}, {Busarello}, {Rawle}, \&
  {Okabe}}]{2013ApJ...775..126H}
{Haines}, C.~P., {Pereira}, M.~J., {Smith}, G.~P., {et~al.} 2013, \apj, 775,
  126

\bibitem[{Harris {et~al.}(2020)Harris, Millman, van~der Walt, Gommers,
  Virtanen, Cournapeau, Wieser, Taylor, Berg, Smith, Kern, Picus, Hoyer, van
  Kerkwijk, Brett, Haldane, del R{'{\i}}o, Wiebe, Peterson,
  G{'{e}}rard-Marchant, Sheppard, Reddy, Weckesser, Abbasi, Gohlke, \&
  Oliphant}]{harris2020array}
Harris, C.~R., Millman, K.~J., van~der Walt, S.~J., {et~al.} 2020, Nature, 585,
  357.
\newblock \url{https://doi.org/10.1038/s41586-020-2649-2}

\bibitem[{{Haynes} {et~al.}(1984){Haynes}, {Giovanelli}, \&
  {Chincarini}}]{1984ARA&A..22..445H}
{Haynes}, M.~P., {Giovanelli}, R., \& {Chincarini}, G.~L. 1984, \araa, 22, 445

\bibitem[{Hunter(2007)}]{Hunter:2007}
Hunter, J.~D. 2007, Computing in Science \& Engineering, 9, 90

\bibitem[{{Hutchings} \& {Balogh}(2000)}]{2000AJ....119.1123H}
{Hutchings}, J.~B., \& {Balogh}, M.~L. 2000, \aj, 119, 1123

\bibitem[{{Jaff{\'e}} {et~al.}(2015){Jaff{\'e}}, {Smith}, {Candlish},
  {Poggianti}, {Sheen}, \& {Verheijen}}]{2015MNRAS.448.1715J}
{Jaff{\'e}}, Y.~L., {Smith}, R., {Candlish}, G.~N., {et~al.} 2015, \mnras, 448,
  1715

\bibitem[{{Jaff{\'e}} {et~al.}(2016){Jaff{\'e}}, {Verheijen}, {Haines}, {Yoon},
  {Cybulski}, {Montero-Casta{\~n}o}, {Smith}, {Chung}, {Deshev},
  {Fern{\'a}ndez}, {van Gorkom}, {Poggianti}, {Yun}, {Finoguenov}, {Smith}, \&
  {Okabe}}]{2016MNRAS.461.1202J}
{Jaff{\'e}}, Y.~L., {Verheijen}, M. A.~W., {Haines}, C.~P., {et~al.} 2016,
  \mnras, 461, 1202

\bibitem[{{Jaff{\'e}} {et~al.}(2018){Jaff{\'e}}, {Poggianti}, {Moretti},
  {Gullieuszik}, {Smith}, {Vulcani}, {Fasano}, {Fritz}, {Tonnesen}, {Bettoni},
  {Hau}, {Biviano}, {Bellhouse}, \& {McGee}}]{2018MNRAS.476.4753J}
{Jaff{\'e}}, Y.~L., {Poggianti}, B.~M., {Moretti}, A., {et~al.} 2018, \mnras,
  476, 4753

\bibitem[{{Joye} \& {Mandel}(2003)}]{2003ASPC..295..489J}
{Joye}, W.~A., \& {Mandel}, E. 2003, in Astronomical Society of the Pacific
  Conference Series, Vol. 295, Astronomical Data Analysis Software and Systems
  XII, ed. H.~E. {Payne}, R.~I. {Jedrzejewski}, \& R.~N. {Hook}, 489

\bibitem[{{Kapferer} {et~al.}(2008){Kapferer}, {Kronberger}, {Ferrari},
  {Riser}, \& {Schindler}}]{2008MNRAS.389.1405K}
{Kapferer}, W., {Kronberger}, T., {Ferrari}, C., {Riser}, T., \& {Schindler},
  S. 2008, \mnras, 389, 1405

\bibitem[{{Kauffmann} {et~al.}(2004){Kauffmann}, {White}, {Heckman},
  {M{\'e}nard}, {Brinchmann}, {Charlot}, {Tremonti}, \&
  {Brinkmann}}]{2004MNRAS.353..713K}
{Kauffmann}, G., {White}, S. D.~M., {Heckman}, T.~M., {et~al.} 2004, \mnras,
  353, 713

\bibitem[{{Kawinwanichakij} {et~al.}(2017){Kawinwanichakij}, {Papovich},
  {Quadri}, {Glazebrook}, {Kacprzak}, {Allen}, {Bell}, {Croton}, {Dekel},
  {Ferguson}, {Forrest}, {Grogin}, {Guo}, {Kocevski}, {Koekemoer}, {Labb{\'e}},
  {Lucas}, {Nanayakkara}, {Spitler}, {Straatman}, {Tran}, {Tomczak}, \& {van
  Dokkum}}]{2017ApJ...847..134K}
{Kawinwanichakij}, L., {Papovich}, C., {Quadri}, R.~F., {et~al.} 2017, \apj,
  847, 134

\bibitem[{{Kennicutt} \& {Evans}(2012)}]{2012ARA&A..50..531K}
{Kennicutt}, R.~C., \& {Evans}, N.~J. 2012, \araa, 50, 531

\bibitem[{{Koopmann} \& {Kennedy}(1999)}]{1999ASNYN...5...22K}
{Koopmann}, R., \& {Kennedy}, J.~D.~P. 1999, News Letter of the Astronomical
  Society of New York, 5, 22

\bibitem[{{Kronberger} {et~al.}(2008){Kronberger}, {Kapferer}, {Ferrari},
  {Unterguggenberger}, \& {Schindler}}]{2008A&A...481..337K}
{Kronberger}, T., {Kapferer}, W., {Ferrari}, C., {Unterguggenberger}, S., \&
  {Schindler}, S. 2008, \aap, 481, 337

\bibitem[{{Lewis} {et~al.}(2002){Lewis}, {Balogh}, {De Propris}, {Couch},
  {Bower}, {Offer}, {Bland -Hawthorn}, {Baldry}, {Baugh}, {Bridges}, {Cannon},
  {Cole}, {Colless}, {Collins}, {Cross}, {Dalton}, {Driver}, {Efstathiou},
  {Ellis}, {Frenk}, {Glazebrook}, {Hawkins}, {Jackson}, {Lahav}, {Lumsden},
  {Maddox}, {Madgwick}, {Norberg}, {Peacock}, {Percival}, {Peterson},
  {Sutherland}, \& {Taylor}}]{2002MNRAS.334..673L}
{Lewis}, I., {Balogh}, M., {De Propris}, R., {et~al.} 2002, \mnras, 334, 673

\bibitem[{{Li} {et~al.}(2009){Li}, {Yee}, \& {Ellingson}}]{2009ApJ...698...83L}
{Li}, I.~H., {Yee}, H.~K.~C., \& {Ellingson}, E. 2009, \apj, 698, 83

\bibitem[{{Lotz} {et~al.}(2019){Lotz}, {Remus}, {Dolag}, {Biviano}, \&
  {Burkert}}]{2019MNRAS.488.5370L}
{Lotz}, M., {Remus}, R.-S., {Dolag}, K., {Biviano}, A., \& {Burkert}, A. 2019,
  \mnras, 488, 5370

\bibitem[{{Martin} {et~al.}(2015){Martin}, {Drissen}, \&
  {Joncas}}]{2015ASPC..495..327M}
{Martin}, T., {Drissen}, L., \& {Joncas}, G. 2015, in Astronomical Society of
  the Pacific Conference Series, Vol. 495, Astronomical Data Analysis Software
  an Systems XXIV (ADASS XXIV), ed. A.~R. {Taylor} \& E.~{Rosolowsky}, 327

\bibitem[{{Mayer} {et~al.}(2006){Mayer}, {Mastropietro}, {Wadsley}, {Stadel},
  \& {Moore}}]{2006MNRAS.369.1021M}
{Mayer}, L., {Mastropietro}, C., {Wadsley}, J., {Stadel}, J., \& {Moore}, B.
  2006, \mnras, 369, 1021

\bibitem[{{Muriel} \& {Coenda}(2014)}]{2014A&A...564A..85M}
{Muriel}, H., \& {Coenda}, V. 2014, \aap, 564, A85

\bibitem[{{Muzzin} {et~al.}(2012){Muzzin}, {Wilson}, {Yee}, {Gilbank},
  {Hoekstra}, {Demarco}, {Balogh}, {van Dokkum}, {Franx}, {Ellingson}, {Hicks},
  {Nantais}, {Noble}, {Lacy}, {Lidman}, {Rettura}, {Surace}, \&
  {Webb}}]{2012ApJ...746..188M}
{Muzzin}, A., {Wilson}, G., {Yee}, H.~K.~C., {et~al.} 2012, \apj, 746, 188

\bibitem[{{Muzzin} {et~al.}(2014){Muzzin}, {van der Burg}, {McGee}, {Balogh},
  {Franx}, {Hoekstra}, {Hudson}, {Noble}, {Taranu}, {Webb}, {Wilson}, \&
  {Yee}}]{2014ApJ...796...65M}
{Muzzin}, A., {van der Burg}, R.~F.~J., {McGee}, S.~L., {et~al.} 2014, \apj,
  796, 65

\bibitem[{{Noble} {et~al.}(2013){Noble}, {Webb}, {Muzzin}, {Wilson}, {Yee}, \&
  {van der Burg}}]{2013ApJ...768..118N}
{Noble}, A.~G., {Webb}, T.~M.~A., {Muzzin}, A., {et~al.} 2013, \apj, 768, 118

\bibitem[{{Owers} {et~al.}(2019){Owers}, {Hudson}, {Oman}, {Bland -Hawthorn},
  {Brough}, {Bryant}, {Cortese}, {Couch}, {Croom}, {van de Sande}, {Federrath},
  {Groves}, {Hopkins}, {Lawrence}, {Lorente}, {McDermid}, {Medling},
  {Richards}, {Scott}, {Taranu}, {Welker}, \& {Yi}}]{2019ApJ...873...52O}
{Owers}, M.~S., {Hudson}, M.~J., {Oman}, K.~A., {et~al.} 2019, \apj, 873, 52

\bibitem[{Pedregosa {et~al.}(2011)Pedregosa, Varoquaux, Gramfort, Michel,
  Thirion, Grisel, Blondel, Prettenhofer, Weiss, Dubourg, Vanderplas, Passos,
  Cournapeau, Brucher, Perrot, \& Duchesnay}]{scikit-learn}
Pedregosa, F., Varoquaux, G., Gramfort, A., {et~al.} 2011, Journal of Machine
  Learning Research, 12, 2825

\bibitem[{{Peng} {et~al.}(2015){Peng}, {Maiolino}, \&
  {Cochrane}}]{2015Natur.521..192P}
{Peng}, Y., {Maiolino}, R., \& {Cochrane}, R. 2015, \nat, 521, 192

\bibitem[{{Peng} {et~al.}(2010){Peng}, {Lilly}, {Kova{\v{c}}}, {Bolzonella},
  {Pozzetti}, {Renzini}, {Zamorani}, {Ilbert}, {Knobel}, {Iovino}, {Maier},
  {Cucciati}, {Tasca}, {Carollo}, {Silverman}, {Kampczyk}, {de Ravel},
  {Sanders}, {Scoville}, {Contini}, {Mainieri}, {Scodeggio}, {Kneib}, {Le
  F{\`e}vre}, {Bardelli}, {Bongiorno}, {Caputi}, {Coppa}, {de la Torre},
  {Franzetti}, {Garilli}, {Lamareille}, {Le Borgne}, {Le Brun}, {Mignoli},
  {Perez Montero}, {Pello}, {Ricciardelli}, {Tanaka}, {Tresse}, {Vergani},
  {Welikala}, {Zucca}, {Oesch}, {Abbas}, {Barnes}, {Bordoloi}, {Bottini},
  {Cappi}, {Cassata}, {Cimatti}, {Fumana}, {Hasinger}, {Koekemoer},
  {Leauthaud}, {Maccagni}, {Marinoni}, {McCracken}, {Memeo}, {Meneux}, {Nair},
  {Porciani}, {Presotto}, \& {Scaramella}}]{2010ApJ...721..193P}
{Peng}, Y.-j., {Lilly}, S.~J., {Kova{\v{c}}}, K., {et~al.} 2010, \apj, 721, 193

\bibitem[{{Pintos-Castro} {et~al.}(2019){Pintos-Castro}, {Yee}, {Muzzin},
  {Old}, \& {Wilson}}]{2019ApJ...876...40P}
{Pintos-Castro}, I., {Yee}, H.~K.~C., {Muzzin}, A., {Old}, L., \& {Wilson}, G.
  2019, \apj, 876, 40

\bibitem[{{Poggianti} {et~al.}(2006){Poggianti}, {von der Linden}, {De Lucia},
  {Desai}, {Simard}, {Halliday}, {Arag{\'o}n-Salamanca}, {Bower}, {Varela},
  {Best}, {Clowe}, {Dalcanton}, {Jablonka}, {Milvang-Jensen}, {Pello},
  {Rudnick}, {Saglia}, {White}, \& {Zaritsky}}]{2006ApJ...642..188P}
{Poggianti}, B.~M., {von der Linden}, A., {De Lucia}, G., {et~al.} 2006, \apj,
  642, 188

\bibitem[{{Poggianti} {et~al.}(2017{\natexlab{a}}){Poggianti}, {Moretti},
  {Gullieuszik}, {Fritz}, {Jaff{\'e}}, {Bettoni}, {Fasano}, {Bellhouse}, {Hau},
  {Vulcani}, {Biviano}, {Omizzolo}, {Paccagnella}, {D'Onofrio}, {Cava},
  {Sheen}, {Couch}, \& {Owers}}]{2017ApJ...844...48P}
{Poggianti}, B.~M., {Moretti}, A., {Gullieuszik}, M., {et~al.}
  2017{\natexlab{a}}, \apj, 844, 48

\bibitem[{{Poggianti} {et~al.}(2017{\natexlab{b}}){Poggianti}, {Jaff{\'e}},
  {Moretti}, {Gullieuszik}, {Radovich}, {Tonnesen}, {Fritz}, {Bettoni},
  {Vulcani}, {Fasano}, {Bellhouse}, {Hau}, \& {Omizzolo}}]{2017Natur.548..304P}
{Poggianti}, B.~M., {Jaff{\'e}}, Y.~L., {Moretti}, A., {et~al.}
  2017{\natexlab{b}}, \nat, 548, 304

\bibitem[{{Postman} {et~al.}(2005){Postman}, {Franx}, {Cross}, {Holden},
  {Ford}, {Illingworth}, {Goto}, {Demarco}, {Rosati}, {Blakeslee}, {Tran},
  {Ben{\'\i}tez}, {Clampin}, {Hartig}, {Homeier}, {Ardila}, {Bartko},
  {Bouwens}, {Bradley}, {Broadhurst}, {Brown}, {Burrows}, {Cheng}, {Feldman},
  {Golimowski}, {Gronwall}, {Infante}, {Kimble}, {Krist}, {Lesser}, {Martel},
  {Mei}, {Menanteau}, {Meurer}, {Miley}, {Motta}, {Sirianni}, {Sparks}, {Tran},
  {Tsvetanov}, {White}, \& {Zheng}}]{2005ApJ...623..721P}
{Postman}, M., {Franx}, M., {Cross}, N.~J.~G., {et~al.} 2005, \apj, 623, 721

\bibitem[{{Ramos-Mart{\'\i}nez} {et~al.}(2018){Ramos-Mart{\'\i}nez},
  {G{\'o}mez}, \& {P{\'e}rez-Villegas}}]{2018MNRAS.476.3781R}
{Ramos-Mart{\'\i}nez}, M., {G{\'o}mez}, G.~C., \& {P{\'e}rez-Villegas}, {\'A}.
  2018, \mnras, 476, 3781

\bibitem[{{Rhee} {et~al.}(2017){Rhee}, {Smith}, {Choi}, {Yi}, {Jaff{\'e}},
  {Candlish}, \& {S{\'a}nchez-J{\'a}nssen}}]{2017ApJ...843..128R}
{Rhee}, J., {Smith}, R., {Choi}, H., {et~al.} 2017, \apj, 843, 128

\bibitem[{{Ricarte} {et~al.}(2020){Ricarte}, {Tremmel}, {Natarajan}, \&
  {Quinn}}]{2020ApJ...895L...8R}
{Ricarte}, A., {Tremmel}, M., {Natarajan}, P., \& {Quinn}, T. 2020, \apjl, 895,
  L8

\bibitem[{{Roediger} {et~al.}(2011){Roediger}, {Br{\"u}ggen}, {Simionescu},
  {B{\"o}hringer}, {Churazov}, \& {Forman}}]{2011MNRAS.413.2057R}
{Roediger}, E., {Br{\"u}ggen}, M., {Simionescu}, A., {et~al.} 2011, \mnras,
  413, 2057

\bibitem[{{Russell} {et~al.}(2019){Russell}, {McNamara}, {Fabian}, {Nulsen},
  {Combes}, {Edge}, {Madar}, {Olivares}, {Salom{\'e}}, \&
  {Vantyghem}}]{2019MNRAS.490.3025R}
{Russell}, H.~R., {McNamara}, B.~R., {Fabian}, A.~C., {et~al.} 2019, \mnras,
  490, 3025

\bibitem[{{S{\'a}nchez} {et~al.}(2012){S{\'a}nchez}, {Kennicutt}, {Gil de Paz},
  {van de Ven}, {V{\'\i}lchez}, {Wisotzki}, {Walcher}, {Mast}, {Aguerri},
  {Albiol-P{\'e}rez}, {Alonso-Herrero}, {Alves}, {Bakos}, {Bart{\'a}kov{\'a}},
  {Bland-Hawthorn}, {Boselli}, {Bomans}, {Castillo-Morales}, {Cortijo-Ferrero},
  {de Lorenzo-C{\'a}ceres}, {Del Olmo}, {Dettmar}, {D{\'\i}az}, {Ellis},
  {Falc{\'o}n-Barroso}, {Flores}, {Gallazzi}, {Garc{\'\i}a-Lorenzo},
  {Gonz{\'a}lez Delgado}, {Gruel}, {Haines}, {Hao}, {Husemann},
  {Igl{\'e}sias-P{\'a}ramo}, {Jahnke}, {Johnson}, {Jungwiert}, {Kalinova},
  {Kehrig}, {Kupko}, {L{\'o}pez-S{\'a}nchez}, {Lyubenova}, {Marino},
  {M{\'a}rmol-Queralt{\'o}}, {M{\'a}rquez}, {Masegosa}, {Meidt},
  {Mendez-Abreu}, {Monreal-Ibero}, {Montijo}, {Mour{\~a}o}, {Palacios-Navarro},
  {Papaderos}, {Pasquali}, {Peletier}, {P{\'e}rez}, {P{\'e}rez}, {Quirrenbach},
  {Rela{\~n}o}, {Rosales-Ortega}, {Roth}, {Ruiz-Lara},
  {S{\'a}nchez-Bl{\'a}zquez}, {Sengupta}, {Singh}, {Stanishev}, {Trager},
  {Vazdekis}, {Viironen}, {Wild}, {Zibetti}, \&
  {Ziegler}}]{2012A&A...538A...8S}
{S{\'a}nchez}, S.~F., {Kennicutt}, R.~C., {Gil de Paz}, A., {et~al.} 2012,
  \aap, 538, A8

\bibitem[{{Schaefer} {et~al.}(2017){Schaefer}, {Croom}, {Allen}, {Brough},
  {Medling}, {Ho}, {Scott}, {Richards}, {Pracy}, {Gunawardhana}, {Norberg},
  {Alpaslan}, {Bauer}, {Bekki}, {Bland-Hawthorn}, {Bloom}, {Bryant}, {Couch},
  {Driver}, {Fogarty}, {Foster}, {Goldstein}, {Green}, {Hopkins},
  {Konstantopoulos}, {Lawrence}, {L{\'o}pez-S{\'a}nchez}, {Lorente}, {Owers},
  {Sharp}, {Sweet}, {Taylor}, {van de Sande}, {Walcher}, \&
  {Wong}}]{2017MNRAS.464..121S}
{Schaefer}, A.~L., {Croom}, S.~M., {Allen}, J.~T., {et~al.} 2017, \mnras, 464,
  121

\bibitem[{{Schaefer} {et~al.}(2019){Schaefer}, {Tremonti}, {Pace}, {Belfiore},
  {Argudo-Fernandez}, {Bershady}, {Drory}, {Jones}, {Maiolino}, {Stark},
  {Wake}, \& {Yan}}]{2019ApJ...884..156S}
{Schaefer}, A.~L., {Tremonti}, C., {Pace}, Z., {et~al.} 2019, \apj, 884, 156

\bibitem[{{Smith} {et~al.}(2010){Smith}, {Lucey}, {Hammer}, {Hornschemeier},
  {Carter}, {Hudson}, {Marzke}, {Mouhcine}, {Eftekharzadeh}, {James},
  {Khosroshahi}, {Kourkchi}, \& {Karick}}]{2010MNRAS.408.1417S}
{Smith}, R.~J., {Lucey}, J.~R., {Hammer}, D., {et~al.} 2010, \mnras, 408, 1417

\bibitem[{{Sobral} {et~al.}(2011){Sobral}, {Best}, {Smail}, {Geach},
  {Cirasuolo}, {Garn}, \& {Dalton}}]{2011MNRAS.411..675S}
{Sobral}, D., {Best}, P.~N., {Smail}, I., {et~al.} 2011, \mnras, 411, 675

\bibitem[{{Vijayaraghavan} \& {Ricker}(2013)}]{2013MNRAS.435.2713V}
{Vijayaraghavan}, R., \& {Ricker}, P.~M. 2013, \mnras, 435, 2713

\bibitem[{{Vulcani} {et~al.}(2017){Vulcani}, {Treu}, {Nipoti}, {Schmidt},
  {Dressler}, {Morshita}, {Poggianti}, {Malkan}, {Hoag}, {Brada{\v{c}}},
  {Abramson}, {Trenti}, {Pentericci}, {von der Linden}, {Morris}, \&
  {Wang}}]{2017ApJ...837..126V}
{Vulcani}, B., {Treu}, T., {Nipoti}, C., {et~al.} 2017, \apj, 837, 126

\bibitem[{{Vulcani} {et~al.}(2018){Vulcani}, {Poggianti}, {Gullieuszik},
  {Moretti}, {Tonnesen}, {Jaff{\'e}}, {Fritz}, {Fasano}, \&
  {Bettoni}}]{2018ApJ...866L..25V}
{Vulcani}, B., {Poggianti}, B.~M., {Gullieuszik}, M., {et~al.} 2018, \apjl,
  866, L25

\bibitem[{{Wang} {et~al.}(2020){Wang}, {Xu}, {Lee}, {Du}, {Overzier}, \&
  {Shao}}]{2020ApJ...903..103W}
{Wang}, J., {Xu}, W., {Lee}, B., {et~al.} 2020, \apj, 903, 103

\bibitem[{{Wegner}(2011)}]{Wegner2011}
{Wegner}, G.~A. 2011, \mnras, 413, 1333

\bibitem[{{Wegner} {et~al.}(2015){Wegner}, {Chu}, \& {Hwang}}]{Wegner2015}
{Wegner}, G.~A., {Chu}, D.~S., \& {Hwang}, H.~S. 2015, \mnras, 447, 1126

\bibitem[{{Wegner} {et~al.}(2017){Wegner}, {Umetsu}, {Molnar}, {Nonino},
  {Medezinski}, {Andrade-Santos}, {Bogdan}, {Lovisari}, {Forman}, \&
  {Jones}}]{Wegner2017}
{Wegner}, G.~A., {Umetsu}, K., {Molnar}, S.~M., {et~al.} 2017, \apj, 844, 67

\bibitem[{{Weinmann} {et~al.}(2009){Weinmann}, {Kauffmann}, {van den Bosch},
  {Pasquali}, {McIntosh}, {Mo}, {Yang}, \& {Guo}}]{2009MNRAS.394.1213W}
{Weinmann}, S.~M., {Kauffmann}, G., {van den Bosch}, F.~C., {et~al.} 2009,
  \mnras, 394, 1213

\bibitem[{{Yee} {et~al.}(1996){Yee}, {Ellingson}, {Abraham}, {Gravel},
  {Carlberg}, {Smecker-Hane}, {Schade}, \& {Rigler}}]{1996ApJS..102..289Y}
{Yee}, H.~K.~C., {Ellingson}, E., {Abraham}, R.~G., {et~al.} 1996, \apjs, 102,
  289

\bibitem[{{Yoon} {et~al.}(2017){Yoon}, {Chung}, {Smith}, \&
  {Jaff{\'e}}}]{2017ApJ...838...81Y}
{Yoon}, H., {Chung}, A., {Smith}, R., \& {Jaff{\'e}}, Y.~L. 2017, \apj, 838, 81

\bibitem[{{Yoshida} {et~al.}(2008){Yoshida}, {Yagi}, {Komiyama}, {Furusawa},
  {Kashikawa}, {Koyama}, {Yamanoi}, {Hattori}, \&
  {Okamura}}]{2008ApJ...688..918Y}
{Yoshida}, M., {Yagi}, M., {Komiyama}, Y., {et~al.} 2008, \apj, 688, 918

\bibitem[{{Yun} {et~al.}(2019){Yun}, {Pillepich}, {Zinger}, {Nelson},
  {Donnari}, {Joshi}, {Rodriguez-Gomez}, {Genel}, {Weinberger}, {Vogelsberger},
  \& {Hernquist}}]{2019MNRAS.483.1042Y}
{Yun}, K., {Pillepich}, A., {Zinger}, E., {et~al.} 2019, \mnras, 483, 1042

\end{thebibliography}

\appendix

\section{Reduction of Interferometric Fringes} \label{appendix:fringe}
    
    Below we describe the low-pass filtering (LPF) technique used to remove fringes on the SITELLE datacube. A source detection with an S/N threshold of 3 is first run on the stack image using the python photometric package \texttt{photutils} to pick out possible sources. These sources are masked before the LPF procedure to avoid them being over-subtracted. Next, LPF is performed on each channel by convolving the image with an elliptical gaussian kernel elongated in the x-axis to construct a low-frequency components image and then subtracting it from the original image. The method is based on the fact that the fringes are high-frequency components that are relatively continuous along the x-axis but may radically change its brightness in the y-axis. In practice, we adopt a kernel size with FWHM of 28 pix in the x-axis and 7 pix in the y-axis. The kernel size is determined by experimenting with different sizes to reach a trade-off between reducing the high-frequency fringes and avoiding over-subtraction of faint candidates that are possibly missed in the initial detection.
    \begin{figure}
      \centering
      \includegraphics[width=\hsize]{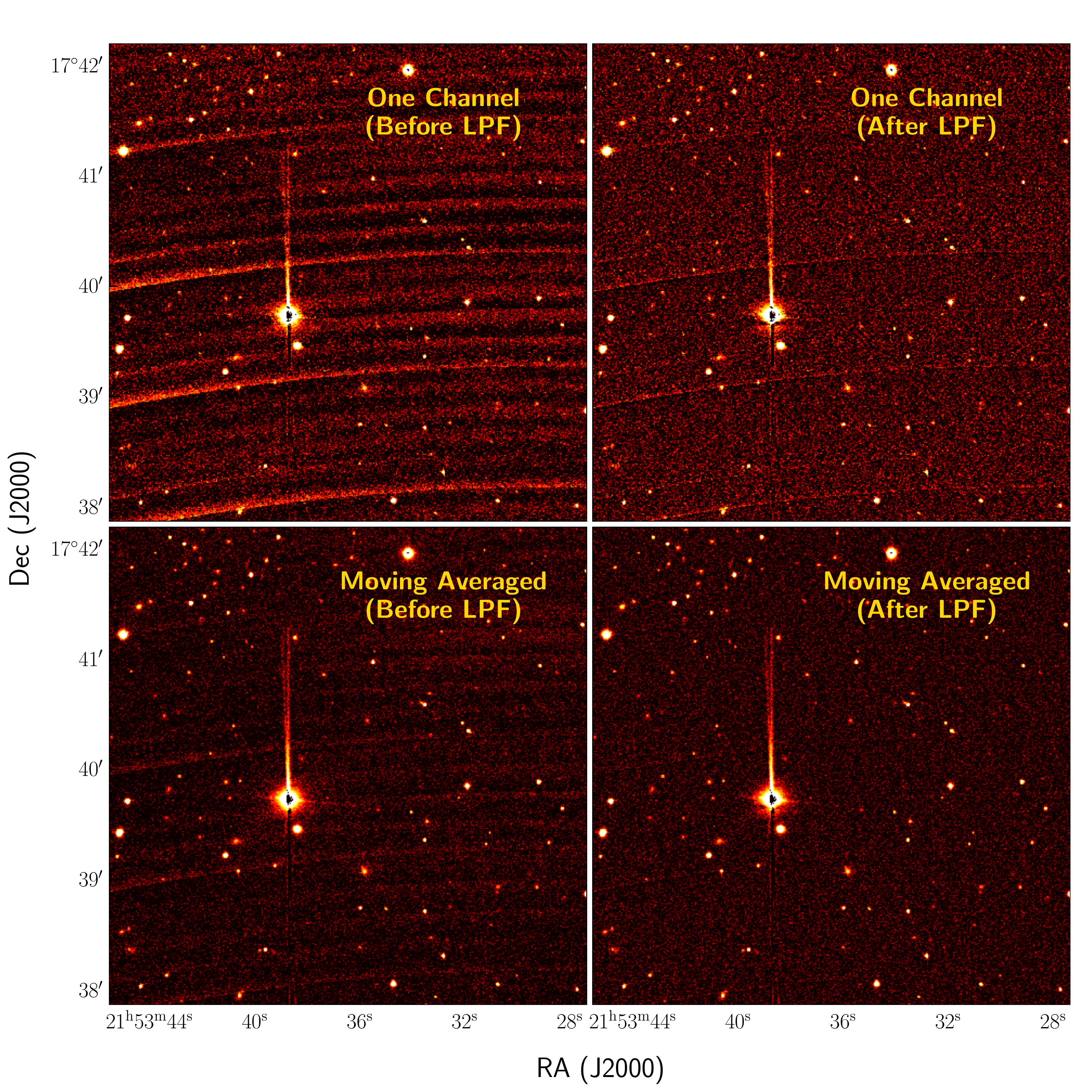}
      \caption{Improvement by post-processing the spectral datacube in the presence of interferometric fringes. Upper left: image of a single channel showing fringes after subtracting the large-scale background. The brightness of the fringes does not follow a regular pattern. Upper right: image of the same channel  processed with the LPF process. Fringes have been mitigated, although some residuals remain in the image. Lower left: image of the same channel but each pixel is manipulated by a moving average using a 3x3x5 box. Lower right: image of the same channel applying both LPF and moving average processing. The background residuals are largely suppressed. For direct visual comparison, the four panels share the same contrast. The image contrast is in arcsinh stretch to visually augment the small difference in the background.}
      \label{fig:LPF_field}
    \end{figure}
    
    \begin{figure}
      \centering
      \includegraphics[width=0.85\hsize]{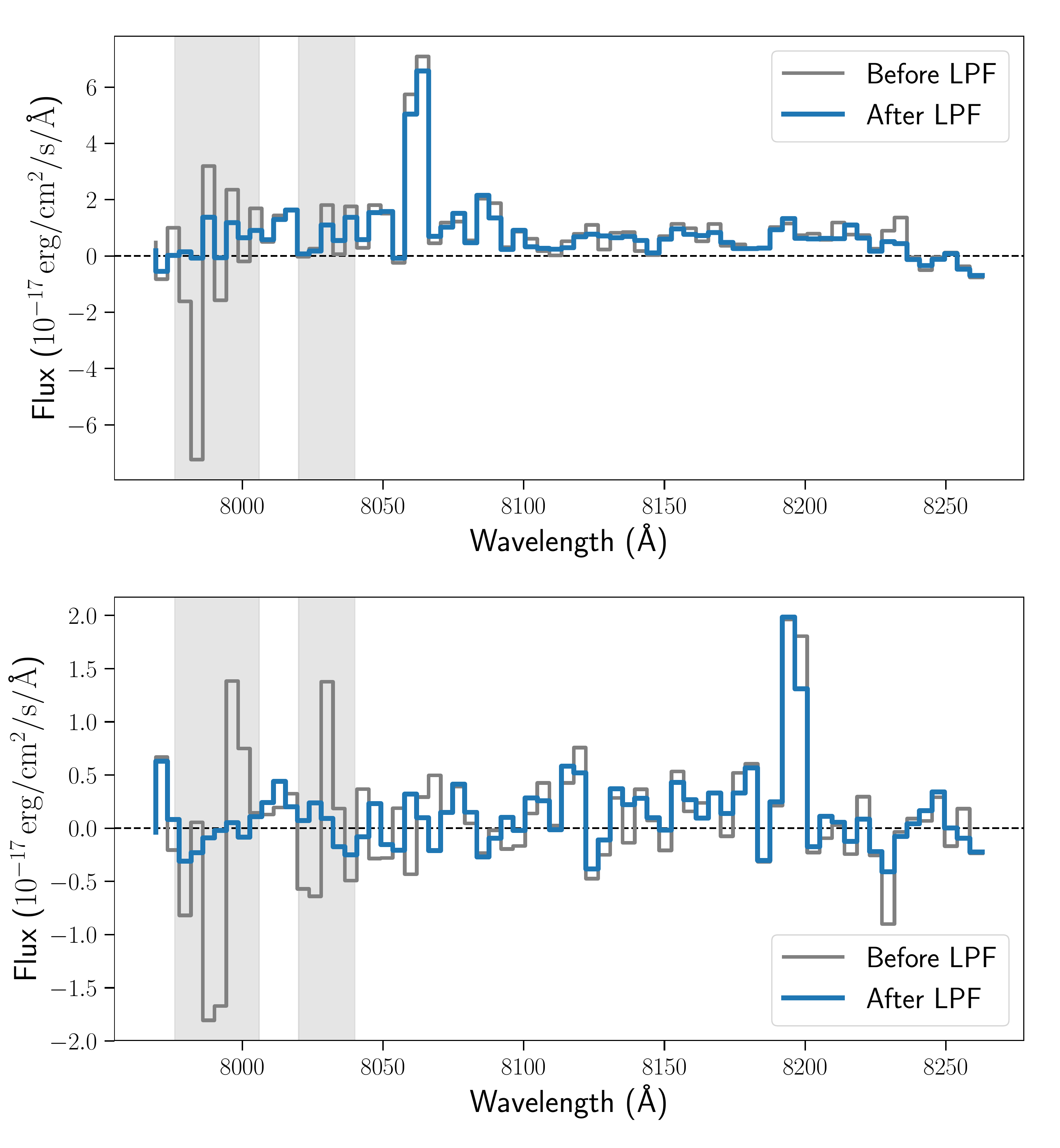}
      \caption{Sample spectra of two ELGs demonstrating the effectiveness of LPF in reducing additional noise caused by fringes (gray: before LPF; blue: after LPF) for channels around strong sky-lines (gray band). Upper: {\nii$\lambda\lambda$6548,6584} at z $\sim0.225$. Lower: a {\oiii$\lambda\lambda$4959,5007} doublet at z $\sim0.634$. The two spectra are identified and extracted from the A2390E field.}
      \label{fig:LPF_spec}
    \end{figure}
    
    After iterating through all the channels, a source detection is run again on the low-pass-filtered datacube, again with all of the detected sources masked. The convolve-subtract-detect-mask process is repeated until the change in the number of detected sources is smaller than 5\% in order to capture as many faint sources as possible. The efficacy of the LPF is clear by comparing the upper two panels of Figure \ref{fig:LPF_field} showing the same cutout of a channel that suffered from fringes before and after LPF, where contamination from fringes occurring in the left panel has been mitigated as shown in the right panel. Two sample spectra extracted in the A2390E field, which are ELGs identified using procedures in Section \ref{sec:ELG:CC}, are shown in Figure \ref{fig:LPF_spec} to demonstrate the effect of the LPF on the extracted spectra. The LPF processing can greatly reduce additional noise caused by fringes in channels between 7980 {\AA} and 8030 {\AA} where strong sky-lines are located, thus facilitating the detection and identification of faint emission-line candidates.
    
    Because of the finite kernel size used, in some channels there exist fringe residuals whose spatial variation is smaller than the kernel. This can be observed in the upper-right panel of Figure \ref{fig:LPF_field}. To further clean the residuals and facilitate candidate detection, we apply a moving average processing to construct a new detection datacube: the low-pass-filtered datacube is convolved with a 3x3x5 averaging kernel (2D spatial + 1D spectral), i.e., for a single spaxel its value is taking the mean of contiguous pixels in 2D and in the adjacent four channels. This takes advantage of the feature of the fringe pattern that it moves across the field as the scanning/wavelength proceeds/increases. As a result, residuals in nearby channels are expected to be more or less cancel out. The moving average processing also reduces the sky noise. Note this new datacube is only used for source detection, not for centroid measurement in Section \ref{sec:cen_measure}.

\section{Uncertainties in Centroid Analysis} \label{appendix:uncertainty}

    \subsection{Uncertainty Propagation from Centroids to Angles and Offsets} \label{appendix:uncertainty_prop}
    The uncertainties of flux-weighted centroids ($\bar{x}, \bar{y}$) are given by error propagation through
    \begin{align}
        \sigma_{\bar{x}}^2 &= \sum_i \sigma^2_{I_i} \left(\frac{\partial \bar{x}}{\partial I_i}\right)^2  = \sum_i \sigma^2_{I_i} \left[\frac{\partial }{\partial I_i}\left(\frac{\sum_j x_j I_j}{\sum_j I_j}\right)\right]^2 \nonumber \\
        &= \sum_i \frac{\sigma^2_{I_i}}{(\sum_j I_j)^2} \left[ -\frac{\sum_j x_j I_j} {\sum_j I_j} \frac{\partial (\sum_j I_j)}{\partial I_i} + \frac{\partial (\sum_j x_j I_j)}{\partial I_i} \right]^2 \, \nonumber \\
        &= \frac{\sum_i \sigma^2_{I_i} \cdot (x_i-\bar{x})^2}{({\sum_j I_j})^2}\,, \\
        \sigma_{\bar{y}}^2 &= \frac{\sum_i \sigma^2_{I_i} \cdot (y_i-\bar{y})^2}{(\sum_j I_j)^2}\,.
    \end{align}
    
    Let $\Delta \bar{x}=\bar{x}_E-\bar{x}_C$ and $\Delta \bar{y}=\bar{y}_E-\bar{y}_C$, the uncertainties in the offset and difference angle are given by error propagation through:
    \begin{align}
        \sigma_{\Delta d} &= \sigma \langle \left[ (\Delta \bar{x})^2 + (\Delta \bar{y})^2\right]^{\frac{1}{2}} \rangle \nonumber \\
        &= \left\{ \left[\frac{\Delta \bar{x} \cdot \sigma_{\Delta \bar{x}}}{\sqrt{(\Delta \bar{x})^2 + (\Delta \bar{y})^2}}\right]^2+\left[\frac{\Delta \bar{y} \cdot \sigma_{\Delta \bar{y}}}{\sqrt{(\Delta \bar{x})^2 + (\Delta \bar{y})^2}}\right]^2 \right\}^{\frac{1}{2}} \nonumber \\
        &= \sqrt{\left((\Delta \bar{x} \cdot \sigma_{\Delta \bar{x}})^2
        + (\Delta \bar{y} \cdot \sigma_{\Delta \bar{y}})^2\right)} \,/\, {\Delta d}\,,
    \end{align} and
    \begin{align}
        \sigma_{\theta_d} &= \sigma \langle \text{atan}(\Delta \bar{y}/\Delta \bar{x}) \rangle \nonumber \\ 
        &= \frac{\sigma \langle \Delta \bar{y}/\Delta \bar{x} \rangle }{(\Delta \bar{y}/\Delta \bar{x})^2+1} \nonumber \\ 
        &= (\Delta \bar{y}/\Delta \bar{x}) \cdot \frac{\sqrt{(\sigma_{\Delta \bar{y}}/\Delta \bar{y})^2+(\sigma_{\Delta \bar{x}}/\Delta \bar{x})^2}}{(\Delta \bar{y}/\Delta \bar{x})^2+1} \nonumber \\ 
        &= (\sqrt{(\Delta \bar{x} \cdot \sigma_{\Delta \bar{y}})^2 + (\Delta \bar{y} \cdot \sigma_{\Delta \bar{x}})^2})\,/\,{\Delta d}^2\,,
    \end{align}
    where $\sigma_{\Delta \bar{x}} = \sqrt{\sigma_{\bar{x}_E}^2+\sigma_{\bar{x}_C}^2}$ and $\sigma_{\Delta \bar{y}} = \sqrt{\sigma_{\bar{y}_E}^2+\sigma_{\bar{y}_C}^2}$.

    \subsection{An Empirical Assessment for Seeing-limited Centroid Measurements} \label{appendix:assessment}
    \textnormal{The measured centroid offsets in emission and continuum are in general small relative to the seeing and therefore suffer from smearing effects. However, it should be noted that the centroid of an object can be measured to a considerably higher precision than the seeing, depending on the S/N of the object. We perform several experiments with a control sample to demonstrate the robustness and effectiveness of the centroid measurements.} 
    
    \textnormal{The control sample is constructed using unsaturated stars cross-matched with PAN-STARRS (\citealt{2016arXiv161205560C}) and is matched with the ELG sample in terms of S/N as follows: we randomly draw a star from the crossmatch and choose a medium wide window in its spectrum with channels in it to represent the pseudo-emission. The rest of the channels are used as the continuum. Edges and channels in the presence of strong sky-lines are excluded. The pseudo-emission and continuum images are constructed from these channels with which centroids are measured in the same approach as Section \ref{sec:cen_measure} except that the continuum is not subtracted from the emission. This process is repeated for 500 times field by field to build a parent sample for each field. We then resample the measurements by their photometric S/N for $N=100$ times according to the distribution of S/N of the ELGs on the emission image. The sample size $N$ is chosen to match the average sample size of ELGs detected in each field for the statistical tests below. We construct 10 different control samples to account for sample variation.}
    
    \textnormal{With the control samples, we test whether there is systematic bias in the centroid measurement and quantify the uncertainties from random noise. Because the measurements are based on stars, the centroids should have no intrinsic offset unless there is any systematic bias.} 
    
    \textnormal{We first perform a Hotelling's $T^2$ test on the location of the distribution of centroid difference $\mathbf{x}=(\Delta x,\,\Delta y$) with the null hypothesis $\rm H_0$: $\mathbf{x}=\mathbf{x_0}=(0,0)$. The $t^2$ statistic is given by}
    \begin{align}
    t^{2}=n({\overline{\mathbf{x}}}-\mathbf{x_0})^{\prime} \hat{\mathbf{S}}^{-1}({\overline{\mathbf{x}}}-\mathbf{x_0}) \sim \chi_{p}^{2}
    \end{align}
    \textnormal{with $\overline{\mathbf{x}}$ representing the sample mean, $\hat{\mathbf{S}} =\frac{1}{N-1} \sum_{i=1}^{N}\left(\mathbf{x}_{i}-\overline{\mathbf{x}}\right)\left(\mathbf{x}_{i}-\overline{\mathbf{x}}\right)^{\prime}$ to be the sample covariance and $p=2$ to be the degree of freedom. The last condition in the equation above satisfies assuming the large sample approximation. If $t^2$ is larger than the critical value, the upper 1-$\alpha$ quantile $\chi_{p}^{2}(\alpha)$, $\rm H_0$ is then rejected at confidence level of $\alpha$. With the computed $t^2$ in each field, we cannot reject $\rm H_0$ at confidence level of $\alpha=0.01$ in 9/8/9/9 out of 10 control samples for A2390C/A2390E/A2390W/A2465C, that is, no significant systematic bias is found for the centroid measurements.}
    
    \begin{figure}
      \centering
      \includegraphics[width=\hsize]{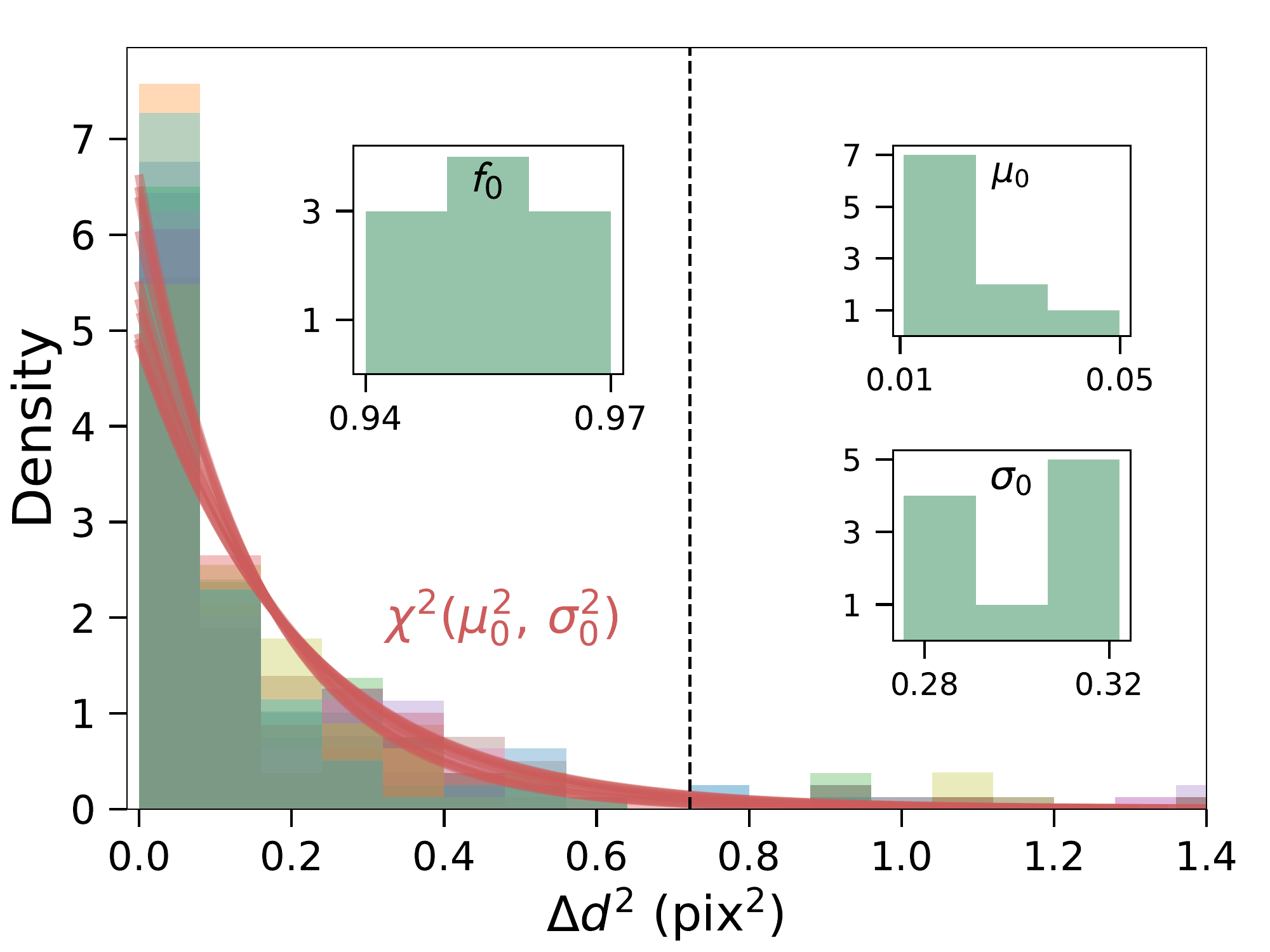}
      \caption{Distributions of $(\Delta d)^2$ measured from 10 control samples made up of stars in the A2390C field. Red curves show the best fits to $\chi^{2}$ distributions. Distributions of fitted centers, standard deviations, and fractions of measurements within the adopted threshold are displayed in small panels.}
      \label{fig:centroid_test_A2390C}
    \end{figure}
    
    \textnormal{The other test is motivated by the fact that ${(\Delta d)}^2=(\Delta x)^2+(\Delta y)^2 \sim \chi_{p=2}^{2}$ under normality and independence assumptions for $\Delta x$ and $\Delta y$. We then fit a $\chi^{2}(\mu_0^2, \sigma_0^2)$ distribution for ${(\Delta d)}^2$ in each field with the center $\mu_0$ and standard deviation $\sigma_0$ indicating the overall bias and degree of deviation. Figure \ref{fig:centroid_test_A2390C} displays the distributions of $(\Delta d)^2$ and their fits (red line) out of the 10 control samples for the A2390C field as an example. $\mu_0$ are close to 0, which further proves the small systematic bias, and $\sigma_0$ are at least two times smaller than the adopted threshold. The contamination level can also be revealed by the fraction of measurements of offset below the threshold (black dashed line), $\mathfrak{f_0}$, where on average $\sim95\%$ of measurements fall within the threshold. The average $\mathfrak{f_0}$ is 93\%/90\%/93\% for A2390E/A2390W/A2465C.}
    
    \textnormal{In summary, we conclude that although our centroid measurements are seeing-limited, the systematic bias and random uncertainties in the process of measurement are small and do not impact our conclusions.}

\section{Results Using Morphological Centroids}
\label{appendix:res_morph_centroid}

    \begin{figure}
      \centering
      \includegraphics[width=\hsize]{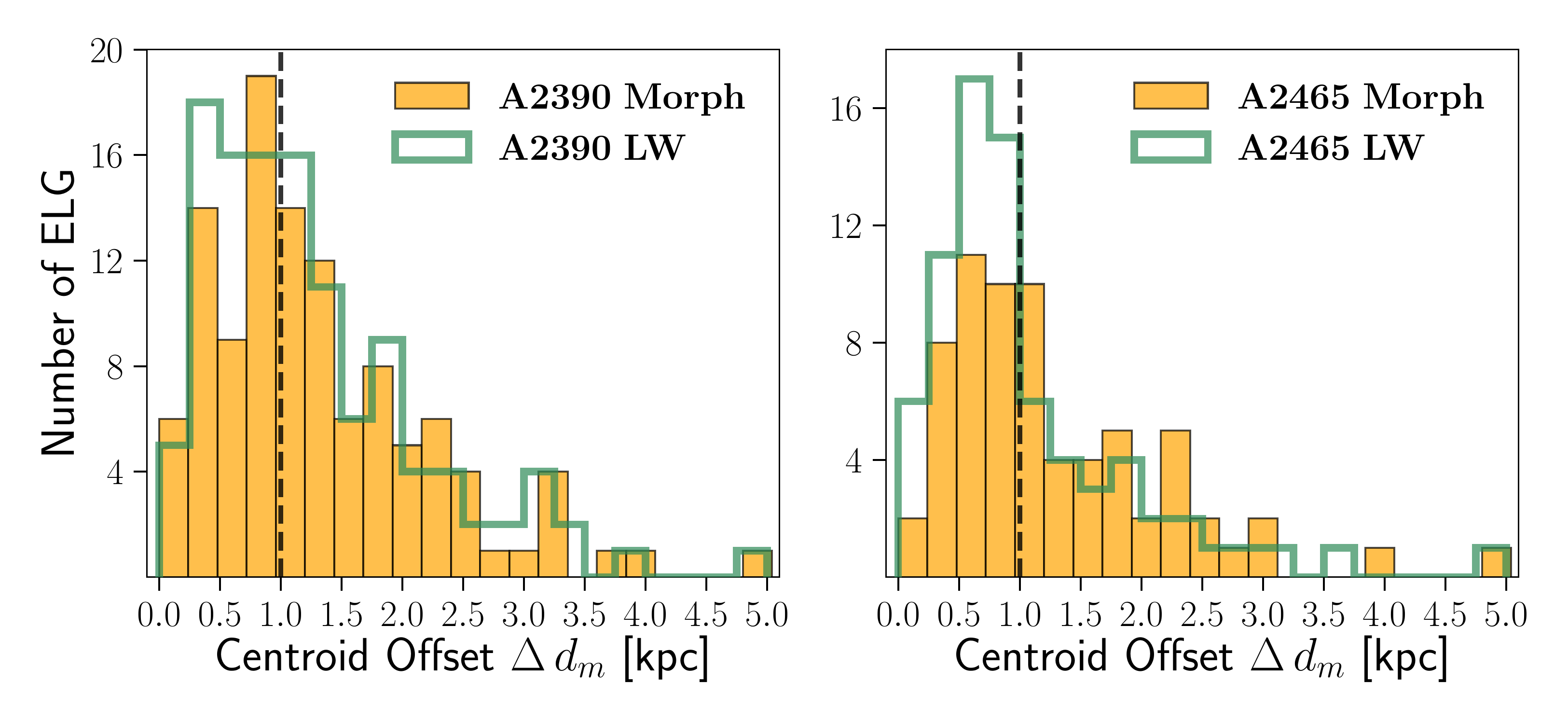}
      \caption{Distribution of morphological centroid offset $\Delta\,d_m$ between the emission and the stellar continuum for ELGs in A2390 (combining three fields, left) and A2465C (right). The distributions of $\Delta\,d$ from light-weighted centroids are overplotted as green outlines. The 1kpc threshold used is marked as the black dashed line in each panel.}
      \label{fig:centroid_offset_morph}
    \end{figure}
        
    \begin{figure}
      \centering
      \includegraphics[width=\hsize]{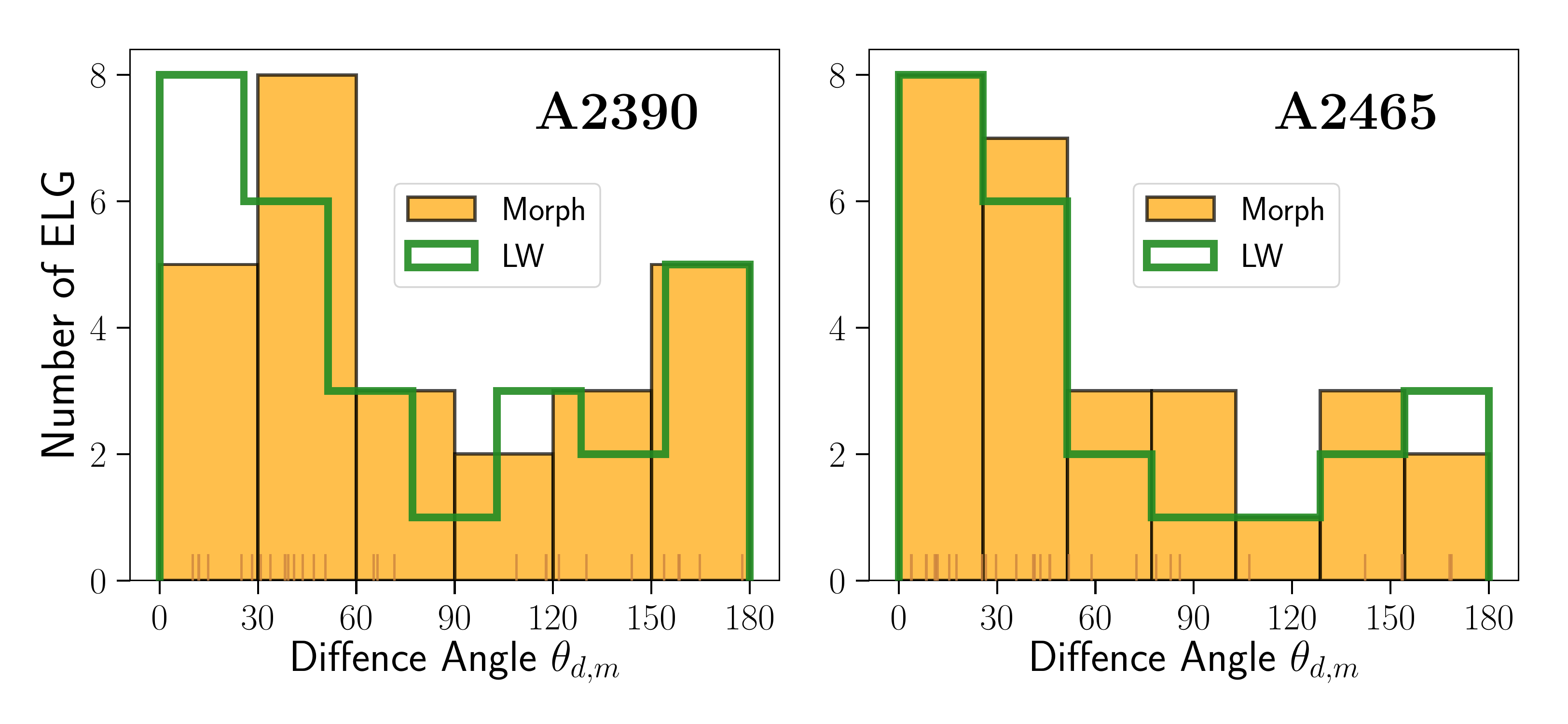}
      \caption{Histogram of difference angle measured from morphological centroids ($\theta_{d,m}$) for cluster member ELGs in A2390 fields (left) and A2465C (right) with $\Delta\,d_m$ $>$ 1 kpc. The bin number is slightly adjusted for A2390 to match the y scale. Individual data points are shown as small sticks at the bottoms of the histograms. The distributions of $\theta_{d}$ from light-weighted centroids are overplotted as green outlines.}
      \label{fig:diff_angle_hist_morph}
    \end{figure}

    In the main text, the emission and continuum centroids are measured in a light-weighted procedure. Alternatively, we can measure the morphological centroids of the output segmentation of emission and continuum by simply giving equal weights to all the pixels within the border:
    \begin{equation}
        ({\bar{x}}_\mathrm{m}, {\bar{y}}_\mathrm{m}) = \left(\frac{1}{N}\sum_i\,x_i, \, \frac{1}{N}\sum_i\,y_i\right)\,,
    \end{equation}
    where $N$ is the number of summed pixels. This is motivated by attempting to give higher weights to the weaker but more spatially distorted ionized gas, e.g. gas tails at the outskirts of the ELG. We use $\sigma_{{\bar{x}}_m}^2=\sigma_{{\bar{y}}_m}^2$ = $\frac{1}{N}$ as an empirical error. The propagation to $\Delta d_m$ and $\theta_{d,m}$ is the same as in Appendix \ref{appendix:uncertainty_prop}. Because the morphological centroids are sensitive to the way how segmentation is performed and they do not have well-defined errors, our preference is to use the light-weighted centroids for our analysis; however, the morphological centroids produce essentially the same conclusions.
    
    We can measure the difference vector $\mathbf{d}$ using the morphological centroids, and accordingly measure the centroid offset $\Delta d_m$ and difference angle $\theta_{d,m}$ in the same manner as in Section \ref{sec:cen_measure}. We show the distributions of $\Delta d_m$ in Figure \ref{fig:centroid_offset_morph}. The distributions of centroid offsets are similar but slightly more skewed to larger values in both clusters. This is as expected as morphological centroids put more stress on the faint outskirts. 
    
    Another selected sample is generated following the same criteria in Section \ref{sec:cen_measure} except for requiring $\Delta\,d_m > 3\,\sigma_{\Delta d_m}$, with 49 ELGs obtained for A2390 and 28 ELGs obtained for A2465. The distributions of $\theta_{d,m}$ are shown in Figure \ref{fig:diff_angle_hist_morph}. Similar to the ones in Figure \ref{fig:diff_angle_hist}, they also visually deviated strongly from a uniform distribution. The difference lies in that the major peak being shifted to a higher value ($30^\circ - 60^\circ$) in A2390, which can be inferred as a result of more influence from the tangential component of the ram pressure. Furthermore, the minor peak toward $180^\circ$ in A2465 is no longer significant. The K-S test on A2465 rejects the uniformity at 1\% confidence level ($p=0.009$), although it cannot reject the uniformity of $\theta_{d,m}$ in A2390 ($p=0.15$). Finally we inspect their distribution in phase space diagram and their continuum normalized emission-line flux, which are consistent with the results using light-weighted centroids.
    
    Overall, the results of morphological centroids also support the scenario where ram pressure stripping takes effect during the infall of gas-rich cluster galaxies and shuts down their star formation through removal of gas.

\end{document}